\documentclass[acmsmall]{acmart}

\AtBeginDocument{%
  \providecommand\BibTeX{{%
    \normalfont B\kern-0.5em{\scshape i\kern-0.25em b}\kern-0.8em\TeX}}}

\acmJournal{JACM}

\usepackage{algorithm}
\usepackage{algorithmicx}
\usepackage{amsmath,amsfonts}
\usepackage{array}
\usepackage{balance}
 \usepackage{booktabs}
\usepackage[skip=1pt,labelfont=bf]{caption}
 \usepackage{calligra}
\usepackage{color}
\usepackage{colortbl}
\usepackage{courier}
\usepackage{csvsimple}
\usepackage{enumitem}
\usepackage{fancybox}
\usepackage{fontenc}
\usepackage{graphicx}
\usepackage{listings}
\usepackage{longtable}
\usepackage{lscape}
\usepackage{makecell}
\usepackage{marvosym}
\usepackage{moreverb}
\usepackage{multicol}
\usepackage{multirow}
\usepackage{pifont}%
\usepackage{rotating}
\usepackage{setspace}
\usepackage{subfigure}
\usepackage[most]{tcolorbox}
\usepackage{threeparttable}
\usepackage{tikz}
\usepackage[normalem]{ulem}
\usepackage{url}
\usepackage{soul}
\usepackage{wasysym}
\usepackage{xspace}

\algnewcommand\algorithmicforeach{\textbf{for each}}
\algdef{S}[FOR]{ForEach}[1]{\algorithmicforeach\ #1\ \algorithmicdo}

\newcolumntype{L}[1]{>{\raggedright\let\newline\\\arraybackslash\hspace{0pt}}m{#1}}
\newcolumntype{C}[1]{>{\centering\let\newline\\\arraybackslash\hspace{0pt}}m{#1}}
\newcolumntype{R}[1]{>{\raggedleft\let\newline\\\arraybackslash\hspace{0pt}}m{#1}}

\definecolor{codegreen}{rgb}{0,0.6,0}
\definecolor{codered}{rgb}{1,0,0}
\definecolor{codegray}{rgb}{0.5,0.5,0.5}
\definecolor{codepurple}{rgb}{0.58,0,0.82}
\definecolor{backcolour}{rgb}{0.95,0.95,0.92}
\definecolor{lightgray}{gray}{0.9}

\lstdefinestyle{mystyle}{
    commentstyle=\color{codegreen},
    keywordstyle=\color{magenta},
    numberstyle=\small\color{black},
    stringstyle=\color{codepurple},
    basicstyle=\scriptsize\ttfamily,
    breakatwhitespace=false,
    breaklines=true,
    captionpos=b,
    keepspaces=true,
    showspaces=false,
    showstringspaces=false,
    showtabs=false,
    tabsize=2
}

\lstdefinelanguage{diff}{
  morecomment=[f][\color{blue}]{@@},     %
  morecomment=[f][\color{red}]-,         %
  morecomment=[f][\color{codegreen}]+,       %
  morecomment=[f][\color{red}]{---}, %
  morecomment=[f][\color{codegreen}]{+++},
}

\setlist{noitemsep} %

\definecolor{darkpastelred}{rgb}{0.76, 0.23, 0.13}
\definecolor{ao(english)}{rgb}{0.0, 0.5, 0.0}

\definecolor{darkpastelred}{rgb}{0.76, 0.23, 0.13}
\definecolor{ao(english)}{rgb}{0.0, 0.5, 0.0}

\definecolor{yellow}{RGB}{255,255,153}
\definecolor{grey}{RGB}{224,224,224}

\newboolean{showcomments}
\setboolean{showcomments}{true}
\ifthenelse{\boolean{showcomments}}
 { \newcommand{\mynote}[2]{
      \fbox{\bfseries\sffamily\scriptsize#1}
        {\small$\blacktriangleright$\textsf{\emph{#2}}$\blacktriangleleft$}}}
        { \newcommand{\mynote}[2]{}}

\definecolor{DarkOrange}{rgb}{0.8,0.3,0.0}
\definecolor{DarkCyan}{rgb}{0.0, 0.55, 0.55}
\definecolor{DarkCyel}{rgb}{1.0, 0.49, 0.0}
\definecolor{yellow-green}{rgb}{0.6, 0.8, 0.2}

\newcolumntype{?}{!{\vrule width 1pt}}

\newcommand{\toolname}{\textsc{Leopard}\xspace}
\newcommand{\tool}{\textsc{Panther}\xspace}

\newcommand{\embeddings}{learned embeddings\xspace}

\newcommand{\find}[1]{
\begin{tcolorbox}[leftrule=1mm,toprule=0mm,bottomrule=0mm,left=1pt,right=2pt,top=2pt,bottom=2pt]%
\em #1
\end{tcolorbox}
}

\RequirePackage[normalem]{ulem} %
\RequirePackage{color}\definecolor{RED}{rgb}{1,0,0}\definecolor{BLUE}{rgb}{0,0,1} %

\begin{document}
\title{The Best of Both Worlds: Combining Learned Embeddings with Engineered Features for Accurate Prediction of Correct Patches}

\author{Haoye Tian}
\email{haoye.tian@uni.lu}
\affiliation{%
   \institution{University of Luxembourg}
 	\country{Luxembourg}
}

\author{Kui Liu}\authornote{Corresponding author.}
\email{brucekuiliu@gmail.com}
\affiliation{%
   \institution{Huawei}
 	\country{China}
}

\author{Yinghua Li}
\email{yinghua.li@uni.lu}
\affiliation{%
   \institution{University of Luxembourg}
 	\country{Luxembourg}
}

\author{Abdoul Kader Kaboré}
\email{abdoulkader.kabore@uni.lu}
\affiliation{%
   \institution{University of Luxembourg}
 	\country{Luxembourg}
}

\author{Anil Koyuncu}
\email{anil.koyuncu@sabanciuniv.edu}
\affiliation{%
 \institution{Sabanci University}
 \city{Istanbul}
 \country{Turkey}
}

\author{Andrew Habib}
\email{andrew.a.habib@gmail.com}
\affiliation{%
   \institution{University of Luxembourg}
 	\country{Luxembourg}
}

\author{Li Li}
\email{li.li@monash.edu}
\affiliation{%
   \institution{Monash University}
   \country{Australia}
}

\author{Junhao Wen}
\email{jhwen@cqu.edu.cn}
\affiliation{%
   \institution{Chongqing University}
   \country{China}
}

\author{Jacques Klein}
\email{jacques.klein@uni.lu}
\affiliation{%
   \institution{University of Luxembourg}
 	\country{Luxembourg}
}
\author{Tegawend\'e F. Bissyand\'e}
\email{tegawende.bissyande@uni.lu}
\affiliation{%
  \institution{University of Luxembourg}
  \country{Luxembourg}
}

\renewcommand{\shortauthors}{Tian and Liu et al.}
\renewcommand{\shorttitle}{Combining Learned Embeddings with Engineered Features for Accurate Prediction of Correct Patches}

\begin{abstract}
A large body of the literature on automated program repair develops approaches where patches are automatically generated to be validated against an oracle (e.g., a test suite). 
Because such an oracle can be imperfect, the generated patches, although validated by the oracle, may actually be incorrect. 
While the state of the art explores research directions that require dynamic information or rely on manually-crafted heuristics, we study the benefit of learning code representations in order to learn deep features that may encode the properties of patch correctness.
Our empirical work investigates different representation learning approaches for code changes to derive embeddings that are amenable to similarity computations of patch correctness identification,
and assess the possibility of accurate classification of correct patch by combining learned embeddings with engineered features.
Experimental results demonstrate the potential of learned embeddings to empower \toolname (a patch correctness predicting framework implemented in this work) with learning algorithms in reasoning about patch correctness: a machine learning predictor with BERT transformer-based learned embeddings associated with XGBoost achieves an AUC value of about 0.803 in the prediction of patch correctness on a new dataset of 2,147 labeled patches that we collected for the experiments. 
Our investigations show that deep learned embeddings can lead to complementary/better performance when comparing against the state-of-the-art, PATCH-SIM, which relies on dynamic information. 
By combining deep learned embeddings and engineered features, \tool (the upgraded version of \toolname implemented in this work) outperforms \toolname with higher scores in terms of AUC, +Recall and -Recall, and can accurately identify more (in)correct patches that cannot be predicted by the classifiers only with learned embeddings or engineered features.
Finally, we use an explainable ML technique, SHAP, to empirically interpret how the learned embeddings and engineered features are contributed to the patch correctness prediction. 
\end{abstract}

\begin{CCSXML}
<ccs2012>
<concept>
<concept_id>10011007.10011074.10011099</concept_id>
<concept_desc>Software and its engineering~Software verification and validation</concept_desc>
<concept_significance>500</concept_significance>
</concept>
<concept>
<concept_id>10011007.10011074.10011099.10011102</concept_id>
<concept_desc>Software and its engineering~Software defect analysis</concept_desc>
<concept_significance>300</concept_significance>
</concept>
<concept>
<concept_id>10011007.10011074.10011099.10011102.10011103</concept_id>
<concept_desc>Software and its engineering~Software testing and debugging</concept_desc>
<concept_significance>100</concept_significance>
</concept>
</ccs2012>
\end{CCSXML}

\ccsdesc[500]{Software and its engineering~Software verification and validation}
\ccsdesc[300]{Software and its engineering~Software defect analysis}
\ccsdesc[100]{Software and its engineering~Software testing and debugging}

\keywords{
Program Repair, Patch Correctness, Distributed Representation Learning, Machine Learning, Embeddings, Features Combination, Explanation
}

\maketitle

\section{Introduction}
\label{sec:intro}

Automatic program repair (APR)~\cite{le2019automated,monperrus2018automatic,liu2021critical}, the process of fixing software bugs automatically, has gained a huge momentum with the ever increasing pervasiveness of software.
While a few APR techniques try to model program semantics and synthesize execution constraints towards producing correct-by-construction patches, they often fail to scale to large programs.
Instead, the large majority of APR research~\cite{monperrus2018living} focuses on generate-and-validate approaches where patch candidates are generated and then validated against an oracle.

In the absence of precise program specifications, test suites provide affordable approximations that are widely used as the oracle in APR. 
In their seminal work on test-based APR, Weimer~{\em et~al.}~\cite{weimer2009automatically} consider that a patch is acceptable as soon as the patched program passes all test cases in the given test suite. 
Since then, a number of studies~\cite{qi2015analysis,smith2015cure} have explored the {\em overfitting problem} in patch validation: an automatically generated patch makes the buggy program pass a given test suite and yet it is incorrect w.r.t. the intended program specification.
Since test suites only weakly approximate program specifications, a patched program can indeed satisfy the requirements encoded in the test suite yet present a behavior that deviates from what is expected by the developer but not specified in the existing test suite.

Overfitting patches constitute a key challenge in generate-and-validate APR approaches. Recent studies~\cite{liu2019avatar,liu2019tbar,jiang2018shaping,saha2019harnessing,wen2018context,liu2018lsrepair,koyuncu2019ifixr,liu2019you,koyuncu2020fixminer,wang2020automated,liu2020efficiency} on APR systems highlight the importance of estimating the correct ratio among the valid patches that can be found. To improve this ratio of correct patches, researchers explore several directions which we categorize in three groups depending on when the processing to detect correct patches is applied: before, during, or after patch generation:

\begin{enumerate}%
	\item {\em Test-suite augmentation:} Yang~{\em et~al.}~\cite{yang2017better} proposed to generate better test cases to enhance the validation of patches, while Xin and Reiss~\cite{xin2017identifying} opted for increasing test inputs. 
	\item {\em Curation of repair operators:} approaches such as CapGen~\cite{wen2018context} demonstrate that carefully-designed repair operators (e.g., fine-grained fix ingredients) can lead to correct patches.
	\item {\em Post-processing of generated patches:} Long and Rinard~\cite{long2016automatic} introduced some heuristics to discard patches that are likely overfitting.
\end{enumerate}

So far, the state-of-the-art works targeting the identification of patch correctness are based on computing the similarity of test case execution traces~\cite{xiong2018identifying}, or using machine learning to identify correct patches based on engineered static code 
features~\cite{ye2019automated}, pre-trained natural language-based embeddings~\cite{csuvik2020utilizing}, and source code trained embeddings~\cite{tian2020evaluating}.

{\bf This paper.}
In this work, we extensively study and evaluate how effective are source code embeddings and engineered features in predicting correct patches.
For example, which set of features: engineered or learned embeddings yield better performance in predicting correct patches? Can a combination of both kinds of feature achieve higher performance?
Our work fills this gap.   

This work builds on and extends our previous work~\cite{tian2020evaluating} in the following manner:
\begin{itemize}
    \item We examine and compare the effectiveness of \emph{code embeddings}, \emph{engineered features}, and their combination for predicting patch correctness.
    \item We present an analysis for detecting which kinds of features contribute to the (in)correct prediction of patch correctness.
\end{itemize}

We investigate in this paper the feasibility of leveraging advances in deep representation learning to produce embeddings for APR-generated patches and their engineered features, that are amenable to reasoning about correctness. 
\begin{itemize}[leftmargin=*]
	\item[\ding{182}] We investigate different representation learning models adapted to natural language tokens and source code tokens that are more specialized to code changes. Our study considers both pre-trained models and the retraining of models.
	\item[\ding{183}] We empirically investigate whether, with learned embeddings, the hypothesis of minimal changes incurred by correct patches remains valid: experiments are performed to check the statistical difference between similarity scores yielded by correct patches and those yielded by incorrect patches. 
	\item[\ding{184}] We run exploratory experiments assessing the possibility to select cutoff similarity scores between learned embeddings of buggy code and patched code fragments for heuristically filtering out incorrect patches.
	\item[\ding{185}] We investigate the discriminative power of learned embeddings in a classification training pipeline (that we named \toolname) aimed at learning to predict patch correctness with learned embeddings. We evaluate our and state of the art approaches by applying a 10-group cross validation in a practical perspective. Comparing against the state of the art, \toolname is complementary to them, even outperforms them on filtering out incorrect patches.
	\item[\ding{186}] We explore the combination of the learned embeddings and the engineered features to improve the performance on identifying patch correctness with more accurate classification, and implement an upgraded version of \toolname, that we named \tool. The exploring examination is supported by our experimental results.
	\item[\ding{187}] We empirically interpret the cause of prediction behind features and classifiers to help aware the essence of identifying patch correctness with an explainable ML technique SHAP.
\end{itemize}

The remainder of this paper is organized as follows. Section~\ref{sec:bg} provides the background of our work, Section~\ref{sec:method} introduces our methodology and study design, Sections~\ref{sec:results} and~\ref{sec:discussion} cover the experimental results and a discussion, and Sections~\ref{sec:relatedWork} and~\ref{sec:conc} discuss related work and conclude the paper.
\section{Background}
\label{sec:bg}
This works leverages learning representation and and machine learning techniques to tackle the problem of identifying correct patches among incorrect and plausible APR-generated patches.
Additionally, we examine the explainability of ML models used to predict correct patches. 
The explainability aspect is of high importance to developers applying APR in their workflow.
Therefore, we begin by providing the necessary background of the four pillars of our work: 
(i) patch correctness, 
(ii) representation learning for code,
(iii) engineered features for predicting patch correctness, and
(iv) the explainability of ML models using SHAP.

\subsection{Patch Plausibility and Correctness}
Defining patch correctness is a non-trivial challenge in the APR community. 
Until the release of empirical investigations by Smith~{\em et~al.}~\cite{smith2015cure}, actual correctness (w.r.t. the intended behavior of program) was seldom used as a performance criterion of APR systems. 
Instead, experimental results were focused on the number of patches that make the program pass all test cases. Such patches are actually only {\bf plausible}. 
Qi~{\em et~al.}~\cite{qi2015analysis} demonstrated that an overwhelming majority of plausible patches generated by GenProg~\cite{le2012genprog}, RSRepair~\cite{qi2014strength} and AE~\cite{weimer2013leveraging} are overfitting the test suite while actually being incorrect.
To improve the practicability of APR systems to generate {\bf correct} patches, researchers have mainly invested in strengthening the validation oracle (i.e., the test suites).
Opad~\cite{yang2017better}, DiffTGen~\cite{xin2017identifying}, UnsatGuided~\cite{yu2019alleviating}, PATCH-SIM/TEST-SIM~\cite{xiong2018identifying} generate new test inputs that trigger behavior cases which are not addressed by APR-generated patches.

More recent works~\cite{ye2019automated,csuvik2020utilizing} are starting to investigate static features and heuristics (or machine learning) to build predictive models of patch correctness. Ye~{\em et~al.}~\cite{ye2019automated} presented the ODS approach which relates to our study since it investigated machine learning with static features (i.e., carefully hand-crafted features~\cite{ye2019automated}) extracted from Java program patches. 
The study of Csuvik~{\em et~al.}~\cite{csuvik2020utilizing} is also closely related to ours since it explores BERT embeddings to define similarity thresholds between buggy and patched code. Their work however remains preliminary (it does not investigate the discriminative power of features) and has been performed at a very small scale (single pre-trained model on 40 one-line bugs from simple programs).

\subsection{Distributed Representation Learning}
Learning distributed representations have been widely used to advance several machine learning-based software engineering tasks~\cite{feng2020codebert, devlin2019bert, pian2022metatptrans, zhao2018deepsim, fang2020functional}. In particular, embedding techniques such as  {\bf Word2Vec}~\cite{le2014distributed}, {\bf Doc2Vec}~\cite{le2014distributed} and {\bf BERT}~\cite{devlin2019bert} have been successfully applied to different semantics-related tasks such as code clone detection~\cite{wei2017supervised}, vulnerability detection~\cite{ndichu2019machine}, 
code recommendation~\cite{zhou2019lancer}, and commit message generation~\cite{hoang2020cc2vec}.

By building on the hypothesis of code naturalness~\cite{hindle2012naturalness,allamanis2018survey}, a number of software engineering research works have also leveraged the aforementioned approaches for learning distributed representations of code~\cite{liu2018mining2,liu2019learning}. 
Alon~{\em et~al.}~\cite{alon2019code2vec} have then proposed {\bf code2vec}, an embedding technique that explores AST paths to take into account structural information in code. 
More recently, Hoang~{\em et~al.}~\cite{hoang2020cc2vec} have proposed {\bf CC2Vec}, which further specializes to code changes.
Our work explores different techniques across the spectrum of distributed representation learning. We therefore consider four variants from the seemingly-least specialized to code (i.e., Doc2Vec) to the state of the art for code change representation (i.e., CC2Vec).

{\bf Doc2Vec}~\cite{le2014distributed} is an unsupervised framework mostly used to learn continuous distributed vector representations of sentences, paragraphs and documents, regardless of their lengths.
It works on the intuition, inspired by the method of learning word vectors~\cite{mikolov2013efficient}, that the document representation should be good enough to predict the words in the document.
Doc2Vec has been applied in various software engineering tasks.
For example, 
Wei and Li~\cite{wei2017supervised} leveraged Doc2Vec to exploit deep lexical and syntactical features for software functional clone detection.
Ndichu~{\em et~al.}~\cite{ndichu2019machine} employed Doc2Vec to learn code structure representation at AST level to predict JavaScript-based attacks.

{\bf BERT}~\cite{devlin2019bert} is a language representation model that has been introduced by an AI language team in Google.
BERT is devoted to pre-train deep bidirectional representations from unlabelled texts. 
Then a pre-trained BERT model can be fine-tuned to accomplish various natural language processing tasks such as question answering or language inference.
Zhou~{\em et~al.}~\cite{zhou2019lancer} employed a BERT pre-trained model to extract deep semantic features from code name information of programs in order to perform code recommendation.
Yu~{\em et~al.}~\cite{yu2020order} even leveraged BERT on binary code to identify similar binaries.

{\bf code2vec}~\cite{alon2019code2vec} is an attention-based neural code embedding model developed to represent code fragments as continuous distributed vectors, by training on AST paths and code tokens. 
 Its embeddings have notably been used to predict the semantic properties of code fragments~\cite{alon2019code2vec}, in order, for instance, to predict method names.  
Compton~{\em et~al.}~\cite{compton2020embedding} recently leveraged code2vec to embed Java classes and learn code structures for the task of variable naming obfuscation.

{\bf CC2Vec}~\cite{hoang2020cc2vec} is a specialized hierarchical attention neural network model which learns vector representations of code changes (i.e., patches)  
guided by the associated commit messages (which is used as a semantic representation of the patch). 
As the authors demonstrated in their large empirical evaluation, CC2Vec presents promising performance on commit message generation, bug fixing patch identification, and just-in-time defect prediction.

\subsection{Engineered Features}%
\label{sec:engineeredFeature}

Engineered features are carefully designed and selected features which represent and capture important properties of the underlying data.
In APR, one possibility is to statically extract those features from the abstract syntax tree (AST) of the buggy code, the AST of the patched path and the related AST edit scripts as proposed by ODS~\cite{ye2019automated}.

ODS extract three kinds of features to detect correct patches:
(i) Code description features, e.g., kinds of specific operators in patch code and kinds of statements, 
(ii) Repair pattern features, whether the repair code has specific patterns according to~\cite{Madeiral2018}, and 
(iii) Contextual syntactic features, e.g., the types of faulty statements and the types of their surrounding statements.
Using these engineered features, ODS trains a series of machine learning classifiers to predict patch correctness. The experimental evaluation on 713 patches shows that ODS can filter out 57\% of overfitting patches and exhibits competitive results when compare with state of the art. 
We adopt ODS engineered features to conduct our study.
Because ODS can not steadily generate all the originally designed engineered features in their research for our patches, we consider to mainly use, in our study, two kinds of engineered features generated by ODS\footnote{We have received the confirmation from the authors about this bug and the effectiveness of these two kinds of features.}: (1) Prophet features (i.e., the re-implementation of Long~{\em et~al.}s' work~\cite{long2016automatic}) and (2) the repair pattern (the related operations of transforming the buggy code to patched code).

\subsection{SHAP - SHapley Additive exPlanations}
SHAP is a unified framework proposed by Lundberg~{\em et~al.}~\cite{NIPS2017_7062} to interpret the output of machine learning models. 
It connects optimal credit allocation with local explanations using the classic Shapley values from the game theory and their related extensions,
thus can provide the importance of each feature for certain particular prediction.
Through SHAP, the positive and negative effect of features on prediction can be generated, which allow practitioners to understand which behaviors lead to the (in)correct prediction. 
Besides, SHAP provides the interaction analysis between features to explore how different features are complementary to each other.
\section{Methodology}
\label{sec:method}

In this section, we first present the methodology of our study and then we introduce the research questions that we aim to answer using the proposed methodology.

\begin{figure}[t]
	\includegraphics[width=1\columnwidth]{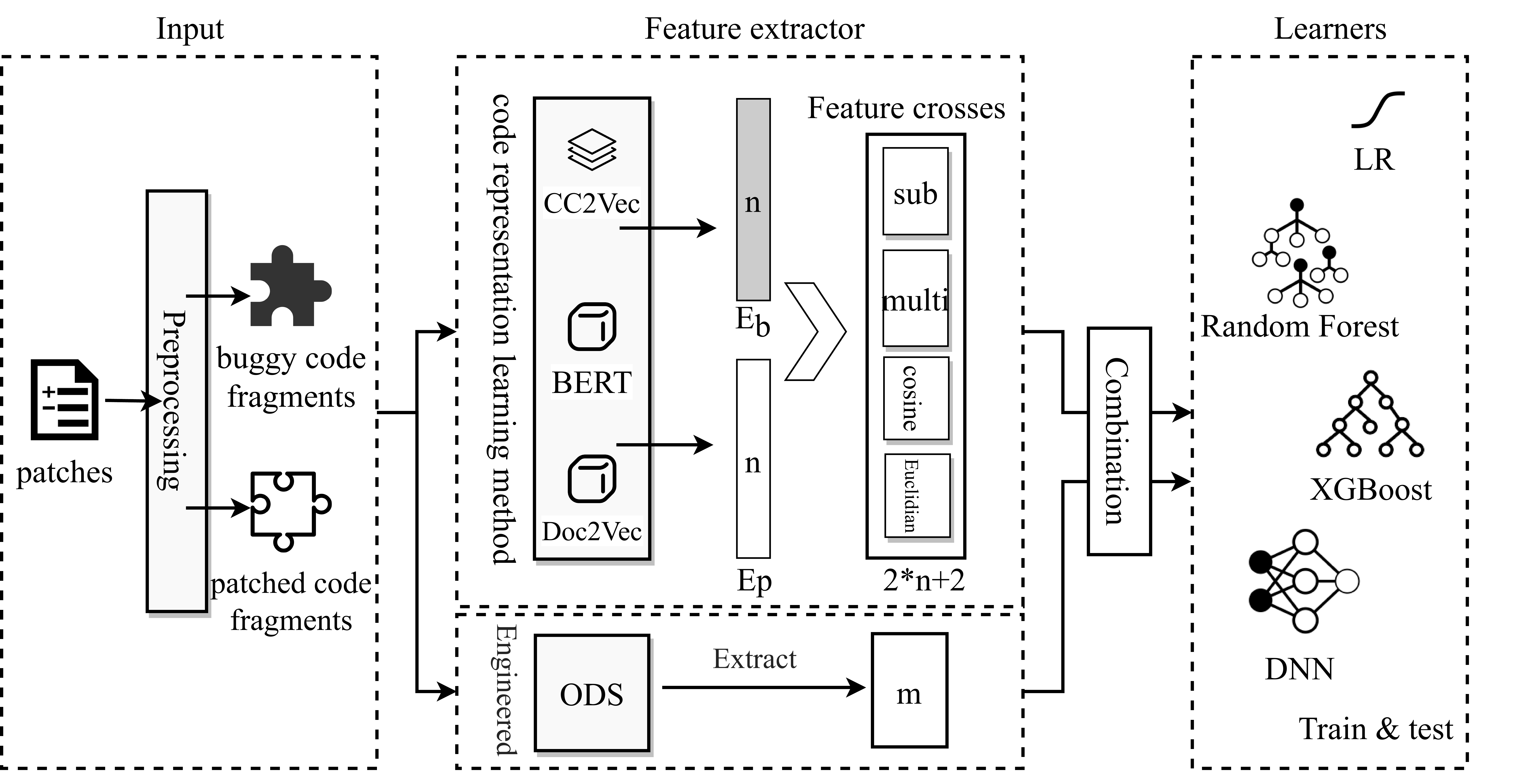}
	\caption{Overview of \tool.}
	\label{fig:pipeline}
\end{figure}

Overall, our goal is to study the effectiveness of different representations of APR-generated patches and codes for the task of predicting which patches are correct. We first investigate a widespread hypothesis that a patch incurring minimal changes is more likely to be correct. To quantify the patch changes, we exploit different code representation learning methods that leverage deep learning techniques to learn features for code. We adapt them to generate the vectors of buggy code and patched code as well as compute the similarity value of vectors. Based on the similarity distribution, we experimentally filter out incorrect APR-generated patches by relying on naively-defined thresholds. 

In the view of learning representation reveals the properties of code related to patch correctness, we propose to further identify patch by training classifiers (learners) on the representation vector of a patch. Figure~\ref{fig:pipeline} provides an overview of such a pipeline and its variants. To represent patches in a format suitable for learning algorithms, we use the aforementioned representation learning methods to generate vectors for buggy code and patched code. Afterwards, we cross the vectors by applying subtraction, multiplication, cosine similarity and euclidean similarity to obtain the deep learned feature of the patches. The resulting patch embedding has 2*n+2 dimensions where n is the dimension of input code fragment embeddings. The values of the dimension n for BERT, Doc2Vec and CC2Vec are set as 1024, 64 and 64, respectively. On the other hand, we also exploit the manually engineered features that are extracted from the given data, the patch in our case, and aim to capture specific information that is thought to be relevant to the patch correctness. The dimension m for ODS is 195. 

Learned and engineered features represent a patch from different perspectives. To improve the identification performance of patch correctness, we further propose three methods ({\em Ensemble learning}, {\em Na{\"i}ve Vector Concatenation}, and {\em Deep Combination.}) to combine the two features for obtaining the informative representation of a patch. After obtaining a vector that represents a given patch, different machine learning algorithms such as random forest or a deep neural network (DNN) are trained as classifiers that distinguish correct from incorrect APR patches. In the end, we provide the SHAP explanation of the features and interaction of different features that contribute to the patch correctness prediction. Our study follows the four parts below:
\begin{enumerate}%
    \item The empirical study of a hypothesis in filtering out incorrect patches (RQs 1 to 2),
    \item The effectiveness of machine and deep learning based classifiers with learned representations and engineered features in predicting patch correctness (RQ 3),
    \item The effectiveness of combining learned representations with engineered features in predicting patch correctness (RQ 4), and
    \item The contribution of features in predicting patch correctness (RQ 5).
\end{enumerate}
In the following, we present the details of each research question.

\subsection{Research Questions}
\begin{description}
	\item {\em {\bf RQ-1:} Do different representation learning models yield comparable distributions of similarity values between buggy code and patched code?} 
	A widespread hypothesis in program repair is that bug fixing generally induce minimal changes~\cite{chen2017testing,xiong2017precise,jiang2019inferring,jiang2018shaping,liu2019avatar,liu2019tbar,wen2018context,weimer2009automatically,liu2018closer,martinez2015mining,barr2014plastic}. We propose to investigate whether learned embeddings can be a reliable means for assessing the extent of changes through computation of cosine similarity between vector representations.
	\item {\em {\bf RQ-2:} To what extent similarity distributions can be generalized for inferring a cutoff value to filter out incorrect patches?} 
	Following up on RQ1, we propose in this research question to experiment ranking patches based on cosine similarity of their vector representations, and rely on naively-defined similarity thresholds to decide on filtering of incorrect patches.
	\item {\em {\bf RQ-3:} Can we learn to identify patch correctness by training predictors with learned embeddings of code input?} 
	We investigate whether deep learned features (i.e., learned embeddings) are indeed relevant for building machine learning predictors for patch correctness. In particular we assess whether such a predictor built with static features can provide comparable performance with dynamic approaches, such as PATCH-SIM, which leverage execution behaviour information. We also compare the performance yielded when using deep learned features against the performance yielded when using the engineered features in the state of the art.
	\item {\em {\bf RQ-4:} Can the combination of learned embeddings and engineered features achieve optimum performance for predicting patch correctness?} 
	We investigate the possibility of ensuring high accuracy in patch correctness identification by combining different representations of patches. 
	\item {\em {\bf RQ-5:} Which features are most useful for predicting patch correctness?} 
	We leverage SHAP explanation models to provide an interpretation of the contribution of different features to the predictions.
\end{description}

\subsection{Datasets}
We collect patch datasets by building on previous efforts in the community. An initial dataset of correct patches is collected by using five literature benchmarks, namely Bugs.jar~\cite{saha2018bugs}, Bears~\cite{madeiral2019bears}, Defects4J~\cite{just2014defects4j}, QuixBugs~\cite{lin2017quixbugs} and ManySStuBs4J~\cite{karampatsis2020how}. These are human-written patches as committed by developers in open-source project repositories.

We also consider patches generated by APR tools integrated into the \texttt{RepairThemAll} framework. We use all patch samples released by Durieux~{\em et~al.}~\cite{durieux2019empirical}. This only includes sample patches that make the programs pass all test cases. They are thus plausible. However, no validation information on correctness was given. In this work, we proceed to manually validate the generated patches, among which we identified 900 correct patches. The correctness validation follows the criteria defined by Liu~{\em et~al.}~\cite{liu2020efficiency}.
In a recent study on the efficiency of program repair, Liu~{\em et~al.}~\cite{liu2020efficiency} released a labeled dataset of patches generated by 16 APR systems for the Defects4J bugs. 
We consider this dataset as well as the labeled dataset that was used to evaluate the PATCH-SIM~\cite{xiong2018identifying} approach.

Overall, Table~\ref{tab:datasets} summarizes the data sets that we used for our experiments. Each experiment in Section~\ref{sec:results} has specific requirements on the data (e.g., large patch sets for training models, labeled datasets for benchmarking classifiers, etc.). For each experiment, we will recall which sub-dataset has been leveraged and why. 

\begin{table}[h!t]
	\centering
	\small
	\caption{Datasets of Java patches used in our experiments.}	
 	\resizebox{1\linewidth}{!}
	{
	\begin{threeparttable}
	\begin{tabular}{l|cccr}
			\toprule
		Subjects &  contains incorrect patches & contains correct patches & labelled dataset & \# Patches \\\hline
		Bears~\cite{madeiral2019bears}  & No & Yes & - & 251 \\
		Bugs.jar~\cite{saha2018bugs}  & No & Yes & - & 1,158 \\
		Defects4J~\cite{just2014defects4j}$^\dagger$  & No & Yes & - & 864 \\
		ManySStubBs4J~\cite{karampatsis2020how}  & No & Yes & - & 34,051 \\
		QuixBugs~\cite{lin2017quixbugs}  & No & Yes & - & 40 \\\midrule
		RepairThemAll~\cite{durieux2019empirical} & Yes & Yes & No$^\ddagger$ & 64,293 \\%67,211 - 2,918
		Liu~{\em et~al.}~\cite{liu2020efficiency} & Yes & Yes & Yes & 1,245 \\	
		Xiong~{\em et~al.}~\cite{xiong2018identifying} & Yes & Yes & Yes & 139 \\\midrule
		{\bf Total} & & & & 102,041 \\
			\bottomrule
	\end{tabular}
	{\small$^\dagger$The latest version 2.0.0 of \protect{Defects4J}\protect\footnotemark is considered in this study.\\
	$^\ddagger$The patches are not labeled in~\cite{durieux2019empirical}. We support the labeling effort in this study by comparing the generated patches against the developers' patches. The 2,918 patches for IntroClassJava in~\cite{durieux2019empirical} are also excluded from our study since IntroClassJava is a lab-built Java benchmark transformed from the C program bugs in small student-written programming assignments from IntroClass~\cite{le2015manybugs}.}
	\end{threeparttable}
	}
\label{tab:datasets}
\end{table}
\footnotetext{\url{https://github.com/rjust/defects4j/releases/tag/v2.0.0}}

\subsection{Model Input Pre-processing}
Samples in our datasets are patches such as the one presented in Figure~\ref{fig:chart1} extracted from the Defects4J dataset.
Our investigations with representation learning however require input data about the buggy and patched code. A straightforward approach to derive those inputs would be to consider the code files before and after the patch. Unfortunately, depending on the size of the code file, the differences could be too minimal to be captured by any similarity measurement. To that end, we propose to focus on the code fragment that appears in the patch. Thus, to represent the buggy code fragment (cf. Figure~\ref{fig:chart1-buggy}) from the initial patch in Figure~\ref{fig:chart1}, we keep all removed lines (i.e., starting with `-') as well as the patch context lines (the code that has not been modified, i.e., those lines not starting with either `-', `+' or `@'). Similarly, the patched code fragment (cf. Figure~\ref{fig:chart1-patched}) is represented by added lines (i.e., starting with `+') as well as the same context lines. 
Since tool support for the representation learning techniques BERT, Doc2Vec, and CC2Vec require each input sample to be on a single line, we flatten multi-line code fragments into a single line.

In contrast to BERT, Doc2Vec, and CC2Vec, which can take as input some syntax-incomplete code fragments, code2vec requires the fragment to be fully parsable in order to extract information on Abstract Syntax Tree paths. Since patch datasets include only text-based diffs, code context is generally truncated and is likely not parsable. 
However, as just explained, we opt to consider only the removed/added lines to build the buggy and patched code input data. By doing so, we substantially improved the success rate of the JavaExtractor\footnote{\url{https://github.com/tech-srl/code2vec/tree/master/JavaExtractor}} tool used to build the tokens in the code2vec pipeline.

\begin{figure}[!t]
    \centering\vspace{2mm}
\begin{lstlisting}[language=diff,linewidth={\linewidth},frame=tb,basicstyle=\ttfamily\footnotesize]
--- source/org/jfree/chart/renderer/category/AbstractCategoryItemRenderer.java
+++ source/org/jfree/chart/renderer/category/AbstractCategoryItemRenderer.java
@@ -1795,6 +1795,6 @@ public abstract class AbstractCategoryItemRenderer
         int index = this.plot.getIndexOf(this);
         CategoryDataset dataset = this.plot.getDataset(index);
-        if (dataset != null) {
+        if (dataset == null) {
             return result;
         }
\end{lstlisting}
    \caption{Example of a patch for the Defects4J bug Chart-1.}
    \label{fig:chart1}
\end{figure}

\begin{figure}[!t]
    \centering
\begin{lstlisting}[language=Java,linewidth={\linewidth},frame=tb,basicstyle=\ttfamily\footnotesize ]
(@\bf 1:@) a/source/org/jfree/chart/renderer/category/AbstractCategoryItemRenderer.java
(@\bf 2:@)         int index = this.plot.getIndexOf(this);
(@\bf 3:@)         CategoryDataset dataset = this.plot.getDataset(index);
(@{\bf 4:}@)         (@{\colorbox{codered}{{if (dataset != null) \{}}@)
(@\bf 5:@)             return result;
(@\bf 6:@)         }
\end{lstlisting}
    \caption{Buggy code fragment associated to patch in Fig.~\ref{fig:chart1}.}
    \label{fig:chart1-buggy}
\end{figure}

\begin{figure}[!t]
    \centering
\begin{lstlisting}[language=Java,linewidth={\linewidth},frame=tb,basicstyle=\ttfamily\footnotesize]
(@\bf 1:@) b/source/org/jfree/chart/renderer/category/AbstractCategoryItemRenderer.java
(@\bf 2:@)         int index = this.plot.getIndexOf(this);
(@\bf 3:@)         CategoryDataset dataset = this.plot.getDataset(index);
(@\bf 4:@)         (@{\colorbox{codegreen}{{if (dataset == null) \{}}@)
(@\bf 5:@)             return result;
(@\bf 6:@)         }
\end{lstlisting}
    \caption{Patched code fragment associated to patch in Fig.~\ref{fig:chart1}.}
    \label{fig:chart1-patched}
\end{figure}

\subsection{Embedding Models Setup}
\label{sec:embeddings}

\begin{figure}[!h]
	\includegraphics[width=0.9\columnwidth]{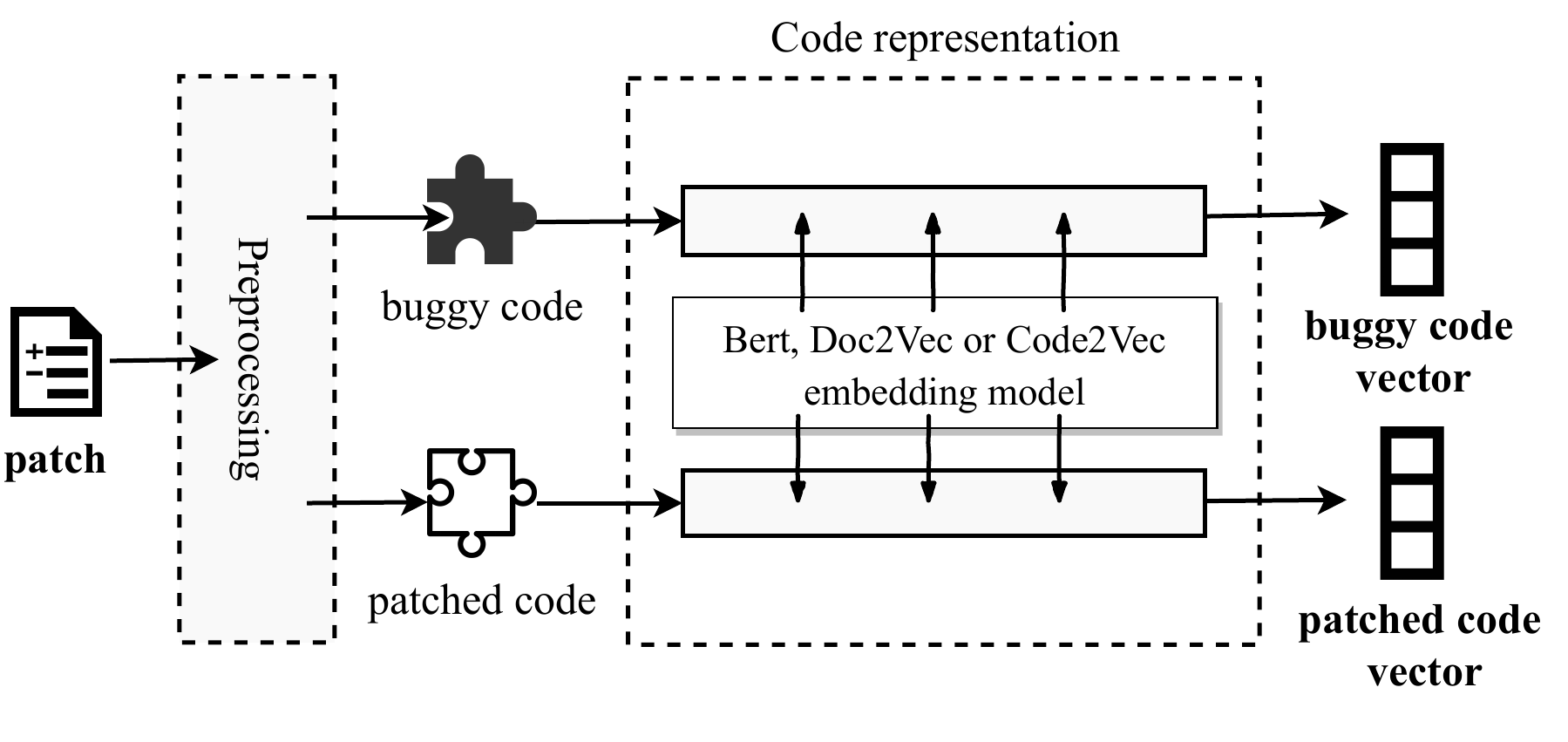}
	\caption{Producing code fragment \embeddings with BERT, Doc2Vec and code2vec.}
	\label{fig:embedding}
\end{figure}

When representation learning algorithms are applied to some training data, they produce {\em embedding models} that have learned to map a set of code tokens in the vocabulary of the training data to vectors of numerical values. These vectors are also referred to as {\em \embeddings}.
Figure~\ref{fig:embedding} illustrates the process of embedding buggy code and patched code for the purpose of our experiments. Considering that the pre-trained embedding models require huge resources (e.g. BERT has 340M parameters to be trained) to fine-tune for our classification task, we resort to directly leverage the pre-trained models to embed the patches, and train the classifiers separately. We propose to use four baseline embedding models from the literature to explore our proposed hypothesis. We consider a variety of models trained on, and targeting, different artifact types (natural language, structured code, code changes).

\begin{figure}[!t]
	\includegraphics[width=0.9\columnwidth]{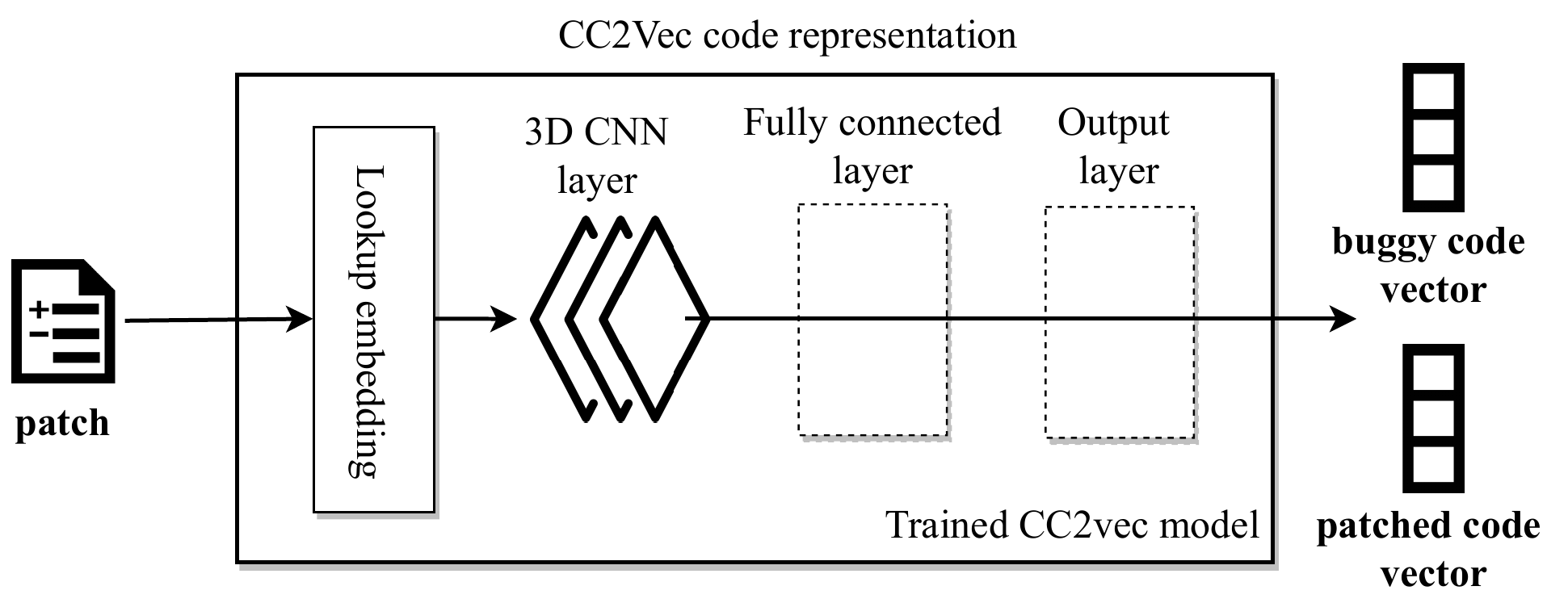}
	\caption{Extracting code fragment \embeddings from CC2Vec pre-trained model.}
	\label{fig:cc2vec}
\end{figure}

\begin{itemize}[leftmargin=*]
	\item {\bf BERT.} In the first scenario, we consider an embedding model that initially targets natural language data, both in terms of the learning algorithm and in terms of training data. We thus select BERT, one of the state of the art models, for the evaluation of our hypothesis. The network structure of BERT, however, is deep, meaning that it requires large datasets for training the embedding model. As is now customary in the literature, we instead leverage a pre-trained 24-layer BERT model, which was trained on a Wikipedia corpus.
	\item {\bf Doc2Vec.} In the second scenario, we consider an embedding model that is trained on code data but using a representation learning technique that was developed for text data. Doc2Vec represents documents as a vector by generalizing the basic model word2vec. The code snippets of patches are able to be seen as documents. Therefore, we have trained the Doc2Vec model with code data of 36,364 patches from the 5 repair benchmarks (Bears, Bugs.jar, etc., cf. Table ~\ref{tab:datasets}).
	\item {\bf code2vec.} In the third scenario, we consider an embedding model that primarily targets code, both in terms of the learning algorithm and in terms of training data. Code2vec was specifically developed for programming languages and trained on a dataset of 14M methods. On the other hand, CodeBert~\cite{feng2020codebert} was trained both on programming and natural language. We thus use in this case a pre-trained model of code2vec, which was trained by the authors using \textasciitilde14 million code examples from Java projects.
	\item {\bf CC2Vec.} Finally, in the fourth scenario, we consider an embedding model that was built in representation learning experiments for code changes. CC2Vec~\cite{hoang2020cc2vec} models the hierarchical structure of the code change and has been applied to the task of patch identification. However, the pre-trained model that we leveraged from the work of Hoang et al. is embedding each patch into a single vector. We investigate the layers and identify the middle CNN-3D layer as the sweet spot to extract embeddings for buggy code and patched code fragments. Figure~\ref{fig:cc2vec} illustrates the process.

\end{itemize}

\section{Experiments and Results}
\label{sec:results}
We first introduce the metrics used in the experiments. Then, we present the experiments that we designed to answer the research questions of our study. For each experiment, we state the objective, overview the execution details, and present the results. 

Our objective is to measure the ability of the approaches in terms of recalling correct patches while filtering out incorrect patches. Thus, we follow the definitions of {\bf Recall} proposed by Tian ~et al. for the evaluation of their BATS~\cite{tian2022predicting} systems:
\begin{itemize}
	\item {\bf +Recall} measures to what extent correct patches are identified, i.e., the percentage of correct patches that are identified from all correct patches.
	\item {\bf -Recall} measures to what extent incorrect patches are filtered out, i.e., the percentage of incorrect patches that are filtered out from all incorrect patches.
\end{itemize}
\vspace{-8mm}
\begin{multicols}{2}
    \begin{equation}\label{Recall_P}
    + Recall=\frac{TP}{TP+FN}
    \end{equation}
    
    \begin{equation}\label{Recall_N}
    - Recall=\frac{TN}{TN+FP}
    \end{equation}
\end{multicols}

\noindent
where $TP$ represents true positive, $FN$ represents false negative, $FP$ represents false positive, $TN$ represents true negative. 

{\bf Accuracy and Precision}. The ratio of positive and negative samples of our dataset is balanced (1.3:1). We thus use accuracy and precision to evaluate the performance of the approaches in classifying the patches.

{\bf Area Under Curve (AUC) and F1-measure}. We train a few machine and deep learning-based classifiers to identify the patch correctness. Therefore, we use two commonly used metrics for evaluating overall performance of the classifiers: AUC and F1 score (harmonic mean between precision and recall for identifying correct patches).

\subsection{[RQ-1: Similarity Measurements for Buggy and Patched Code using Embeddings]}
\label{subsec:rq1}

\paragraph{\bf Objective:} We investigate the capability of different \embeddings to capture the (dis)similarity between buggy code fragments and the (in)correctly-patched ones. The experiments are performed towards providing answers for two sub-questions:
\begin{itemize}%
	\item RQ-1.1 {\em Is correctly-patched code actually similar to buggy code based on \embeddings?}
	\item RQ-1.2 {\em To what extent is buggy code more similar to correctly-patched code than to incorrectly-patched code?}
\end{itemize}

{\bf Experimental Design for RQ-1.1:} Using the four embedding models considered in our study (cf. Section~\ref{sec:embeddings}), we produce the \embeddings for buggy and patched code fragments associated to 36k patches from five repair benchmarks shown in Table~\ref{tab:data1}. 
In this case, the patched code fragment is the correctly-patched code fragment since it comes from labeled benchmark data (generally representing human-written patches). 
Given those \embeddings (i.e., deep learned representation vectors of code), we compute the cosine similarity between the vectors representing the buggy and correctly-patched code fragments.

\begin{table}[!t]
	\centering
	\caption{Datasets used for assessing the similarity between buggy code and correctly-patched code.}
	\label{tab:data1}
	{
	\begin{threeparttable}
		\begin{tabular}{lccccc|C{18mm}}
			 \toprule
			 & {\bf {Bears}} & {\bf {Bugs.jar}} & {\bf {Defects4J}} & {\bf ManySStuBs4J} & {\bf {QuixBugs}} & {\bf Total}\\
			\hline %
			\# Patches & 251 & 1,158 & 864 & 34,051 & 40 & 36,364\protect\footnotemark\\
			\bottomrule
		\end{tabular}
	\end{threeparttable}
	}
\end{table}
\footnotetext{Due to parsing failures, code2vec \embeddings are available for 21,135 patches.}

\begin{figure}[!h]
\centering
	\includegraphics[width=0.95\linewidth]{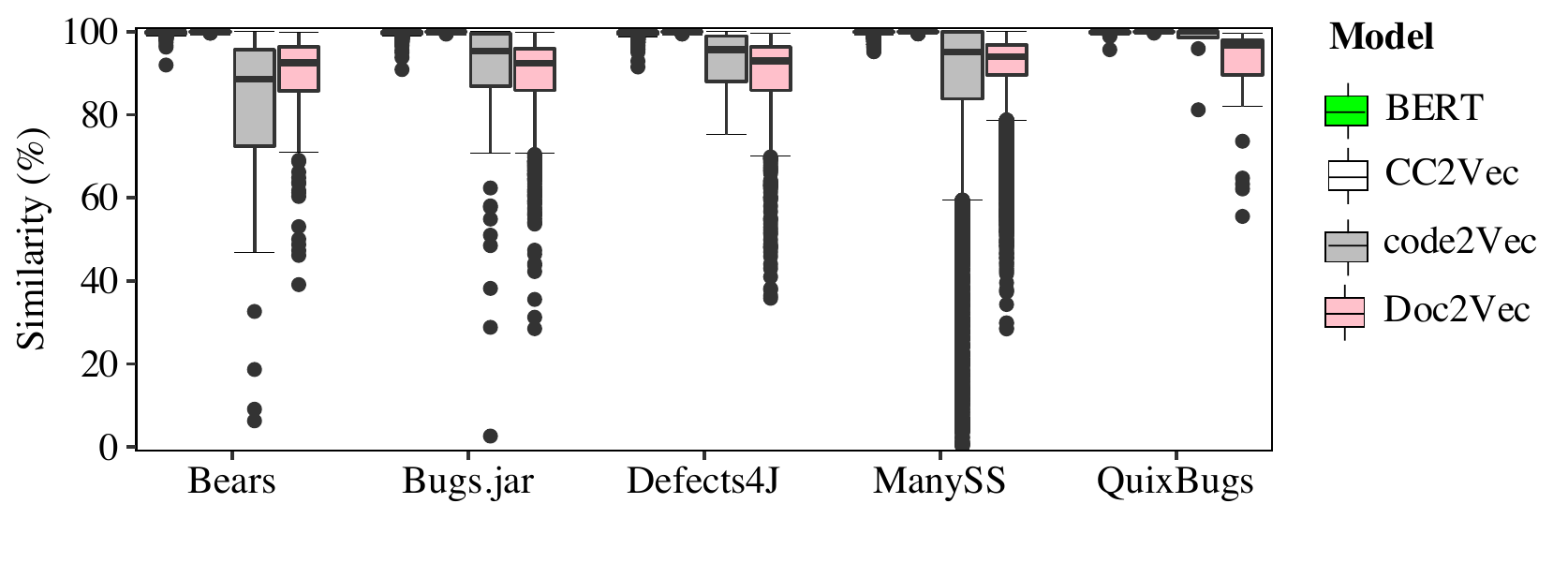}
	\vspace{-3mm}
	\caption{Distributions of similarity scores between correctly-patched code fragments and buggy ones.}
	\label{fig:all-sim-correct}
\end{figure}

{\bf Results for RQ-1.1:} 
Figure~\ref{fig:all-sim-correct} presents the boxplots of the similarity distributions with different embedding models and for samples in different datasets. Doc2Vec and code2vec models appear to yield similarity values that are lower than BERT and CC2Vec models.  

\begin{figure}[!ht]
	\includegraphics[width=1\linewidth]{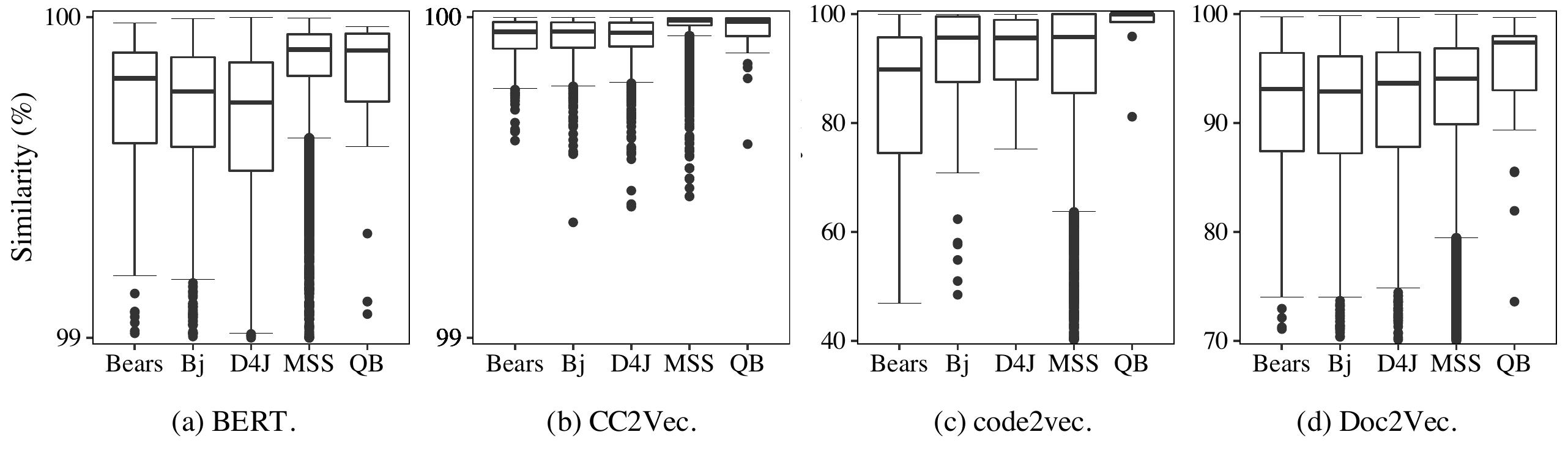}
	\caption{Zoomed views of the distributions of similarity scores between correct and buggy code fragments.}
	\label{fig:results-1}
\end{figure}

Figure~\ref{fig:results-1} zooms in the boxplot region for each embedding model experiment to overview the differences across different benchmark data.
We obverse that, when embedding the patches with BERT, the similarity distribution for the patches in Defects4J dataset is similar to Bugs.jar and Bears dataset, but is different from the dataset ManySStBs4J and QuixBugs. The Mann-Whitney-Wilcoxon (MWW) tests~\cite{wilcoxon1945individual,mann1947test} confirm that the similarity of median scores for Defects4J, Bugs.jar and Bears is indeed statistically significant. MWW tests further confirms the statistical significance of the difference between Defects4J and ManySStBs4J/QuixBugs scores.

Defects4J, Bugs.jar and Bears include diverse human-written patches for a large spectrum of bugs from real-world open-source Java projects. In contrast, ManySStuBs4J only contains patches for single statement bugs. Quixbugs dataset is further limited by its size and the fact that the patches are built by simply mutating the code of small Java implementation of 40 algorithms (e.g., quicksort, levenshtein, etc.).

While CC2Vec and Doc2Vec exhibit roughly similar performance patterns with BERT (although at different scales), the experimental results with code2vec present different patterns across datasets. Note that, due to parsing failures of code2vec, we eventually considered only 118 Bears patches, 123 Bugs.jar patches, 46 Defects4J patches, 20,840 ManySStuBs4J patches and 8 QuixBugs. The change of dataset size could explain the difference with the other embedding models.

\find{{\bf \ding{45} RQ-1.1 }$\blacktriangleright$ \embeddings of buggy and correctly-patched code fragments exhibit high cosine similarity scores. Median scores are similar for patches that are collected with similar heuristics (e.g., in-the-wild patches vs single-line patches vs debugging example patches). The pre-trained BERT natural language model captures more similarity variations than the CC2Vec model, which is specialized for code changes.$\blacktriangleleft$}

{\bf Experimental Design for RQ-1.2:} To compare the similarity scores of correctly-patched code fragment vs incorrectly-patched code fragment to the buggy one, we consider combining datasets with correct patches and datasets with incorrect patches. 
Note that, all patches in our experiments are plausible since we are focused on correctness: plausibility is straightforward to decide based on test suites. 
Correct patches are provided in benchmarks. However, all the benchmarks in our study do not contain incorrect patches. Therefore, we rely on the dataset released by Liu~{\em et~al.}~\cite{liu2020efficiency}: 674 plausible but incorrect patches generated by 16 repair tools for 184 Defects4J bugs are considered from this dataset. 
Those 674 incorrect patches are selected within a larger set of incorrect patches by adding the constraint that the incorrect patch should be changed the same code location as the developer-provided patch in the benchmark: such incorrect patch cases may indeed be the most challenging to identify with heuristics.

We consider three scenarios to select correct patches for the comparison of the similarity scores. (1) Imbalanced-all, a quick intuition is that we compare the 674 incorrect patches against all correct patches from 5 benchmarks. (2) Imbalanced-Defects4J, we only use the correct patches from Defects4J. We design the second scenario because the correct patches from other benchmarks may create a sample bias. (3) Balanced-Defects4J, we use the correct patches for the 184 Defects4J bugs that the 674 incorrect patches target. In this scenario, incorrect and correct sets have the same number of patches. We design this to avoid the underlying bias of imbalanced sets. The comparison is done with different scenarios specified in Table~\ref{tab:data2}.

\begin{table}[!t]
	\centering
	\caption{Scenarios for similarity distributions comparison.}
	\label{tab:data2}
 	\resizebox{1\linewidth}{!}
	{
	\begin{threeparttable}
		\begin{tabular}{l|c|c}
			\toprule
			{\bf Scenario} & {\bf Incorrect patches} & {\bf Correct patches}\\
			\hline
			Imbalanced-all\protect\footnotemark &  \multirow{3}{*}{\makecell[l]{674 incorrect patches \\by 16 APR tools~\cite{liu2020efficiency} \\for 184 Defects4J bugs.}}
		  & 36,364 correct patches from 5 benchmarks in Table~\ref{tab:data1}.  \\\cline{1-1}\cline{3-3}
			Imbalanced-Defects4J &   & 864 correct patches from Defects4J.  \\\cline{1-1}\cline{3-3}
			Balanced-Defects4J & & 184 correct patches for the 184 Defects4J bugs.  \\
			\bottomrule
		\end{tabular}
	\end{threeparttable}
	}
	\vspace{1mm}
\end{table}
\footnotetext{Except for Defects4J, there are no publicly-released incorrect patches for APR datasets.}

\begin{figure}[!ht]
	\includegraphics[width=1\linewidth]{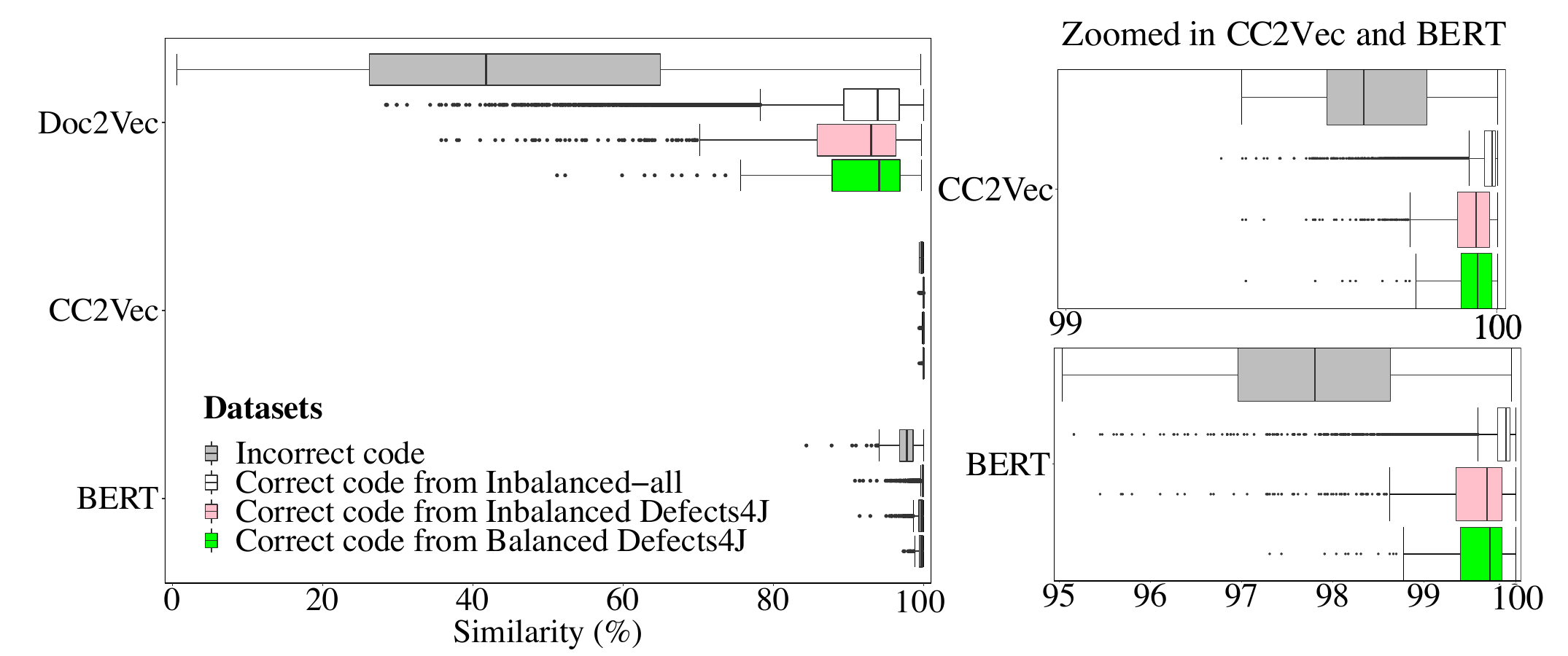}
	\caption{Comparison of similarity score distributions for code fragments in incorrect and correct patches.}
	\label{fig:c-vs-i-3}
\end{figure}

{\bf Results for RQ-1.2:}
In this experiment, we further assess whether incorrectly-patched code exhibits different similarity score distributions than correctly-patched code. Figure~\ref{fig:c-vs-i-3} shows the distributions of cosine similarity scores for correct patches (i.e., similarity between buggy code fragments and correctly-patched ones) and incorrect patches (i.e., similarity between buggy code fragments and incorrectly-patched ones). The comparison is done with different scenarios specified in Table~\ref{tab:data2}.%

The comparisons do not include the case of \embeddings for code2vec. 
Indeed, unlike the previous experiment where code2vec was able to parse enough code fragments, for the considered 184 correct patches of Defects4J, code2vec failed to parse most of the relevant code fragments. Hence, we focus the comparison on the other three embedding models (pre-trained BERT, trained Doc2Vec and pre-trained CC2Vec). Overall, we observe that the distribution of cosine similarity scores is substantially different for correctly-patched and incorrectly-patched code fragments.

We observe that the similarity distributions of buggy code and patched code from incorrect patches are significantly different from the similarities for correct patches. The difference of median values is confirmed to be statistically significant by an MWW test.
Note that the difference remains high for BERT, Doc2Vec and CC2Vec whether the correctly-patched code is the counterpart of the incorrectly-patched ones (i.e., the scenario of Balanced-Defects4J) or whether the correctly-patched code is from a larger dataset (i.e., Imbalanced-Defects4J scenarios). As for the comparison with the dataset of Imbalanced-all, the heuristic remains valid but note it may be affected by other benchmarks, i.e., the different bugs caused the results.

\find{{\bf \ding{45} RQ-1.2 }$\blacktriangleright$ \embeddings of code fragments with BERT, CC2Vec and Doc2Vec yield similarity scores that, given a buggy code, substantially differ between correctly-patched code and incorrectly-patched one. This result suggests that similarity score can be leveraged to discriminate correct patches from incorrect patches.$\blacktriangleleft$}
\subsection{[RQ-2: Filtering of Incorrect Patches based on Similarity Thresholds]}
\label{subsec:rq2}

\paragraph{\bf Objective:} Following up on the findings related to the first research question, we investigate the selection of cut-off similarity scores to decide on which APR-generated patches are likely incorrect. 
Results from this investigation will provide insights to guide the exploitation of code \embeddings in program repair pipelines.
 
{\bf Experimental Design:}
To select threshold values, we consider the distributions of similarity scores from the above experiments (cf. Section~\ref{subsec:rq1}). Table~\ref{tab:sim2} summarizes relevant statistics on the distributions on the similarity scores distribution for correct patches. Given the differences that were exhibited with incorrect patches in previous experiments, we use, for example, the 1$^{st}$ quartile value as an inferred threshold value.  

\begin{table}[!h]
	\centering
	\caption{Statistics on the distributions of similarity scores for correct patches of Bears+Bugs.jar+Defects4J.}
	\label{tab:sim2}
	{
	\begin{threeparttable}
		\begin{tabular}{l|cccccc}
			\toprule
			{\bf Subjects} & {\bf Min.} & {\bf 1st Qu.} & {\bf Median} & {\bf 3rd Qu.} & {\bf Max.} & {\bf Mean}\\
			\hline
			BERT   & 90.84 & 99.47 & 99.73 & 99.86 & 100 & 99.54 \\
			CC2Vec & 99.36 & 99.91 & 99.95 & 99.98 & 100 & 99.93 \\
			Doc2Vec& 28.49 & 85.80 & 92.60 & 96.10 &99.89& 89.19 \\
			code2vec& 2.64 & 81.19 & 93.63 & 98.87 & 100 & 87.11 \\
			\bottomrule
		\end{tabular}
	\end{threeparttable}
	}
\end{table}

Given our previous findings that different datasets exhibit different similarity score distributions, we also consider inferring a specific threshold for the QuixBugs dataset (cf. statistics in  Table~\ref{tab:sim3}).

\begin{table}[!h]
	\centering
	\caption{Statistics on the distributions of similarity scores for correct patches of QuixBugs.}
	\label{tab:sim3}
	{
	\begin{threeparttable}
		\begin{tabular}{l|cccccc}
			\toprule
			{\bf Subjects} & {\bf Min.} & {\bf 1st Qu.} & {\bf Median} & {\bf 3rd Qu.} & {\bf Max.} & {\bf Mean}\\
			\hline
			BERT   & 95.63 & 99.69 & 99.89 & 99.95 & 99.97 & 99.66 \\
			CC2Vec & 99.60 & 99.94 & 99.99 & 100   & 100   & 99.95 \\
			Doc2Vec& 55.51 & 89.56 & 96.65 & 97.90 & 99.72 & 91.29 \\
			code2vec& 81.16& 98.	53 & 100   & 100   & 100   & 97.06 \\
			\bottomrule
		\end{tabular}
	\end{threeparttable}
	}
\end{table}

\begin{table}[!h]
\centering
\caption{Filtering incorrect patches by generalizing thresholds inferred from Section~\ref{subsec:rq1}.Results.}
	\label{tab:filtering}
	{
    \begin{threeparttable} 
        \begin{tabular}{L{19mm}|C{19mm}|C{20mm}|C{20mm}|C{20mm}|C{20mm}}
        \toprule
        \multicolumn{2}{l|}{\textbf{Dataset}}  & \multicolumn{2}{c|}{Bears, Bugs.jar and Defects4J} & \multicolumn{2}{c}{QuixBugs} \\ \hline
        \multicolumn{2}{l|}{\bf \# Correct Patches} & \multicolumn{2}{c|}{893} & \multicolumn{2}{c}{7} \\ \hline
        \multicolumn{2}{l|}{\bf\# Incorrect Patches} & \multicolumn{2}{c|}{61,932} & \multicolumn{2}{c}{1,461} \\\hline
        \multicolumn{2}{l|}{\textbf{Model/Metric/Threshold}} & 1st Qu. & Mean & 1st Qu. & Mean \\ \hline & \# +CP & 57 & 49 & 4 & 4 \\ \cline{2-6} & \# -IP & 48,846 & 51,783 & 1,387 & 1,378 \\ \cline{2-6} & +Recall & 6.4\% & 5.5\% & 57.1\% & 57.1\% \\ \cline{2-6} \multirow{-4}{*}{\textbf{BERT}} & -Recall & {\color[HTML]{000000} \cellcolor{black!25}78.9\%} & \cellcolor{black!25}83.6\% & \cellcolor{black!25}94.9\% & \cellcolor{black!25}94.3\% \\ \hline & \# +CP & 797 & 789 & 4 & 4 \\ \cline{2-6} & \# -IP & 19,499 & 23,738 & 1,198 & 1,255 \\ \cline{2-6} & +Recall  & \cellcolor{black!25}89.2\% & \cellcolor{black!25}88.4\% & 57.1\% & 57.1\% \\ \cline{2-6} \multirow{-4}{*}{\textbf{CC2Vec}} & -Recall  & 31.5\% & 38.3\%  & 82.0\% & 85.9\% \\ \hline & \# +CP & 794 & 771 & 7 & 7 \\ \cline{2-6} & \# -IP & 25,192 & 33,218 & 1,226 & 1,270 \\ \cline{2-6} & \ +Recall & 88.9\% & 86.3\% & \cellcolor{black!25}100\%   & \cellcolor{black!25}100\%  \\ \cline{2-6} \multirow{-4}{*}{\textbf{Doc2Vec}} & \ -Recall & 40.7\% & 53.6\%       & 83.9\% & 86.9\% \\ \hline
        \end{tabular}
		{\footnotesize``{\bf \# +CP}'' means the number of correct patches that can be ranked upon the threshold, while ``{\bf \# -IP}'' means the number of incorrect patches that can be filtered out by the threshold. ``{\bf +Recall}'' and ``{\bf -Recall}'' represent the recall of identifying correct patches and filtering out incorrect patches, respectively.}
    \end{threeparttable}
    }
\end{table}

Our test data is constituted of 64,293 patches generated by 11 APR tools in the empirical study of Durieux~{\em et~al.}~\cite{durieux2019empirical}. First, we use the four embedding models to generate \embeddings of buggy code and patched code fragments and compute cosine similarity scores. Second, for each bug, we rank all generated patches based on the similarity scores between the patched code and the buggy one, where we consider that the higher the score, the more likely the correctness. 
Finally, to filter incorrect candidates, we consider two experiments: 
\begin{enumerate}[leftmargin=*]
\item Patches that lead to similarity scores that are lower to the inferred threshold (i.e., 1$^{st}$ quartile in previous experimental data) will be considered as incorrect. Patches where patched code exhibit higher similarity scores than the threshold are considered correct. 
\item Another approach is to consider only the top-1 patches with the highest similarity scores as correct patches. Other patches are considered incorrect.
\end{enumerate}

In all cases, we systematically validate the correctness of all 64,293 patches to have the correctness labels, for which the dataset authors did not provide (all plausible patches having been considered as valid). First, if the file(s) modified by a patch are not the same buggy files in the benchmark, we systematically consider it as incorrect: with this simple scheme, 33,489 patches are found incorrect. Second, with the same file, if the patch is not making changes at the same code locations, we consider it to be incorrect: 26,386 patches are further tagged as incorrect with this decision (cf. Threats to validity in Section~\ref{sec:discussion}). Finally, for the remaining 4,418 plausible patches in the dataset, we manually validate  correctness by following the strict criteria enumerated by Liu~{\em et~al.}~\cite{liu2020efficiency} to enable reproducibility. Overall, we could label 900 correct patches. The remainders are considered as incorrect.

{\bf Results:} By considering the patch with the highest (top-1) similarity score between the patched code and buggy code as correct, we were able to identify a correct patch for 10\% (with BERT), 9\% (with CC2Vec) and 10\% (with Doc2Vec) of the bug cases. Overall we also misclassified 96\% correct patches as incorrect. However, only 1.5\% of incorrect patches were misclassified as correct patches.

Given that a given bug can be fixed with several correct patches, the top-1 criterion may not be adequate. Furthermore, this criterion makes the assumption that a correct patch indeed  exists among the patch candidates. By using filtering thresholds inferred from previous experiments (which do not include the test dataset in this experiment), we can attempt to filter all incorrect patches generated by APR tools. Filtering results presented in Table~\ref{tab:filtering} show the recall scores that can be reached. We provide experimental results when we use 1$^{st}$ quartile and Mean values of similarity scores in the ``training'' set as threshold values. The thresholds are also applied by taking into account the datasets: thresholds learned on QuixBugs benchmark are applied to generated patches for QuixBugs bugs.

\find{{\bf \ding{45} RQ-2 }$\blacktriangleright$Building on cosine similarity scores, code fragment \embeddings can help to filter out between 31.5\% with CC2Vec and 94.9\% with BERT of incorrect patches. While BERT achieves the highest recall of filtering incorrect patches, it produces \embeddings that lead to a lower recall (at 5.5\%) at identifying correct patches.$\blacktriangleleft$}
\subsection{[RQ-3: Classification of Correct Patches with Supervised Learning]}
\label{sec:classifiers}

\paragraph{\bf Objective:} Cosine similarity between learned embeddings (which was used in the previous experiments) considers every deep learned feature as having the same weight as the others in the embedding vector. 
We investigate the feasibility to infer, using machine learning, the weights that different features may present with respect to patch correctness.
To this end, we build a patch correctness prediction framework, \toolname (\ul{LE}arn t\ul{O} \ul{P}redict p\ul{A}tch co\ul{R}rectness with embe\ul{D}dings), with the embedding models and machine learning algorithms. 
We compare the prediction evaluation results of \toolname with the achievements of related approaches in the literature.
The experiments are performed towards providing insights for the three sub-questions:
\begin{itemize}
	\item RQ-3.1 {\em Can \toolname learn to predict patch correctness by training classifiers based on the \embeddings of code ?}
	\item RQ-3.2 {\em Can \toolname be as reliable as a dynamic state-of-the-art approach such as PATCH-SIM in the patch correctness identification task?}
	\item RQ-3.3 {\em To what extent \embeddings of \toolname are providing different prediction results than the engineered features?}
\end{itemize}

{\bf Experimental Design for RQ-3.1:} 
To perform our machine learning experiments, we first require a ground-truth dataset. 
To that end, we rely on labeled datasets in the literature. 
Since incorrect patches generated by APR tools are only available for the Defects4J bugs, we focus on labeled patches provided by three independent teams (Liu~{\em et~al.}~\cite{liu2020efficiency}, Ye~{\em et~al.}~\cite{ye2019automated-dataset} and Xiong~{\em et~al.}~\cite{xiong2018identifying}) and other patches generated by APR tools. Very few patches generated by the different tools are actually labeled as correct, which leads to an imbalanced dataset. To reduce the imbalance issue, we supplement the dataset with developer (correct) patches as supplied in the Defects4J benchmark. 
Note that one developer patch could include multiple fixing hunks for different files, but the extraction of engineered features only work on the patches with respect to changing single file. Thus, we split such patches into sub patches by their changed files to ensure that one sub patch is only involved with one code file. 
In total, we collect 2,687 patches. After removing duplicates, 2,244 patches are remained.
97 patches are failed to obtain their engineered feature.
Eventually, the ground-truth dataset is built with 2,147 patches, shown in Table~\ref{tab:data3}. 

\begin{table}[!ht]
	\centering
	\caption{Dataset for evaluating ML-based predictors of patch correctness.}
	\label{tab:data3}
	{
	\begin{threeparttable}
		\begin{tabular}{l|c|c|r}
			\toprule
			& {\bf Correct patches}& {\bf Incorrect patches} & {\bf Total}\\
			\hline
			 Liu~{\em et~al.}~\cite{liu2020efficiency} &94& 366&460\\
			 Ye~{\em et~al.}~\cite{ye2019automated-dataset} &242& 452&694\\
			 Xiong~{\em et~al.}~\cite{xiong2018identifying} &30 &109 &139\\
			 Defects4J (developers)~\cite{just2014defects4j}&969 &0 & 969\\ 
			 Other APR tools  & 263  & 162  & 425 \\
			 \midrule
			\bf Dataset  & 1,598  &1,089  &2,687 \\
			\bf Dataset (deduplicated) & 1,288  &956  &2,244 \\
			\midrule
			\bf Dataset (final, with available features) & \bf1,199  &\bf948  &\bf2,147 \\
			\bottomrule
		\end{tabular}
	\end{threeparttable}
	}
\end{table}

Our ground truth dataset patches are then fed to our embedding models in \toolname to produce embedding vectors. As for previous experiments, the parsability of Defects4J patch code fragments prevented the application of code2vec: \toolname uses pre-trained models of BERT (trained with natural language text) and CC2Vec (trained with code changes) as well as a retrained model of Doc2Vec (trained with patches). 
Since the representation learning models are applied to code fragments inferred from patches (and not to the patch themselves), \toolname collects the embeddings of both buggy code fragment and patched code fragment for each patch. 
Then \toolname must merge these vectors back into a single input vector for the classification algorithm. We follow an approach that was demonstrated by Hoang et al.~\cite{hoang2020cc2vec} in a recent work on bug fix patch prediction: the classification model performs best when features of patched code fragment and buggy code fragment are crossed together. 

At first, and following related works in the literature, we used a 10-fold cross validation scheme to evaluate and compare our approach against the state of the art. However, we found that, with this scheme, a patch set generated for the same bug can be split into both the training and testing sets. Such a scenario is actually unrealistic (and biased) since we should not train the model with some labeled patches of a bug that we intend to repair (test set). To address this bias, we propose instead a 10-group cross validation scheme: First, we randomly distribute all bugs into 10 groups. Every group contains unique bugs and their associated patches. Then, we use 9 groups as train data and the remaining group as the test data. Finally, we repeat the selection of train and test groups for ten rounds and obtain the average score of the metrics.

{\bf Results for RQ-3.1:} We compare the performance of different embedding models using different classification algorithms. Table ~\ref{tab:ML-all} presents the results with 10-group cross validation setup. All classical metrics used for assessing predictors are reported: Accuracy, Precision, Recall, F1-Measure, Area Under Curve (AUC). 
XGBoost applied to BERT embeddings yields the best performance on the most of metrics (e.g. AUC with 0.803 and F1-measure with 0.765), while DNN achieves the best performance on precision of 0.744.

Our previous work~\cite{tian2020evaluating} was conducted through a 5-fold cross validation. To evaluate performance change of the approach on the new augmented dataset, we re-conduct a 5-fold cross validation experiment. The results show that after increasing the number of training examples (1,147 more patches), the performance of the decision tree, logistic regression and naive bayes classifiers are improved.
For instance, applying the three classifiers with BERT embeddings, their accuracy, precision, recall and F1-measure are improved with 3 to 23.6 points (except the recall of Naive bayes + BERT embedding is decreased).
Their AUC values are increased with 0.067, 0.06, 0.126, respectively. These results provide us the possibility of evolving the patch identification through datasets augmentation. 
Note that, for the following experiment, we proceed to focus on using 10-group cross validation because of its effectiveness for evaluating the approaches in practice.

\begin{table}[!t]
	\centering
	\caption{Evaluation of \embeddings on six ML classifiers in \toolname.}
	\label{tab:ML-all}
	{
	\begin{threeparttable}
		\begin{tabular}{l|c|ccccC{11mm}}
			\toprule
			{\bf Learner} & {\bf Embedding} &{\bf Accuracy} & {\bf Precision} & {\bf Recall} & {\bf F1-measure} & {\bf AUC} \\
			\hline
			\multirow{3}{*}{Decision Trees} & BERT  & 62.1 & 64.7 & 70.8 & 67.6 & 0.611 \\
			& CC2Vec  & 58.0 & 61.7 & 65.5 & 63.5 & 0.572 \\
			& Doc2Vec & 58.7 & 62.0 & 67.6 & 64.6 & 0.576 \\
			\hline
			\multirow{3}{*}{Logistic regression} & BERT & 72.2 & 73.5 & 78.7 & 76.0 & 0.796 \\
			& CC2Vec  & 61.8 & 64.8 & 68.9 & 66.8 & 0.679 \\
			& Doc2Vec & 65.8 & 66.6 & 77.7 & 71.7 & 0.717 \\
			\hline
			\multirow{3}{*}{Naive bayes} & BERT& 66.5 & 72.5 & 57.6 & 65.7 & 0.726 \\
			& CC2Vec  & 57.6 & 70.1 & 31.9 & 45.7 & 0.670 \\
			& Doc2Vec & 55.9 & 63.0 & 51.0 & 56.4 & 0.610 \\
			\hline
			\multirow{3}{*}{Random forest} & BERT& 69.4 & 68.3 & 77.9 & 75.5 & \cellcolor{black!25}0.793 \\
			& CC2Vec  & 62.1 & 63.9 & 74.1 & 68.6 & 0.705 \\
			& Doc2Vec & 64.9 & 63.5 & 87.6 & 73.6 & 0.705 \\
			\hline
			\multirow{3}{*}{XGBoost} & BERT& \cellcolor{black!25}71.8 & 71.6 & \cellcolor{black!25}82.1 & \cellcolor{black!25}76.5 & \cellcolor{black!25}0.803 \\
			& CC2Vec  & 65.3 & 66.4 & 76.6 & 71.1 & 0.729 \\
			& Doc2Vec & 63.2 & 63.5 & 80.2 & 70.8 & 0.693 \\
			\hline
			\multirow{3}{*}{DNN} & BERT& 70.3 & \cellcolor{black!25}74.4 & 71.3 & 72.8 & 0.767 \\
			& CC2Vec  & 51.8 & 55.5 & 69.0 & 61.6 & 0.503 \\
			& Doc2Vec & 63.2 & 64.7 & 75.1 & 69.5 & 0.679 \\
			\bottomrule
		\end{tabular}
	\end{threeparttable}
	}
	\vspace{2mm}
\end{table}

\find{{\bf \ding{45} RQ3.1 }$\blacktriangleright$ Tree-based boosting classifiers (Random forest and XGBoost) and Deep learning classifier (DNN) with BERT embeddings yield the promising performance on predicting the patch correctness for APR tools (e.g., F1-measure at 76.5\% and AUC at 80.3\%).$\blacktriangleleft$}

{\bf Experimental Design for RQ-3.2:} 
PATCH-SIM~\cite{xiong2018identifying} is the state-of-the-art work on predicting the patch correctness for APR tools. 
It is a dynamic-based approach, which generates execution traces of patched programs with new generated tests, and compares the execution traces across test cases to assess the correctness of APR-generated patches. 
We propose to apply PATCH-SIM to our collected patches (cf. Table~\ref{tab:data3}). 
Unfortunately, PATCH-SIM is implemented to run on Defects4J-v1.2.0\footnote{\url{https://github.com/rjust/defects4j/releases/tag/v1.2.0}}. Therefore, it failed to process 476 patches generated for some bugs (e.g., JSoup bugs) in the latest version of Defects4J (i.e., Defects4J-v2.0.0). Furthermore, even when PATCH-SIM can run, we observe that it does not yield any prediction output for 1,022 patches\footnote{We have reported the issue to the authors but have not yet been made aware of any solution to address it. Note that in their original paper the authors transparently informed readers that the tool indeed is sensitive to the datasets.}.
Eventually, we were able to assess the performance of PATCH-SIM on 649 patches. 
To avoid a potential bias in comparisons, we also conduct the ML-based classification experiments for \toolname on the 649 patches.

\begin{figure}[t]
	\includegraphics[width=0.65\columnwidth]{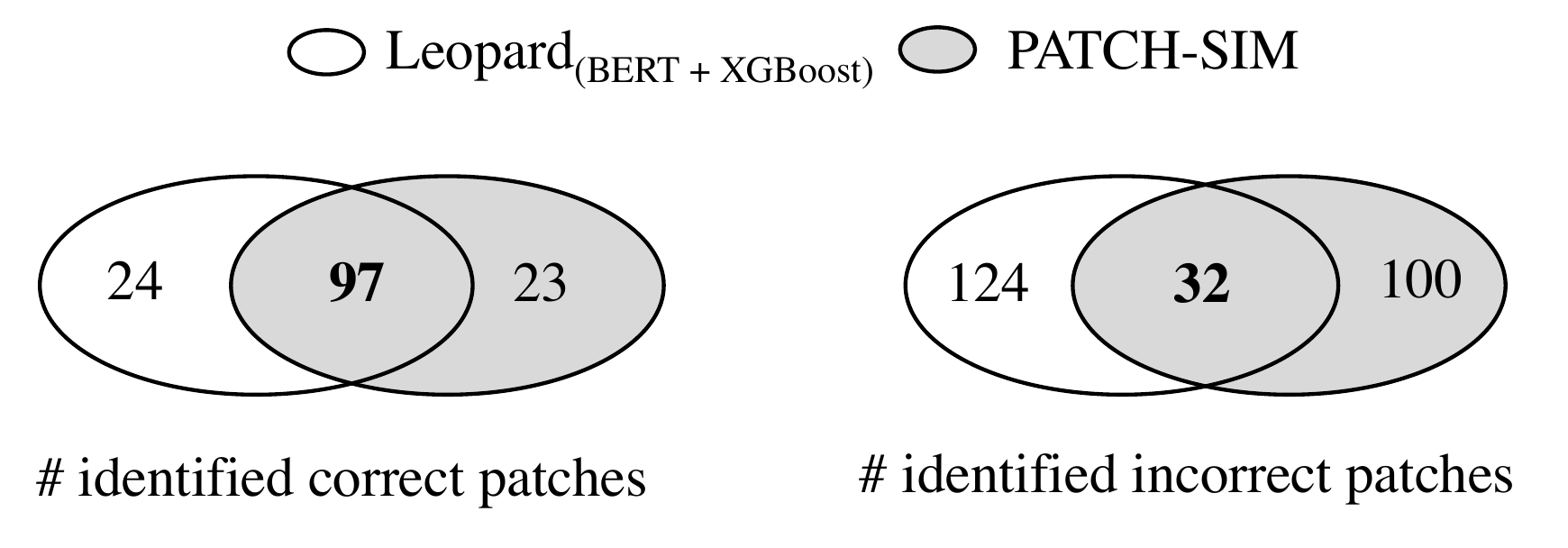}
	\caption{Comparison on the number of (in)patches correctly identified by \toolname (with the BERT embeddings + the XGBoost learner) against PATCH-SIM.}
	\label{fig:rq-3.2}
\end{figure}

{\bf Results for RQ-3.2:} 
Table~\ref{tab:patchsim} provides the comparing results on predicting patch correctness. 
In terms of Recall, PATCH-SIM achieved 78.9\% that is a bit higher than the BERT embedding + Random forest of \toolname, which demonstrates its ability of recalling correct patch from plausible patches as reported in~\cite{xiong2018identifying} by its authors.  
However, the accuracy, precision and AUC measurements are just 38.8\%, 24.7\% and 52.8\%, respectively. 
These results underperform the three ML classifiers of \toolname.
It indicates the many incorrect patches are wrongly identified as correct by PATCH-SIM. 
Figure~\ref{fig:rq-3.2} further gives an example on comparing the BERT embedding + the XGBoost classifier of \toolname and PATCH-SIM in terms of the number of (in) patches correctly identified by them.
XGBoost classifier of \toolname can recall more correct and incorrect patches than the PATCH-SIM, and the 24 correct patches and 124 incorrect patches are exclusively correctly predicted by it.

\noindent
{\em Time cost.} 
Note that we have recorded that, on average, PATCH-SIM takes $\sim$17.5 minutes to predict the correctness of each patch. In contrast, each of the ML classifiers of \toolname takes less than 1 minute for prediction. However, note that the training of \toolname requires the input of the learned embeddings of patches generated by pre-trained models (e.g. BERT). Such models, which are available on-the-shelf, have been trained using hundreds of TPUs that were run for several hours on a large corpus.

\begin{table}[!t]
	\centering
	\caption{Comparing evaluation of \toolname (BERT embedding + ML classifiers) against PATCH-SIM.}
	\label{tab:patchsim}
	{
	\begin{threeparttable}
		\begin{tabular}{l|l|ccccC{11mm}}
			\toprule
			\multicolumn{2}{l|}{\bf Approach}  &{\bf Accuracy} & {\bf Precision} & {\bf Recall} & {\bf F1-measure} & {\bf AUC} \\
			\midrule
			\multicolumn{2}{l|}{PATCH-SIM} & 38.8 & 24.7 & 78.9 & 37.7 & 0.528 \\
			\midrule
			\multirow{3}{*}{\rotatebox{90}{\toolname}} & {BERT + Random forest} & 41.3 & 25.5 &  78.3 & 38.4 & 0.594 \\
			 & {BERT +  XGBoost} & {\bf 42.7} & {\bf 26.2} & 79.6 & 39.4 & {\bf 0.614} \\
			 & {BERT +  DNN} & 40.0 & 26.1 &  {\bf 85.5} & {\bf 40.0} & 0.546 \\
		    \bottomrule
		\end{tabular}
	\end{threeparttable}
	}
\end{table}

\find{{\bf \ding{45} RQ-3.2 }$\blacktriangleright$ ML predictors of \toolname trained on \embeddings can be complementary to the state-of-the-art PATCH-SIM. They can also outperform PATCH-SIM in filtering out more patches generated by APR tools.}

\vspace{-2mm}
{\bf Experimental Design for RQ-3.3:}
As reported by Ye {\em et al.}~\cite{ye2019automated} in a recent study, post-processing APR-generated patches through engineered features achieves promising results.
Therefore, in this study, we also use some of the engineered features (Prophet features and repair pattern) in\cite{ye2019automated} to predict correct patches on a larger dataset: overall, our study is based on 2,147 patches while Ye {\em et al.} applied only 713 patches. Results in this study are given based on 10-group cross validation.

{\bf Results for RQ-3.3:} 
Table~\ref{tab:ods-all-2} presents the results of predicting patch correctness with the engineered features.
The naive bayes learning algorithm achieves a unusual performance compared to the other five learners.
It yields the highest precision, but leads to a much lower recall than others. This suggests that a very small number of correct patches can be recalled via using this learner.
The Random Forest and XGBoost learners achieve similarly high performance (e.g., F1-measure at 74.7\%/74.1\% and AUC at 76.9\%/77.6\%), and are followed by the DNN learner.
Overall, the performance reached with engineered features is generally comparable (in terms of global metrics) to that yielded by \toolname using \embeddings, except when using the Naive Bayes and Decision Trees learning algorithm.

\begin{table}[!h]
	\centering
	\caption{Evaluation of engineered feature on six ML classifiers.}
	\label{tab:ods-all-2}
	{
	\begin{threeparttable}
		\begin{tabular}{l|cC{15mm}C{15mm}cC{15mm}}
			\toprule
			{\bf Learner} &{\bf Accuracy} & {\bf Precision} & {\bf Recall} & {\bf F1-measure} & {\bf AUC} \\
			\hline
			\multirow{1}{*}{DecisionTree}   & 66.6 & 68.6 & 73.9 & 71.1 & 0.666 \\
			\hline
			\multirow{1}{*}{Logistic regression} & 70.0 & 72.7 & 74.1 & 73.4 & 0.773 \\
			\hline
			\multirow{1}{*}{Naive bayes} & 49.6 & 74.6 & 14.7 & 24.5 & 0.689 \\
			\hline
			\multirow{1}{*}{Random forest} & 70.7 & 72.1 & 77.5 & 74.7 & 0.769 \\
			\hline
			\multirow{1}{*}{XGBoost} & 70.5 & 72.6 & 79.9 & 74.1 & 0.776 \\
			\hline
			\multirow{1}{*}{DNN} & 69.8 & 72.1 & 74.8 & 73.4 & 0.777 \\
			\bottomrule
		\end{tabular}
	\end{threeparttable}
	}
\end{table}

\begin{figure}[h]
	\includegraphics[width=0.65\columnwidth]{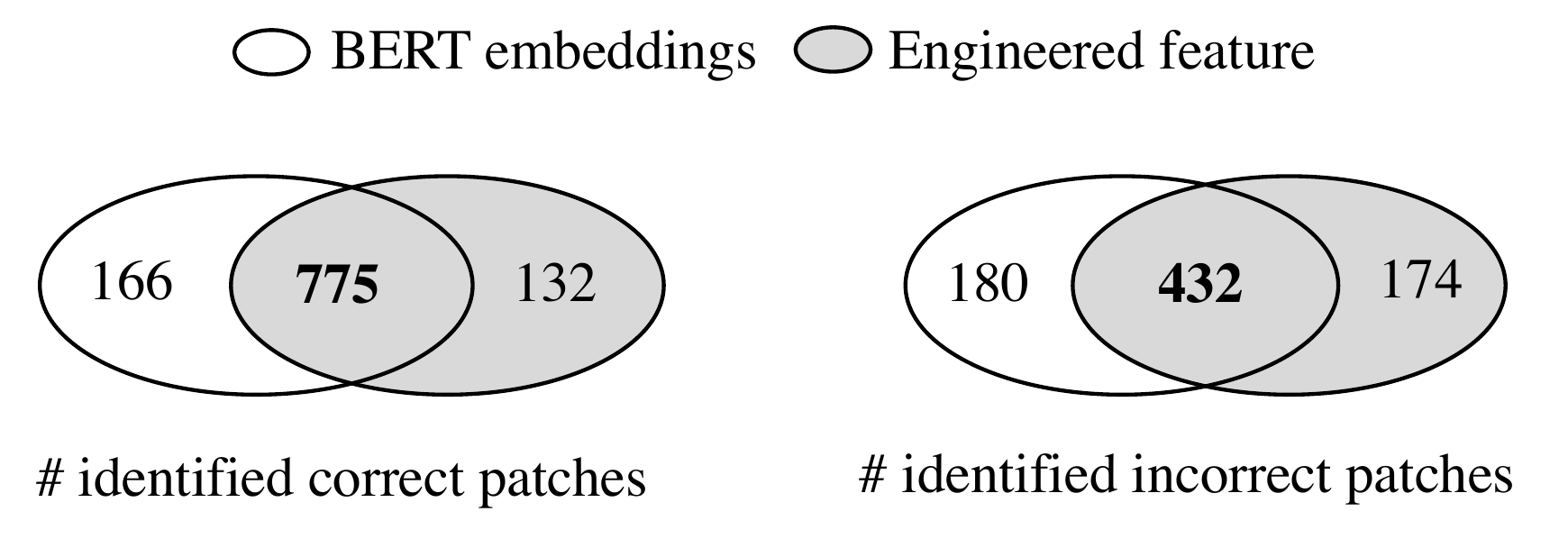}
	\caption{Comparison on the number of (in)patches correctly identified by the XGBoost classifier with the BERT embeddings and the engineered features.}
	\label{fig:rq-3.3}
\end{figure}

Figure~\ref{fig:rq-3.3} further illustrates the differences between the XGBoost classifier with the BERT embeddings and the engineered features in terms of the number of identified (in)correct patches.
More (in)correct patches can be correctly identified by the XGBoost classifier with both two scenarios.
Nevertheless, there still is a big complementary space of identifying the patch correctness for the two scenarios.

\find{{\bf \ding{45} RQ-3.3 }$\blacktriangleright$ The ML classifiers fed with the engineered features (from static code) can achieve comparable performance to \embeddings based classifiers in identifying patch correctness.
There is nevertheless the possibility to improve the prediction performance in both cases since their correct predictions are not perfectly overlapping: \embeddings lead to the identification of correct/incorrect patches that are not recalled with engineered features and vice versa. $\blacktriangleleft$}

\subsection{[RQ-4: Combining Learned Embeddings and Engineered Features for more Accurate Classification of Correct Patches]}
\label{sec:combination}

\paragraph{\bf Objective:} Following up on the insights from the previous research question, which compared engineered features against \embeddings, we investigate the potential of leveraging both feature sets to improve the classification of correct patches.

\begin{figure}[!t]
    \centering
    \subfigure[Ensemble learning.]{
    \includegraphics[width=0.3\textwidth]{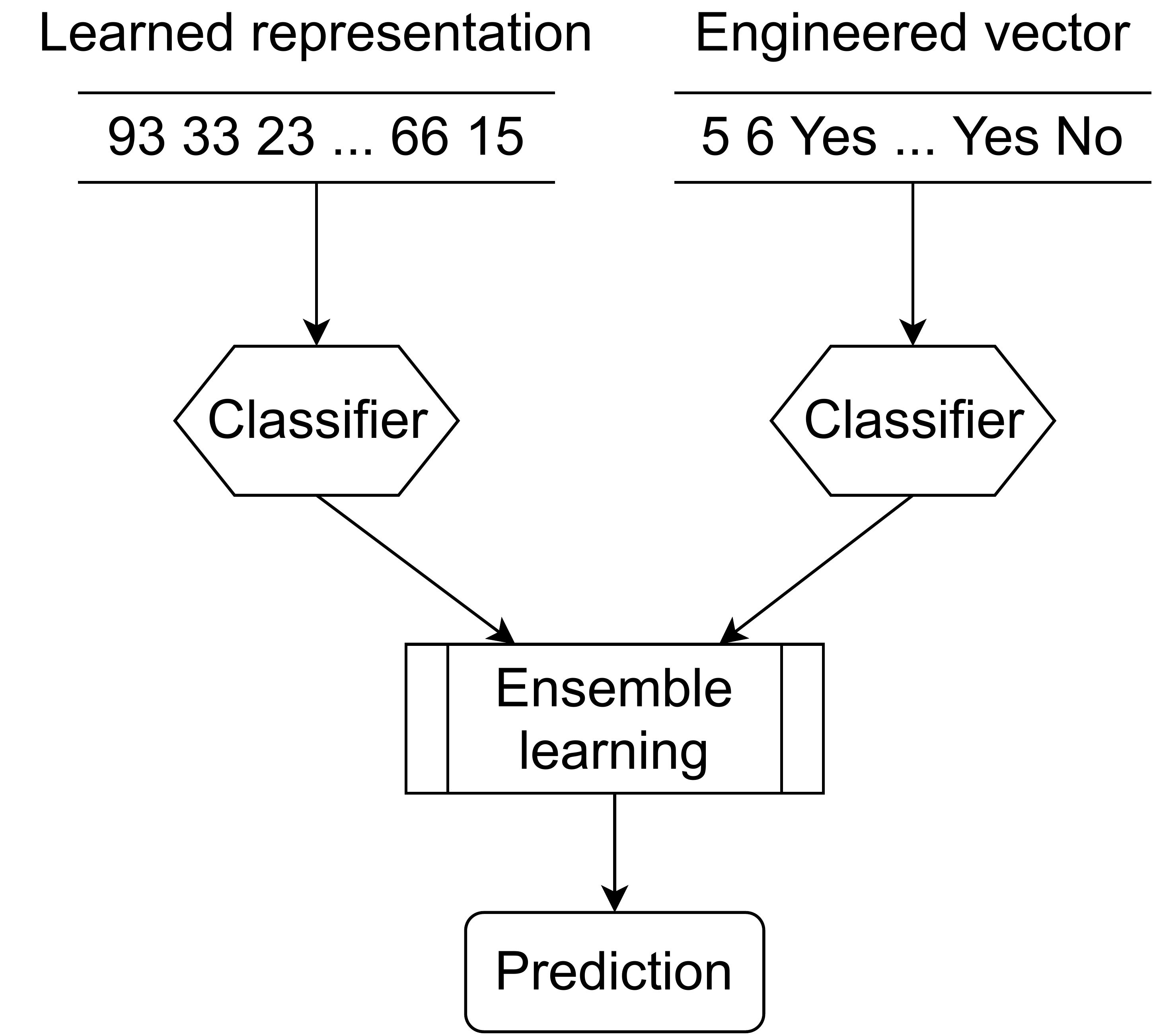}
    }
    \subfigure[Na{\"i}ve Vector Concatenation.]{
    \centering
    \includegraphics[width=0.3\linewidth, ]{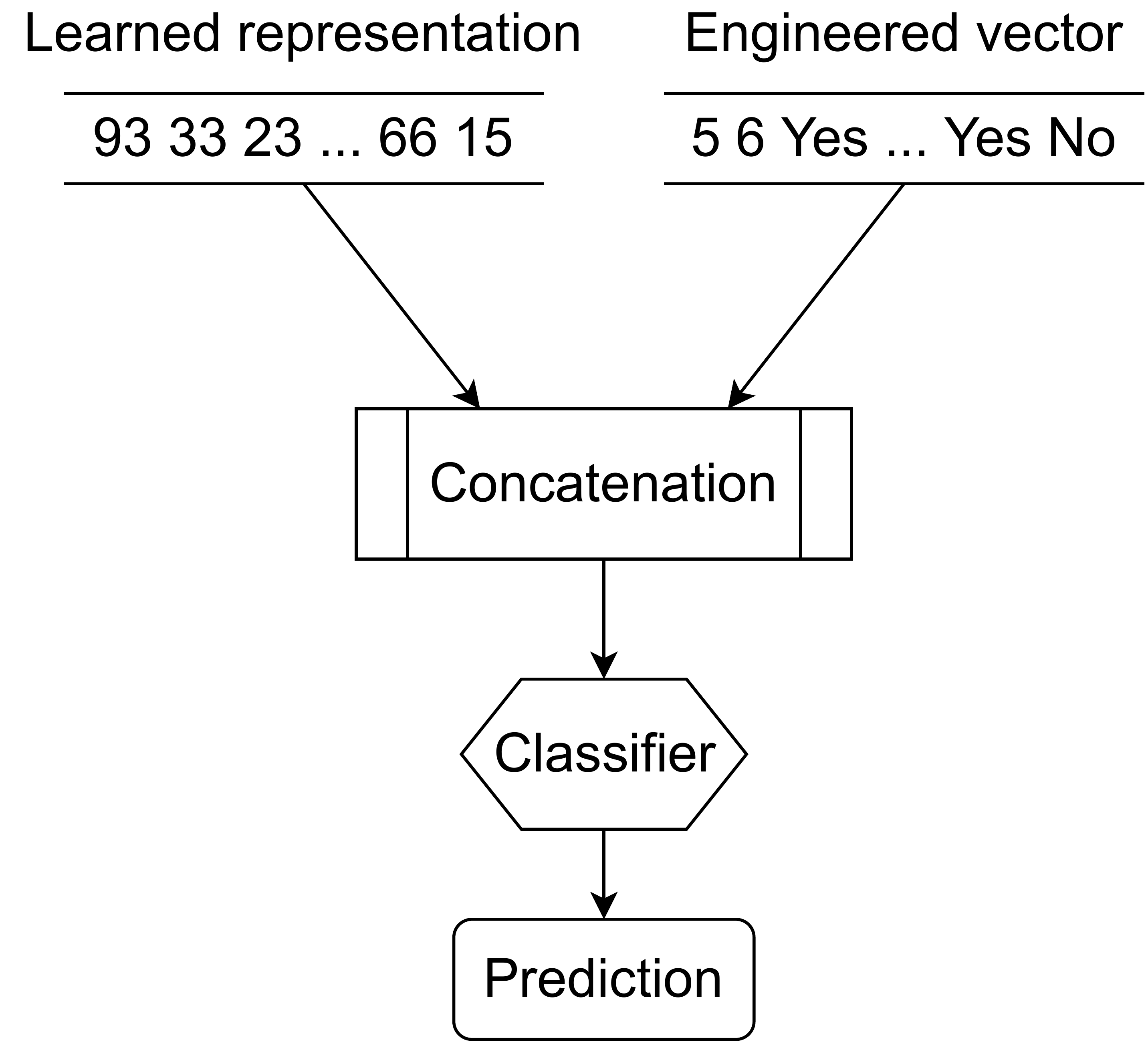}
    }
    \subfigure[Deep Combination.]{
    \centering
    \includegraphics[width=0.3\linewidth, ]{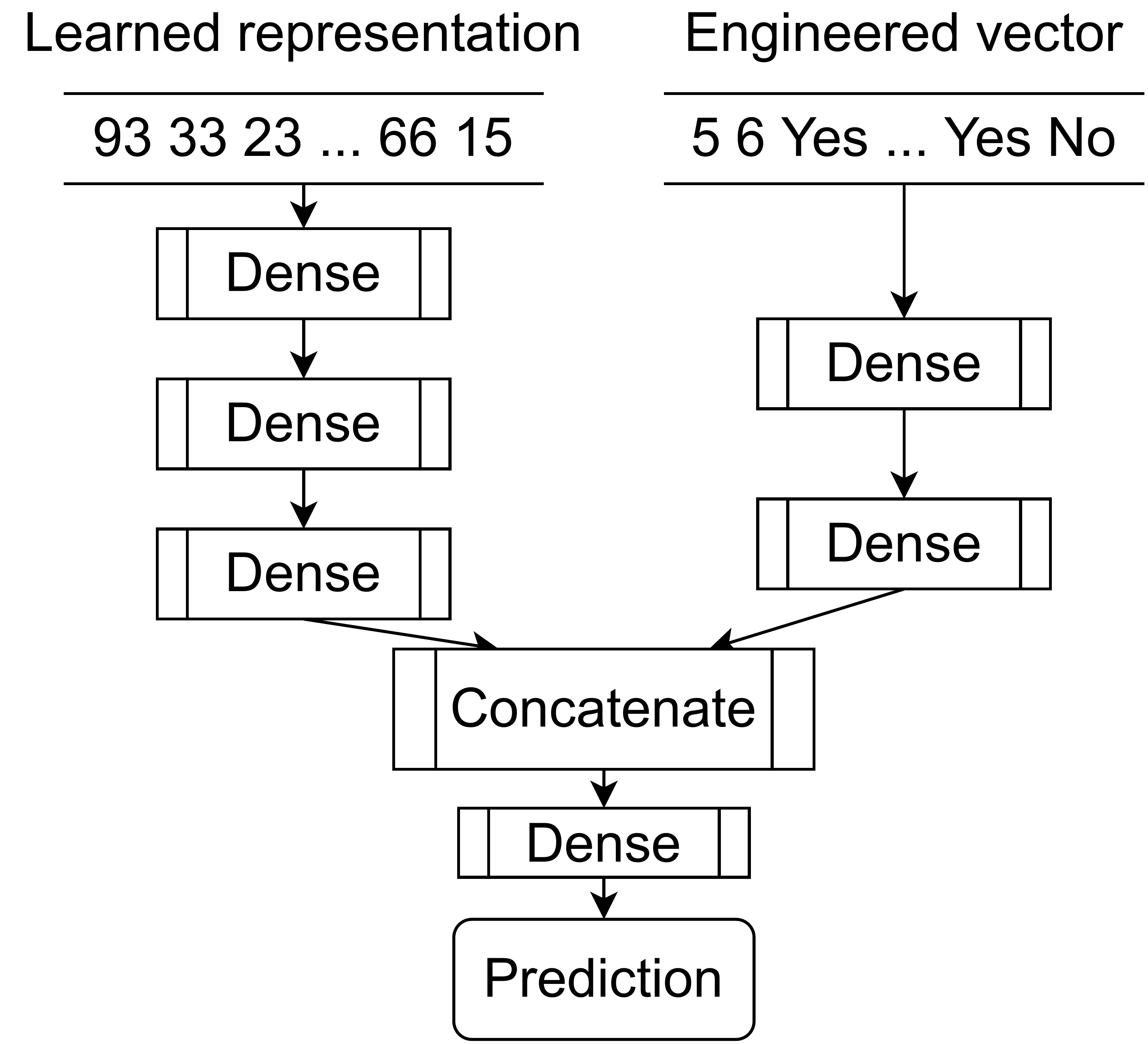}
    }
    \caption{Combination options of features for patch classification in \tool.}
    \label{fig:combination}
\end{figure}

{\bf Experimental Design:}
Leveraging different feature sets can be achieved in several ways, e.g., by concatenating feature vectors or by performing ensemble learning. 
In this study, we investigate three different methods which are implemented in the upgraded version of \toolname, \tool (\ul{P}redict p\ul{A}tch correct\ul{N}ess wi\ul{TH} the learned \ul{E}mbbeddings and enginee\ul{R}ed features), as illustrated in Figure~\ref{fig:combination}:
\begin{enumerate}%
	\item {\em Ensemble learning.}
	We rely on the six learning algorithms (cf. Tables~\ref{tab:ML-all} and~\ref{tab:ods-all-2}) to predict the correctness of patches based either on the \embeddings or on the engineered features. Eventually, to combine both, we
	simply compute the average prediction probability provided by a pair of classifiers (one trained with \embeddings and the other with engineered features), and use this probability to decide on patch correctness. 
	\item {\em Na{\"i}ve Vector Concatenation.}
	In the second method, we ignore the fact that \embeddings vectors and engineered feature vectors are not from the same space and propose to Na{\"i}vely concatenate them into a single representation. Our intuition, indeed, is that both representations capture different features of patches and can therefore offer, together, a better representation. The yielded concatenated vectors are then used to train the classifiers (with the usual learning algorithms).
	\item {\em Deep Combination.} In the last method, we consider that \embeddings and engineered features are from different spaces. Therefore, we must learn their different weights as well as the common representations for them before concatenation. We resort thus to deep neural networks to attempt a deep combination of feature sets before classification.
\end{enumerate}
In this RQ, given the performance of BERT in previous experiments (cf. Table~\ref{tab:ML-all}), we focus on the BERT embedding model to learn the \embeddings of patches. Similarly, we only consider Random forest and XGBoost as the best learners to be applied (cf. Table~\ref{tab:ML-all} and Table~\ref{tab:ods-all-2}). 
The {\em Deep Combination} method is based on the work of Cheng~{\em et~al.}~\cite{cheng2016wide} who proposed a deep learning fusion structure which combined layers that were specialized to explore memorization and generalization of features. 
Following up this idea of fusion, we design a Double-DNN-fusion structure where \embeddings are considered useful for generalization and engineered features are considered for memorization.
Eventually, we conduct 10-group cross validation for the experimental assessment.

{\bf Results:} 
Table~\ref{tab:combine-ml} presents the performance comparison for correctness identification when using combined features vs using single feature sets. The comparison is done in terms of three main metrics: +Recall (to what extent correct patches can be identified), -Recall (to what extent incorrect patches can be filtered out), and AUC (area under the ROC curve, i.e. comprehensive performance of the predictor).
Overall, the performance of classifying correct patches is improved after using each of the three combination strategies (except the -Recall of the random forest classifier with the {\em Na{\"i}ve Vector Concatenation}) for the learned (BERT) and engineered (ODS) feature.
With respect to {\bf +Recall} (i.e., recalling the correct patches), the Random forest and XGBoost based classifier with {\em Ensemble Learning} achieve the highest value at 83.7\%,  improving by 1 to 6 percentage points the performance with single feature sets. 
With respect to {\bf -Recall} (i.e., filtering out the incorrect patches), the best classifier is DNN-based with the {\em Deep Combination} of features: it achieves the highest recall in correctly excluding 69.6\% of the incorrect patches.
With respect to {\bf AUC}, the XGBoost-based classifier with the {\em Ensemble Learning} present the best performance at 82.2\%, improving by 2 to 5 percentage points the performance with single feature sets.
To sum up, combining the BERT embeddings of patches with their ODS features does improve the performance of identifying patch correctness.
Note that the results show that, in general, Ensemble Learning applied to independently trained classifiers yields the highest performance gains. 
The McNemar’s statistical hypothesis test~\cite{dietterich1998approximate} further confirms that the gains are statistically significant for the {\em Ensemble Learning} and {\em Deep Combination} while it is not the case for the {\em Na{\"i}ve Vector Concatenation}.
This suggest that the features (learned and engineered) are from different spaces and are best exploited when applied standalone to model patch correctness, and can complement each other in terms of prediction.

\begin{table}[!t]
	\centering
	\caption{Comparing results of classifying correct patches with combined feature against the single feature.}
	\label{tab:combine-ml}
	\resizebox{1\linewidth}{!}
	{
	\begin{threeparttable}
		\begin{tabular}{l|l|cccccC{11mm}}
			\toprule
			{\bf Tool} & {\bf Feature}  & {\bf Accuracy} & {\bf Precision} & {\bf +Recall} & {\bf -Recall} & {\bf F1-measure} &{\bf AUC}  \\
			\midrule
			\multicolumn{8}{c}{Random Forest}\\
			\midrule
			\multirow{2}{*}{\toolname} & {BERT embeddings} & 0.694 & 0.683 & 0.779 & 0.624 & 0.755 & 0.793  \\
			\cline{2-8}
			 & {Engineered feature} & 0.707 & 0.721 & 0.775 & 0.620 & 0.747 & 0.769  \\
			\cline{1-8}
			\multirow{2}{*}{\tool} & {\em Ensemble Learning} & 0.745 & 0.740 & {\bf 0.837} & {\bf 0.629} & 0.786 & {\bf 0.818} \\
			\cline{2-8}
			& {\em Na{\"i}ve Vector Concatenation} & 0.708 & 0.693 & 0.786 & 0.629 & 0.766 & 0.799  \\\midrule
			
			\multicolumn{8}{c}{XGBoost}\\
			\midrule
			\multirow{2}{*}{\toolname} & {BERT embeddings} & 0.718 & 0.716 & 0.821 & 0.588 & 0.765 & 0.803 \\
			\cline{2-8}
			& {Engineered feature} & 0.705 & 0.726 & 0.799 & 0.596 & 0.741 & 0.776 \\\cline{1-8}
			\multirow{2}{*}{\tool} & {\em Ensemble Learning} & 0.757 & 0.754 & {\bf0.837} & {\bf0.655} & 0.794 & {\bf0.822} 
			\\\cline{2-8}
			& {\em Na{\"i}ve Vector Concatenation} & 0.730 & 0.725 & 0.833 & 0.600 & 0.775 & 0.811 \\\midrule
			
			\multicolumn{8}{c}{DNN}\\
			\midrule
			\multirow{2}{*}{\toolname} & {BERT embeddings} & 0.703 & 0.744 & 0.713 & 0.690 & 0.728 & 0.767 \\
			\cline{2-8}
			& {Engineered feature} & 0.698 & 0.721 & 0.748 & 0.634 & 0.734 & 0.777  \\\cline{1-8}
			{\tool} & {\em Deep Combination} & 0.730 & 0.760 & {\bf0.757} & {\bf0.696} & 0.758 & {\bf0.798}\\
			\bottomrule
		\end{tabular}
	\end{threeparttable}
	}
\end{table}

\begin{figure}[!t]
    \centering
    \subfigure[Identified correct patches.]{
    \centering
    \includegraphics[width=0.4\textwidth]{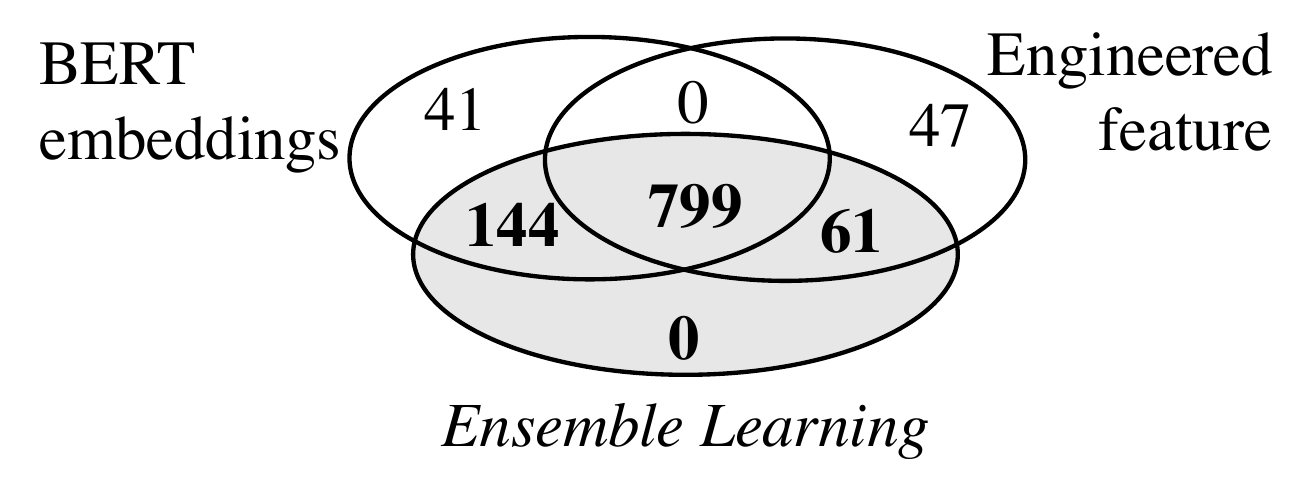}
    }
    \subfigure[Identified incorrect patches]{
    \centering
    \includegraphics[width=0.4\linewidth, ]{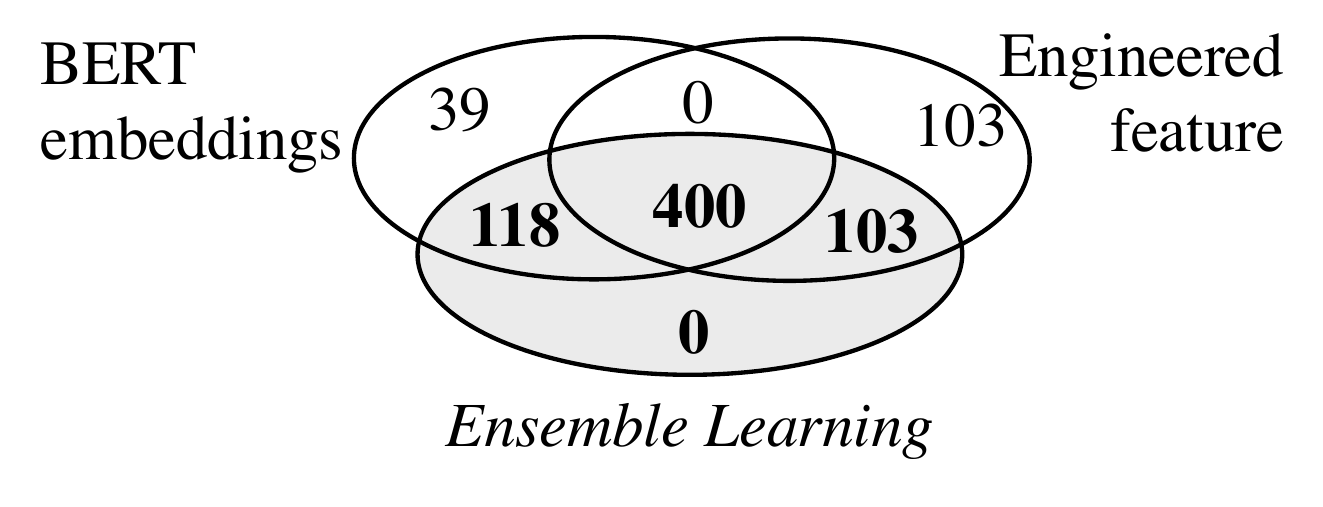}
    }
    \centering
    \subfigure[Identified correct patches.]{
    \centering
    \includegraphics[width=0.4\textwidth]{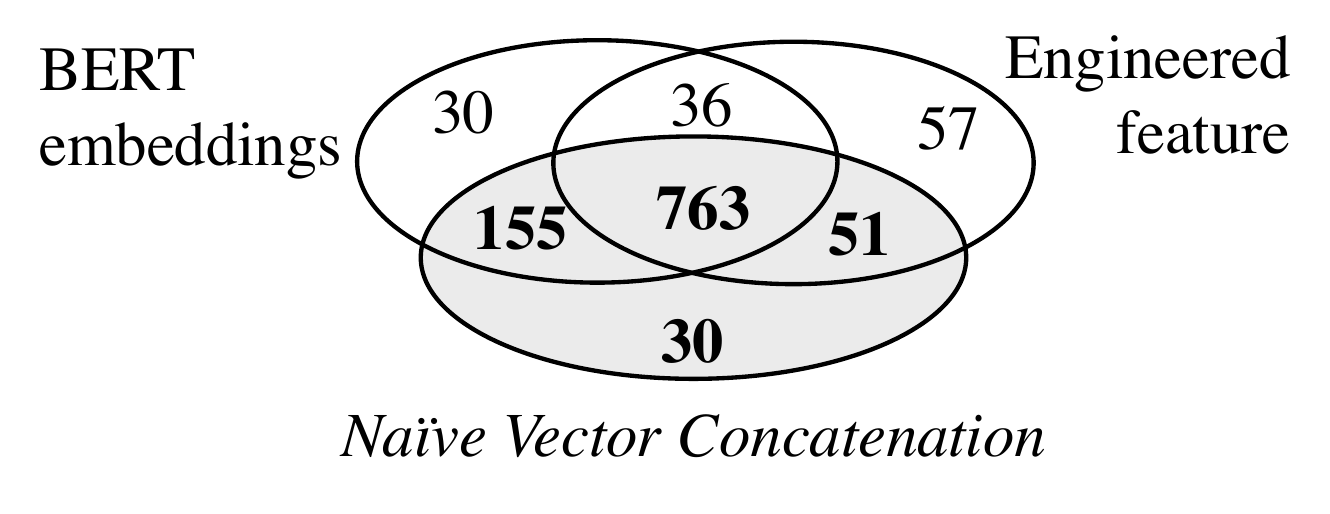}
    }
    \subfigure[Identified incorrect patches.]{
    \centering
    \includegraphics[width=0.4\linewidth, ]{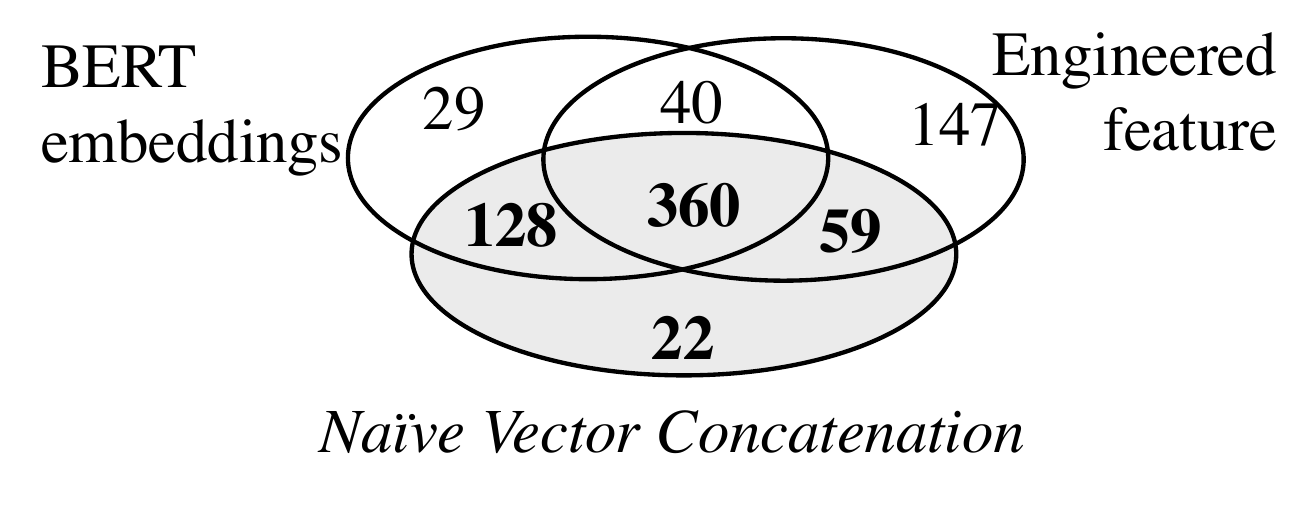}
    }
    \centering
    \subfigure[Identified correct patches.]{
    \centering
    \includegraphics[width=0.4\textwidth]{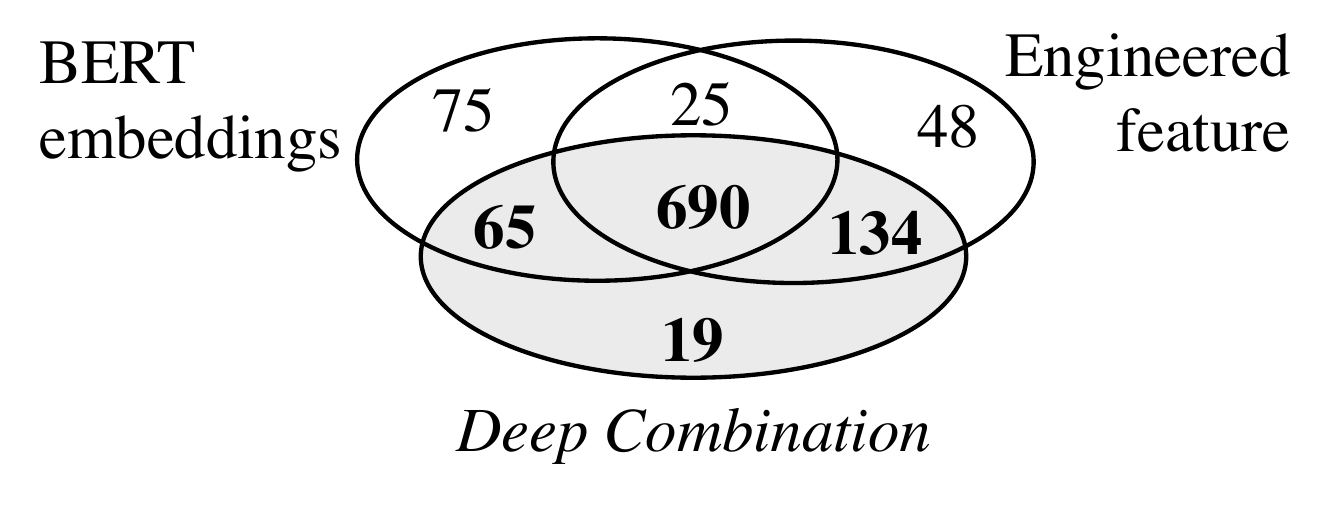}
    }
    \subfigure[Identified incorrect patches.]{
    \centering
    \includegraphics[width=0.4\linewidth, ]{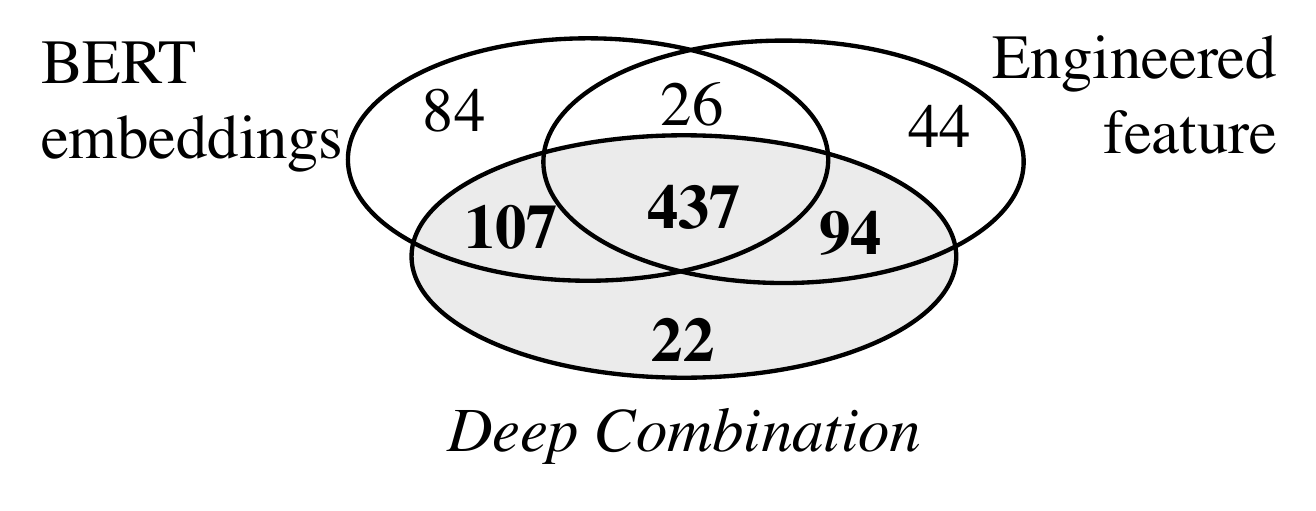}
    }
    \caption{Comparison on the number of patches identified with the combined feature vs. the simple feature.}
    \label{fig:Venn}
\end{figure}

Figure~\ref{fig:Venn} further highlights the number of (in)correct patches identified based on BERT embeddings, engineered features and the combined features, respectively. 
Since the ``Random forest'' learner presents a similar performance with ``XGBoost'', Figure~\ref{fig:Venn} focuses on the latter.

From a {\bf qualitative point of view}, with the {\em Ensemble Learning}, more (in)correct patches can be identified than each single feature set (i.e., BERT embeddings or engineered features). However, this combination does not help to identifying patches that were not identified using at least one feature set.
In contrast, with {\em Na{\"i}ve Vector Concatenation} and the {\em Deep Combination}, which combine features before classification, we can identify some (in)correct patches that could not be identified using either feature set alone.

From a {\bf quantitative point of view}, the {\em Na{\"i}ve Vector Concatenation} helps to identify slightly more correct patches (among those that could not be identified by each feature set alone) than the {\em Deep Combination}.
As for new identified incorrect patches, they achieve the same metrics. Nevertheless, overall, the {\em Ensemble Learning} method helps to identify more correct patches while the {\em Deep Combination} helps to identify more incorrect patches.

\find{{\bf \ding{45} RQ-4 }$\blacktriangleright$ Leveraging \embeddings (BERT) and engineered features (ODS) contributes to improve the performance in predicting patch correctness for APR tools. Merging independently trained classifiers achieves higher performance compared to each separate classifier, but does not lead to the identification of correct/incorrect patches that could not be identified by at least one of the classifier.  
In contrast, feature combination (i.e., {Na{\"i}ve Vector Concatenation} and {Deep Combination}) before classification training appears to provide more information to discriminate some patches that were not correctly classified based on their \embeddings or their engineered features alone. $\blacktriangleleft$}

\subsection{[RQ-5: Explanation of Improvements of Combination]}

\paragraph{\bf Objective:} 
The experimental results for previous RQs show that ML classifiers built based on \embeddings,  or on engineered features, or on both, yield promising performance in predicting patch correctness. %
The fact remains, however, that the classifier is a black box model for practitioners. In particular, when leveraging combined feature sets, it may be helpful to investigate the impact of different features on the identification of patch correctness. To that end, we propose to build on Explainable ML techniques to explore how the models are built. In this work, we focus on Shapley Values, which compute the contributions of each feature in a given prediction. Shapley values originate from the field of game theory and have been implemented in the SHAP framework~\cite{NIPS2017_7062}, which is widely used in the AI community.

{\bf Experimental Design:} Our experiments are focused on the classifier yielded with  the {\em Na{\"i}ve Vector Concatenation} method since it managed to recall more correct patches through combining \embeddings and engineered features (cf. RQ-3.3 in Section~\ref{sec:classifiers}). We consider the case where the classifier is trained with the XGBoost learning algorithm.
Using SHAP values as metric of feature importance, we investigate the top most important features that contribute to the combined model predictions. We further compare those important features against the features that are most contributing when the classifier is trained only with \embeddings or only with engineered features.
Finally, we present three specific patches that identified by different feature sets to observe the contribution of the features to prediction.

\begin{figure}[!h]
    \centering
    \includegraphics[width=0.75\textwidth]{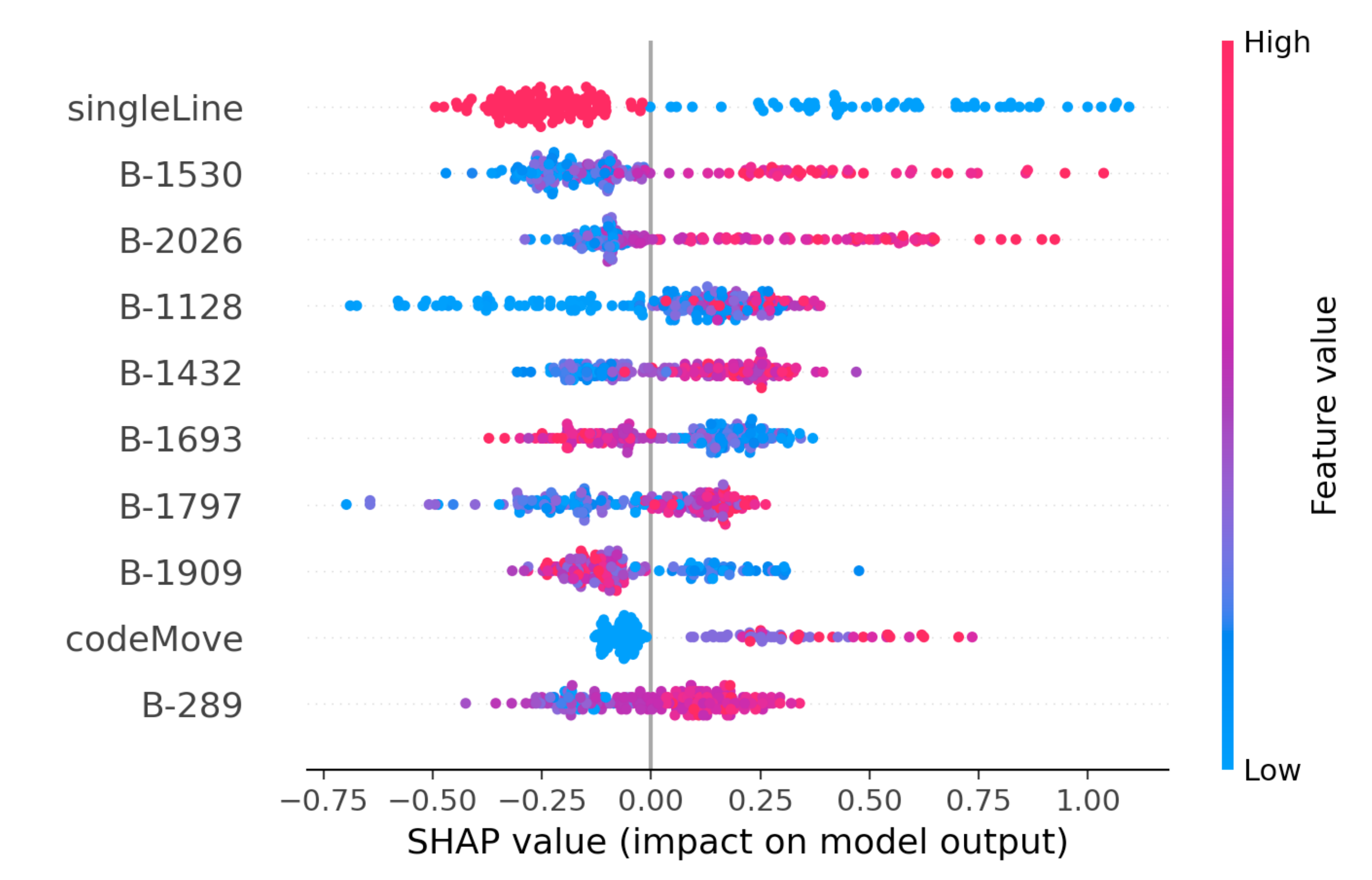}
    \caption{Top-10 Contributing Features (based on SHAP values) for the Classifier built by combining \embeddings and engineered features.}
    \label{fig:shap_c}
\end{figure}

{\bf Results:} Figure~\ref{fig:shap_c} illustrates the top-10 most contributing features: a feature named B-{\em i} refers to the i$^{th}$ feature learned with BERT. Others (e.g., $singleLine$ and $codeMove$) refer to engineered features. The appearance of features from learned and engineered feature sets among the most contributing features suggests that both types of features are not only relevant but are also exploited in the yielded classifier.

{\em Reading a SHAP explanation graph:} In a given SHAP graph, each row is a distribution values for a given feature (Y-axis), where each data point is associated to one sample input data (i.e., a patch in our case). The color indicates the feature value, which is normalized: the more red, the higher the value. The X-axis represents the SHAP values, which indicate to what extent a given feature impacted the model output for a given patch. For example, most patches with high value (red) for feature {\em singleLine} are located on the left (negative SHAP value), which suggests negative impact of {\em singleLine} on correctness prediction. It should be noted that, eventually, it is the contributions of different features that will be merged to yield the final prediction for each sample.

\begin{figure}[!h]
    \centering
    \includegraphics[width=0.8\textwidth]{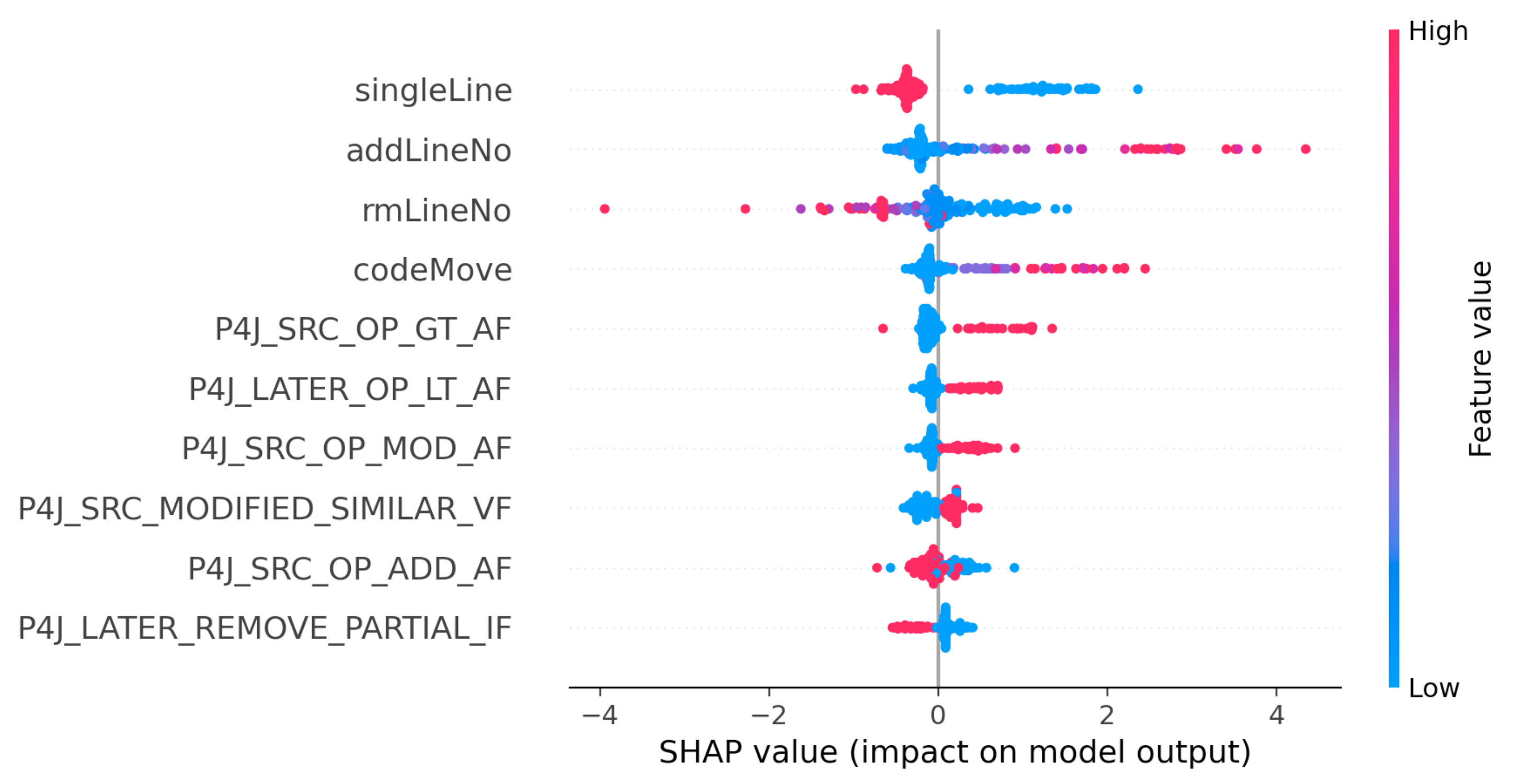}
    \caption{Top-10 contributing features (based on SHAP values) for the Classifier built only by the engineered features.}
    \label{fig:shap_o}
\end{figure}

In Figure~\ref{fig:shap_c}, we note that {\em singleLine} and {\em codeMove} are the top contributing engineered features among the combined feature sets. As we see from the figure, their red (high value) points and blue (low value) points are clearly separated to two sides, which demonstrates their values have obvious positive or negative effects on the model output. 
In Figure~\ref{fig:shap_o}, when leveraging only engineered features, {\em singleLine} and {\em codeMove} also have significant contributions and are appearing in the 1st and 4th positions among the top contributing features. This indicates that the engineered features must be high-contributors to the decision (e.g., in terms of information gain) as shown in Figure~\ref{fig:shap_o},  in order to obtain an efficient combination with learned features. Therefore, in practice we suggest that the research community should focus more on devising few but effective engineered features instead of massive but inefficient features to improve the performance of models.

Overall, the SHAP explanations suggest that engineered features have an important effect on model prediction (because they appear among the top contributing features) but are complementary to the learned feature set. 
Indeed, the combination with {\em Naive Vector Concatenation} enables classifiers to identify correct patches that could not be identified when each feature set was used without the other. Therefore, we conclude that it is the interaction among the features that yields such a performance improvement. We propose to further investigate the interaction among pairs of features (one from the engineered features set and the other from the learned features set). 

\begin{figure}[!h]
    \centering
    \subfigure[Interaction between singleLine and B-1530.]{
    \centering
    \includegraphics[width=0.45\textwidth]{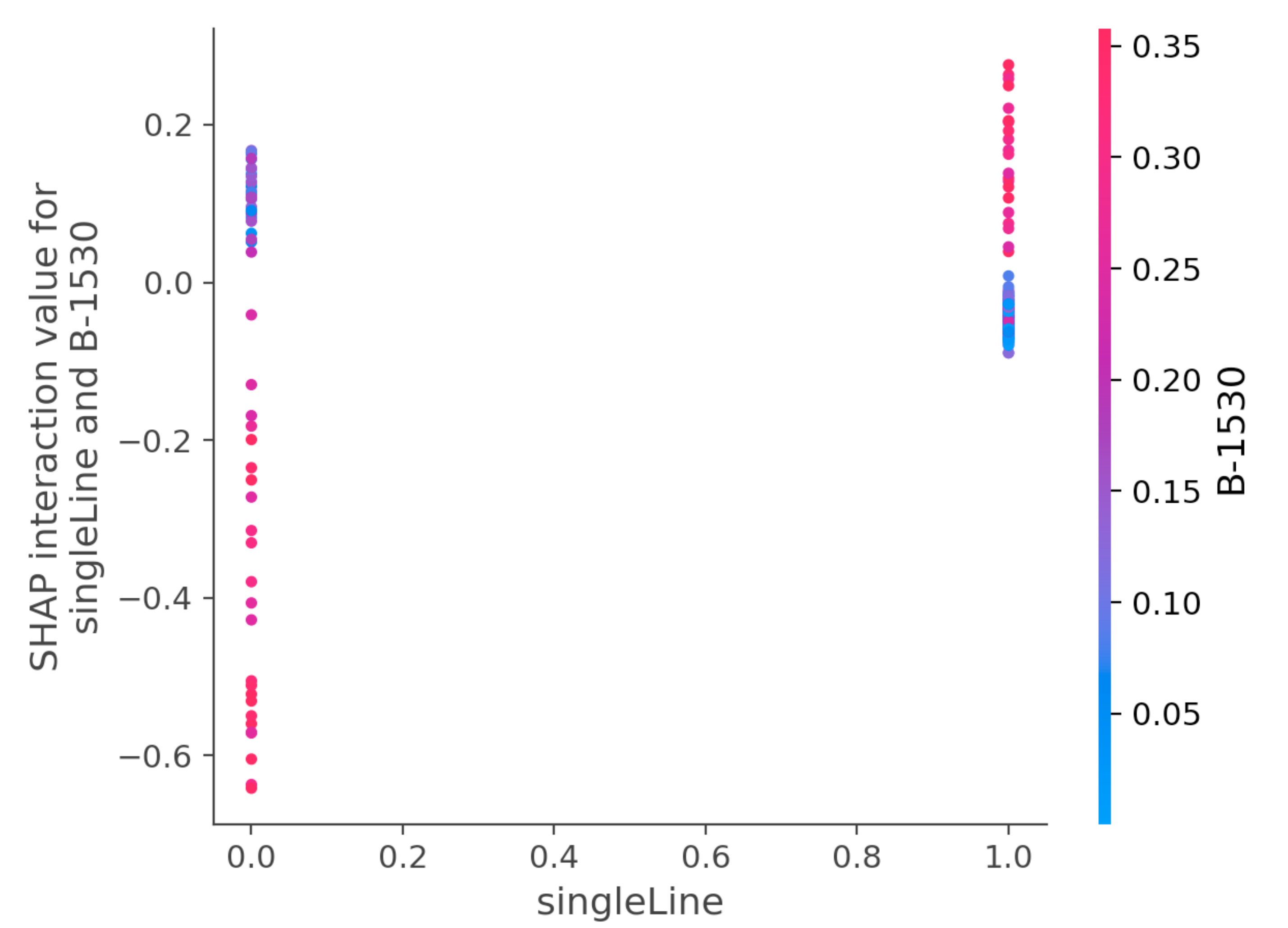}
    \label{fig:singleline_1530}
    }
    \subfigure[Interaction between singleLine and B-2026.]{
    \centering
    \includegraphics[width=0.45\linewidth,]{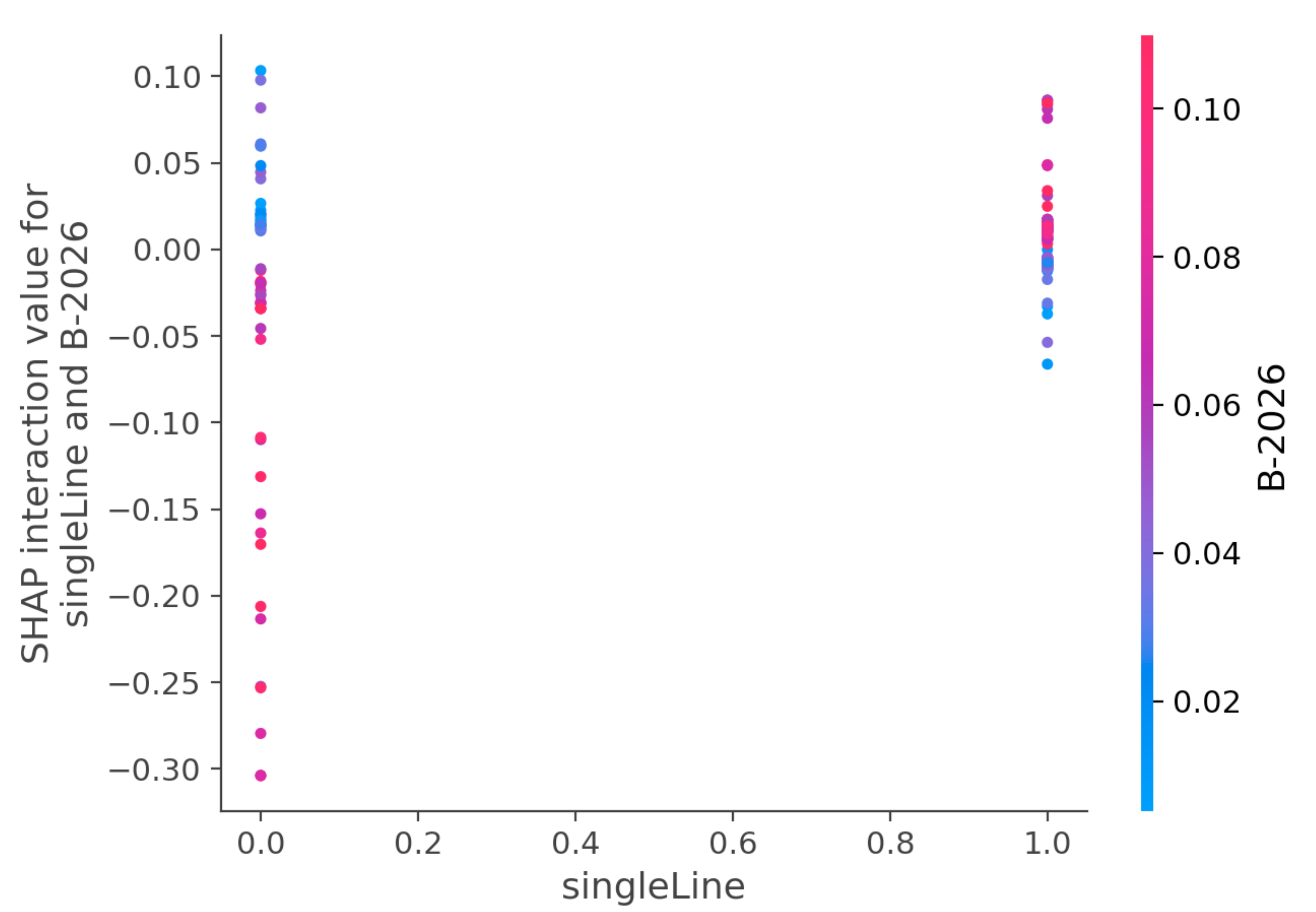}
    \label{fig:singleline_2026}
    }
    \centering
    \subfigure[Interaction between codeMove and B-1530.]{
    \centering
    \includegraphics[width=0.45\textwidth]{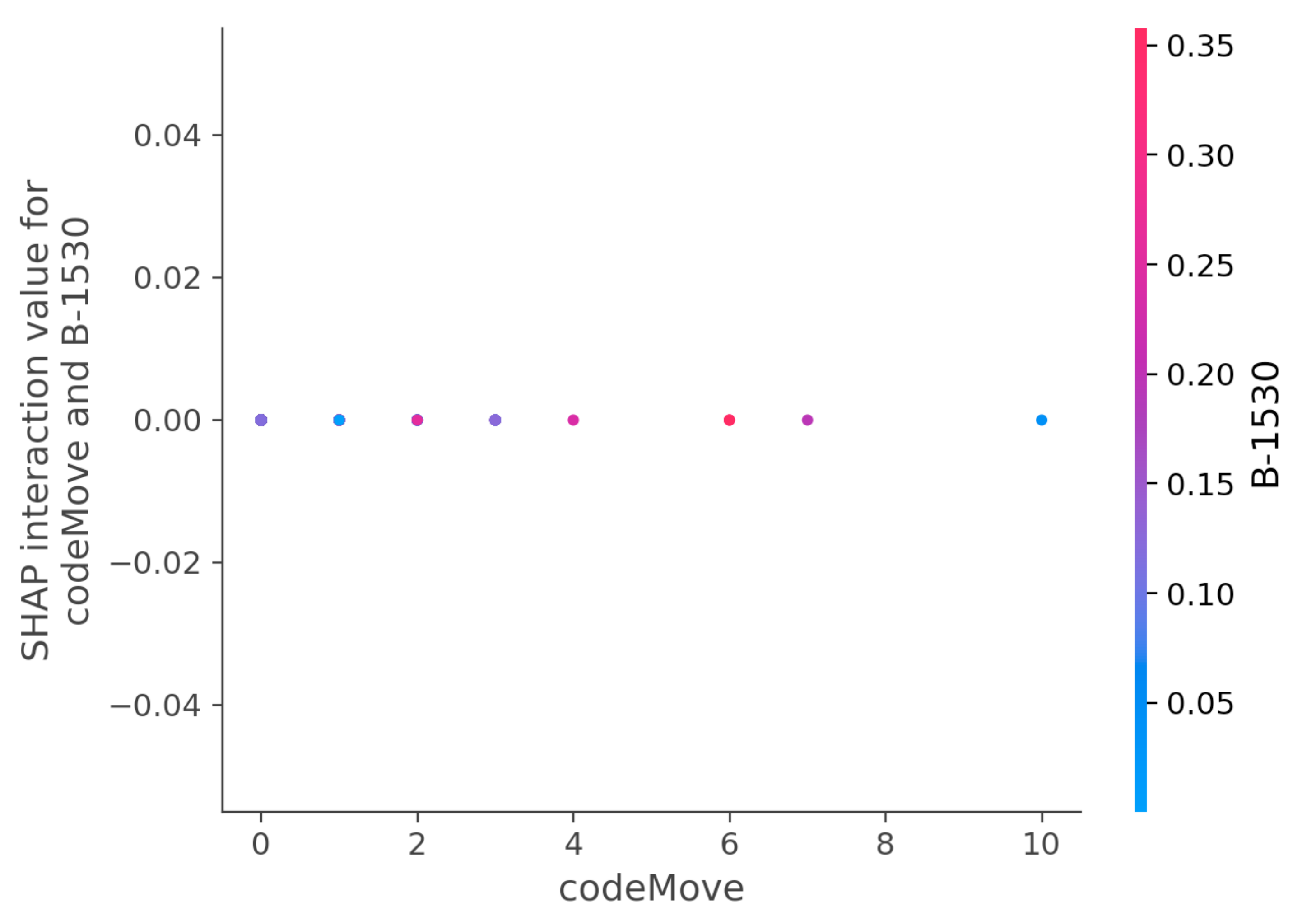}
    \label{fig:codemove_1530}
    }
    \subfigure[Interaction between codeMove and B-2026.]{
    \centering
    \includegraphics[width=0.45\linewidth,]{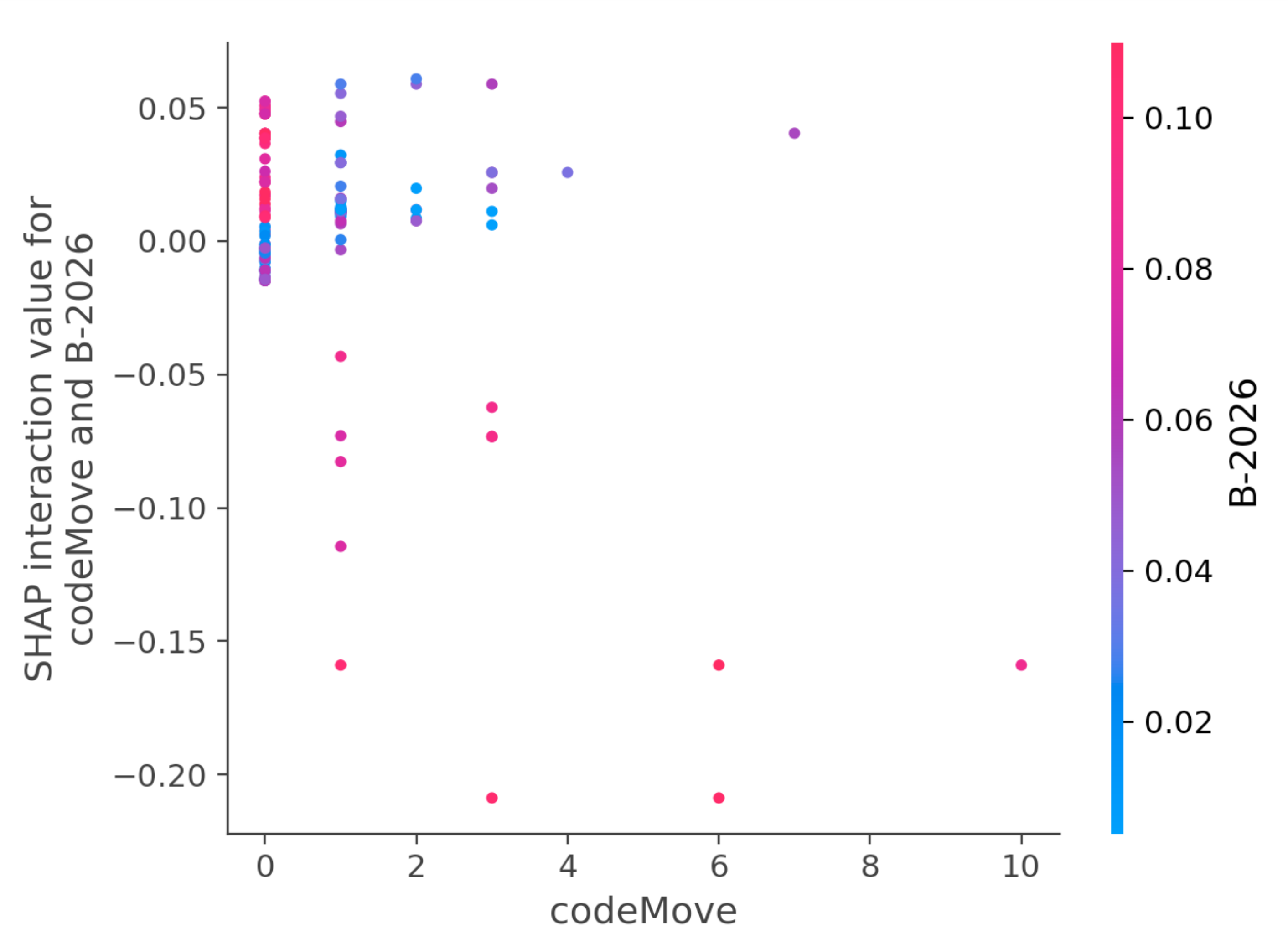}
    \label{fig:codemove_2026}
    }
    \caption{Feature Interaction.}
    \label{fig:feature_interaction}
\end{figure}

Figure~\ref{fig:feature_interaction} illustrates the interactions information provided by SHAP among {\em singleLine}, {\em codeMove} and {\em B-1530}. As it can be seen, in Figure \ref{fig:singleline_1530}, when the feature value of {\em singleLine} is 0, higher (redder) feature values of {\em B-1530} will lead to a more negative SHAP value for {\em singleLine} (i.e., it has negative impact on patch correctness prediction). In contrast, when the feature value of {\em singleLine} is 1, the same higher feature values of {\em B-1530} will tend to draw a positive SHAP value (i.e., positive impact). This example illustrates how learned and engineered features can interact to balance their contributions for the final predictions based on their respective feature values.
Figure~\ref{fig:singleline_2026} and Figure~\ref{fig:codemove_2026} exhibit effective interaction while Figure~\ref{fig:codemove_1530} cannot because not enough of the test data are reaching both the two feature nodes in the tree-based boosting classifier. In the same direction, we cannot present the SHAP interaction between {\em singleLine} and {\em codeMove}. 
Overall, Figure~\ref{fig:feature_interaction} provides evidence for the impact of the interaction between learned and engineered features on the model prediction. In contrast, merging classifiers through {\em Ensemble Learning} does not allow for features interaction and thus fails to identify patches that were not identified using one feature set. This motivates model trainers to combine different types of features through tree-based classifiers or deep neural networks to obtain efficient deep information for identifying previously-unidentified correct patches.

\begin{figure}[!t]
    \centering
    \subfigure[Patch for Closure-57.]{
    \centering
    \includegraphics[width=1\textwidth]{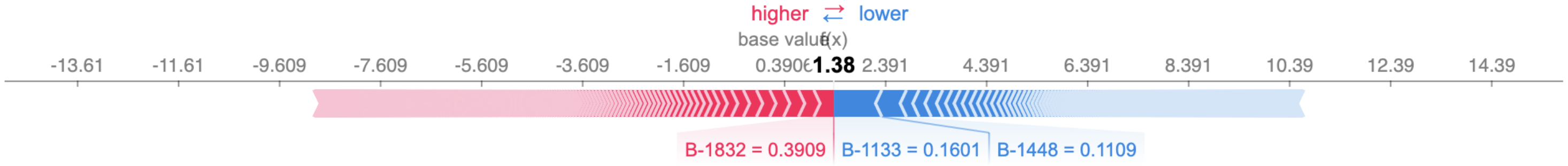}
    \label{fig:shap_sample_b}
    }
    \vspace{-2mm}
    \subfigure[Patch for Math-85.]{
    \centering
    \includegraphics[width=1\textwidth]{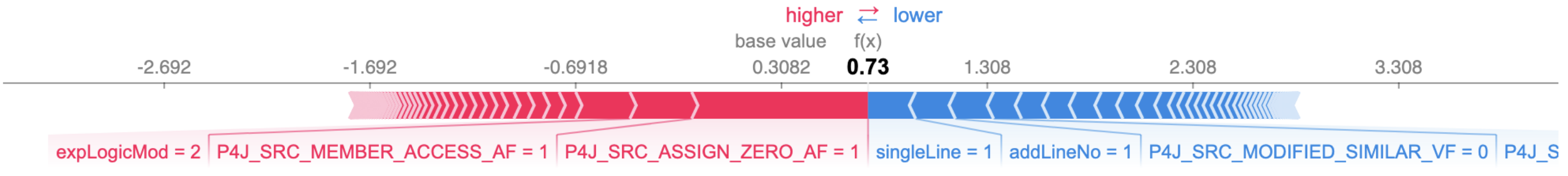}
    \label{fig:shap_sample_o}
    }
    \subfigure[Patch for Math-56.]{
    \centering
    \includegraphics[width=1\linewidth, ]{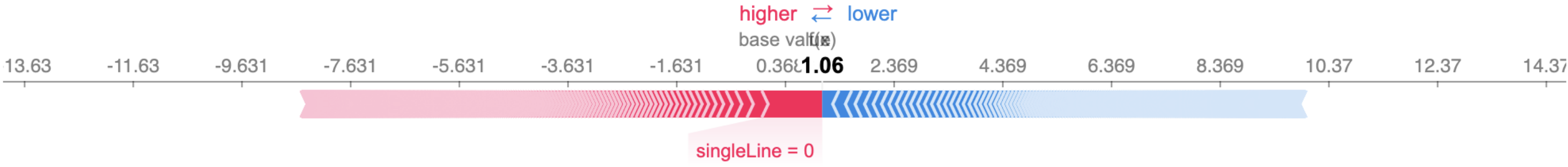}
    \label{fig:shap_sample_c}
    }
    \caption{SHAP Analysis on Patches.}
    \label{fig:shap_patch}
\end{figure}

Finally, Figure~\ref{fig:shap_patch} presents the SHAP analyses of three patches that are exclusively identified by classifiers built based either on learned feature set (a), or on engineered feature set (b), or on combined feature set (c).
We note that contributions of each learned feature is small and it is the sum of contributions that lead to a prediction. In contrast, contributions of engineered features are significantly larger for several features. When the sets are combined, engineered features are contributing in the top, their contributions are impactful, while learned features still contribute, each, to a lesser extent. 
Overall, few engineered features make most of the contributions for good prediction which unsurprisingly imply that the quality and relevance of engineered features are more important than the number of features.

\find{{\bf \ding{45} RQ-5 }$\blacktriangleright$ Thanks to SHAP explanations, we were able to confirm that combining engineered and learned feature sets creates interactions that impact the prediction of classifiers, leading to improved precision and recall in correctness prediction. $\blacktriangleleft$}

\section{Discussions}
\label{sec:discussion}
We enumerate a few insights from our experiments with representation learning models and discuss some threats to validity.

\subsection{Insights from the Experimental Setup}
{\em Code-oriented embedding models may not yield the best embeddings for training patch correctness classifiers.} Our experiments have revealed that the BERT model, which was pre-trained on Wikipedia, is yielding the best recall in the identification of incorrect patches. There are several possible reasons for this situation: BERT implements the deepest neural network and builds on the largest training data. Its performance suggests to researchers that code-oriented embedding models should be trained on large code datasets or fine-tuned on specific target tasks in order to become competitive against BERT.
While we were completing the experiments, a pre-trained CodeBERT~\cite{feng2020codebert} model has been released. In future work, we will investigate its relevance for producing embeddings that may yield higher performance in patch correctness prediction. In any case, we note that CC2Vec provided the best embeddings for yielding the best recall in identifying correct patches (using similarity thresholds). This finding suggests we use the embedding model built for code changes (e.g., CC2Vec) for the objective of having a high recall in identifying correct patches.

\begin{figure}[!h]
	\includegraphics[width=0.85\linewidth]{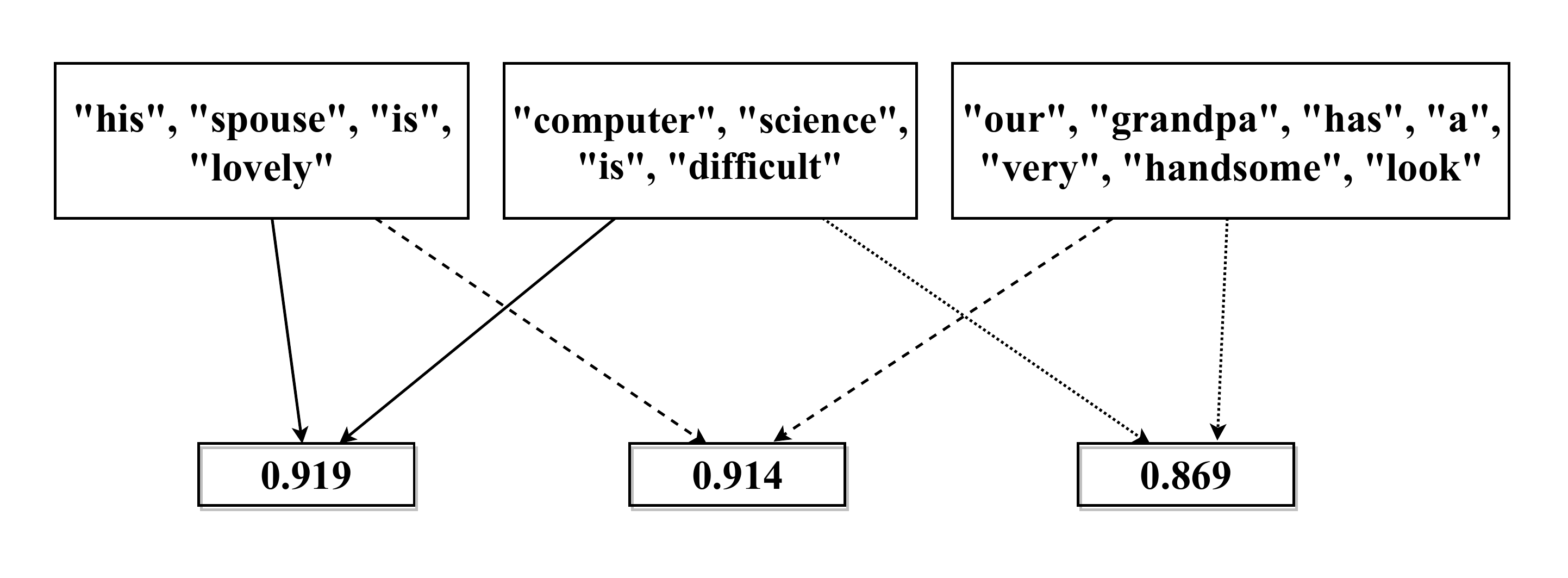}
	\caption{\normalsize Close cosine similarity scores with small-sized inputs for BERT embedding model.}
	\label{fig:input-size}
\end{figure}

{\em The small sizes of the code fragments lead to similar embeddings.}
Figure~\ref{fig:input-size} illustrates the different cosine similarity scores that can be obtained for the BERT embeddings of different pairs of short sentences. Although the sentences are semantically (dis)similar, the cosine similarity scores are quite close. This explains why recalling correct patches based on a similarity threshold was a failed attempt ($\sim 5\%$ for APR-generated patches for Defects4J+Bears+Bugs.jar bugs). Nevertheless, experimental results demonstrated that deep learned features are relevant for learning to discriminate.
Considering the different sizes of code fragments contained in each patch may affect the similarity computation, we suggest that researchers control the size of the code fragments of the patch when investigating the hypothesis in RQ-2 for patch correctness.

{\em Refutation of literature assumption that ``patches with fewer changes are more likely to be correct''.}
In RQ-2, we leveraged similarity between buggy code and patched code to filter out incorrect patches. The hypothesis is the more similar they are, the more likely to be correct the patch is. The best performance appears in QuixBugs that only contain bug on one single line. However, regarding Bears, Bugs.jar and Defects4j, while a large number of incorrect patches are filtered out (cf. -Recall in Table~\ref{tab:filtering}), correct patches are recalled in low numbers (cf. +Recall in Table~\ref{tab:filtering}). Or, -Recall is low while keeping high +Recall. In the RQ-5, we use ground-truth labeled developer's patches and generated patches with balanced numbers for Defects4j to avoid bias. We use SHAP to interpret the impact of feature and find the most important feature is ``singleLine''. The feature analysis implies that patch with one single line (fewer change) is more likely to be incorrect, which is against the hypothesis. This demonstrates correct code normally require more than one-line change.

\subsection{Threats to Validity}

Our empirical study carries a number of threats to validity that we have tried to mitigate.

{\sc \em Threats to External Validity.} There are a variety of representation learning models in the literature. A threat to validity of our study is that we may have a selection bias by considering only four embedding models. We have mitigated this threat by considering representative models in different scenarios (pre-trained vs retrained, code change specific vs natural language oriented).

Another threat to validity is related to the use of Defects4J data in evaluating the ML classifiers. This choice however was dictated by the data available and the aim to compare against related work.

{\sc \em Threats to Internal Validity.}
A major threat to internal validity lies in the manual assessment heuristics that we applied to the RepairThemAll-generated dataset. We may have misclassified some patches due to mistakes or conservatism. 
This threat however holds for all APR work that relies on manual assessment. 
We mitigate this threat by following clear and reproducible decision criteria, and by further releasing our labelled datasets for the community to review\footnote{see: \url{https://github.com/HaoyeTianCoder/Panther}}. 
Besides, we supplement the dataset with developer patches to mainly relieve the imbalance problem of the dataset. This may make the sample distribution of our experiment different from the real APR patches world. This threat however also holds for some current works~\cite{xiong2018identifying, ye2019automated} that focus on patch correctness validation.

{\sc \em Threats to Construct Validity.}
For our experiment, the considered embedding models are not perfect and they may have been under-trained for the prediction task that we envisioned. For this reason, the results that we have reported are likely
an under-estimation of the capability of representation learning models to capture discriminative features for the prediction of patch correctness.
Our future studies on representation learning will
address this threat by considering different re-training experiments.
\section{Related Work}
\label{sec:relatedWork}

\paragraph{\bf Analyzing Patch Correctness:}
To assess the performance of fixing bugs of repair tools and approaches, 
checking the correctness of patches is key, but not trivial.
However, this task was largely ignored or unconcerned in the community until the analysis study of patch correctness conducted by Qi~{\em et~al.}~\cite{qi2015analysis}.
Thanks to their systematic analysis of the patches reported by three generate-and-validate program repair systems (i.e., GenProg, RSRepair and AE), they shown that the overwhelming majority of the generated patches are not correct but just overfit the test inputs in the test suites of buggy programs.
In another study, Smith~{\em et~al.}~\cite{smith2015cure} uncover that patches generated with lower coverage test suites overfit more.
Actually, these overfitting patches often simply break under-tested functionalities, and some of them even make the ``patched'' program worse than the un-patched program. 
Since then, the overfitting issue has been widely studied in the literature.
For example, Le~{\em et~al.}~\cite{le2018overfitting} revisit the overfitting problem in semantics-based APR systems.
In~\cite{le2019reliability}, they further assess the reliability of authors and automated annotations in assessing patch correctness. They recommend to make  publicly available to the community the  patch correctness evaluations of the  authors. 
Yang~\cite{yang2020exploring} explore the difference between the runtime behavior of programs patched with developer's patches and those by APR-generated plausible patches. 
They unveil that the majority of the APR-generated plausible patches lead to different runtime behaviors compared to correct patches.

\paragraph{\bf Predicting Patch Correctness:}
To predict the correctness of patches, one of the first explored research directions relied on the idea of augmenting test inputs, i.e., more tests need to be proposed. 
Yang~{\em et~al.}~\cite{yang2017better} design a framework to detect overfitting patches. %
This framework leverages fuzz strategies on existing test cases in order to automatically generate new test inputs. In addition, it leverages additional oracles (i.e., memory-safety oracles) to improve the validation of APR-generated patches.
In a contemporary study, Xin and Reiss~\cite{xin2017identifying} also explored to generate new test inputs, with the syntactic differences between the buggy code and its patched code, for validating the correctness of APR-generated patches.
As complemented by Xiong~{\em et~al.}~\cite{xiong2018identifying}, they proposed to assess the patch correctness of APR systems by leveraging the automated generation of new test cases and measuring behavior similarity of the failing tests on buggy and patched programs.

Through an empirical investigation, Yu~{\em et~al.}~\cite{yu2019alleviating} summarized two common overfitting issues: incomplete fixing and regression introduction.
To assist alleviating the overfitting issue for synthesis-based APR systems,
they further proposed \texttt{UnsatGuided} that relies on additional generated test cases to strengthen patch synthesis, and thus reduce the generation of incorrect overfitting patches.

The success of predicting patch correctness using an augmented set of test cases, as it is done in prior work, depends on the quality of the tests. In practice, however, tests with high coverage are often unavailable~\cite{ye2019automated}.
To overcome this limitation, our approach does not rely on new test cases, but instead leverages learning techniques to build representation vectors for buggy and patched code of APR-generated patches. Patch correctness prediction is therefore conducted without the constraints in the availability of test cases. 

To predict overfitting patches yielded by APR tools, Ye~{\em et~al.}~\cite{ye2019automated} propose ODS, an overfitting detection system. 
ODS first statically extracts 4,199 code features at the AST level from the buggy code and generated patch code of APR-generated patches. 
Those features are fed into 
three machine learning algorithms (logistic regression, KNN, and random forest) 
to learn an ensemble probabilistic model for classifying and ranking potentially overfitting patches. 
To evaluate the performance of ODS, the authors considered 
19,253 training samples and 713 testing samples from the  Durieux~{\em et~al.} empirical study~\cite{durieux2019empirical}. With these settings,
ODS is capable of detecting 57\% of overfitting patches.
The ODS approach relates to our study since both leverage machine learning and static features. However, ODS only relies on manually identified features which may not generalize to other programming languages or even other datasets. 

In a  recent work, Csuvik~{\em et~al.}~\cite{csuvik2020utilizing} exploit the textual and structural similarity between the buggy code and the APR-patched code with two representation learning models (BERT~\cite{devlin2019bert} and Doc2Vec~\cite{le2014distributed}) by considering three patch code representation (i.e., source code, abstract syntax tree and identifiers).
Their results show that the source code representation is likely to be more effective in correct patch identification than the other two representations, 
and the similarity-based patch validation can filter out incorrect patches for APR tools.
However, to assess the performance of the approach, only 64 patches from QuixBugs~\cite{ye2019comprehensive} have been considered (including 14 in-the-lab bugs). This low number of considered patches raises questions about the generalization of the approach for fixing bugs in the wild. 
Moreover, unlike our study, new representation learning models (code2vec~\cite{alon2019code2vec} and CC2Vec~\cite{hoang2020cc2vec}) dedicated to code representation have not been exploited.
In our work, we first improve the evaluation of the approaches in the real-world by designing a 10-group cross validation on a large labeled deduplicated dataset of 2,147 patches. Then, we propose an extension of our previous works on predicting patch correctness by combining engineered features~\cite{ye2019automated} and representation learning~\cite{tian2020evaluating} (BERT, Doc2Vec, CC2Vec) together and assessing the effectiveness of each and their combination as well as the improvement of the combination. Our study aims to show how the combinations can be carried out to ensure that patches that could not be identified by either set of features are not identifiable by the combined set.
More recently, Yan et al.~\cite{yan2022crex} proposed to predict the patch correctness of fixing C program bugs through the transfer learning of execution semantics.
Tian et al.~\cite{tian2022change} explored the relationship between the bug descriptions carried by bug reports and code changes to identify the correctness of patches for the given Java program bugs.

\paragraph{\bf Representation Learning for Program Repair Tasks:}
In the literature, representation learning techniques have been widely explored to boost program repair tasks.
Long and Rinard explored the topic of learning correct code for patch generation~\cite{long2016automatic}. 
Their approach learns code transformation for three kinds of bugs from their related human-written patches.
After mining the most recent 100 bug-fixing commits from each of the 500 most popular Java projects, Soto and Le Goues~\cite{soto2018using} have built a probabilistic model to predict bug fixes for program repair.
To identify stable Linux patches, Hoang~{\em et~al.}~\cite{hoang2019patchnet} proposed a hierarchical deep learning-based method with features extracted from both commit messages and commit code. 
Liu~{\em et~al.}~\cite{liu2018mining2} and Bader~{\em et~al.}~\cite{bader2019getafix} proposed to learn recurring fix patterns from human-written patches and suggest fixes. %
Our paper does not propose a new automated patch generation approach. Instead, we fill a gap in the literature by proposing a comprehensive assessment of the effectiveness of different representation learning models on predicting the correctness of patches generated by program repair tools.

\section{Conclusion}
\label{sec:conc}
In this paper, we investigated the feasibility of statically predicting patch correctness by leveraging representation learning models and supervised learning algorithms. The objective is to provide insights for the APR research community towards improving the quality of repair candidates generated by APR tools.
To that end, we, first investigated the use of different distributed representation learning to capture the similarity/dissimilarity between buggy and patched code fragments.
These experiments gave similarity scores that substantially differ for across embedding models such as BERT, Doc2Vec, code2vec and CC2Vec.
Building on these results and in order to guide the exploitation of code embeddings in program repair pipelines, we investigated in subsequent experiments the selection of cut-off similarity scores to decide which APR-generated patches are likely incorrect.
We then implemented a patch correctness predicting framework, \toolname, to investigate the discriminative power of the deep learned features by training machine learning classifiers to predict correct Patches. 
Decision Trees, Logistic Regression, Na{\"i}ve Bayes, Random Forest, XGBoost, and DNN are tried with code embeddings from BERT, Doc2Vec and CC2Vec.
With BERT embeddings, \toolname (with XGBoost) yielded very promising performance on patch correctness prediction with metrics like Recall at 82.1\% and F-Measure at 76.5\%, \toolname (with DNN) achieved the highest score with the metric Precision at 0.744 on a labeled deduplicated dataset of 2,147 patches. 
We further showed that the performance of these models on learned embedding features is promising when comparing against the state of the art (PATCH-SIM~\cite{xiong2018identifying}), which uses dynamic execution traces.
We further implemented \tool (an upgraded version of \toolname) to explore the combination of the learning embeddings and the engineered features to improve the performance on identifying patch correctness with more accurate classification. 
Finally, leveraging SHAP, we analyzed the cause of prediction behind features and classifiers to help aware the essence of identifying patch correctness. Since our approach is able to swiftly predict patch correctness, future work should investigate how to incorporate it with APR tools to explore large patch space more efficiently.

\vspace{1mm}
\noindent
{\bf Availability.} All artifacts of this study are available in the following public repository:
\begin{center}
	\url{https://github.com/HaoyeTianCoder/Panther}
\end{center}

\section*{Acknowledgements}{
This work was supported by funding from the European Research Council (ERC) under the European Union's Horizon 2020 research and innovation program (grant agreement No. 949014).
Kui Liu was also supported by the National Natural Science Foundation of China (Grant No. 62172214), the Natural Science Foundation of Jiangsu Province, China (Grant No. BK20210279), and the Open Project Program of the State Key Laboratory of Mathematical Engineering and Advanced Computing (No. 2020A06).
}

\bibliographystyle{ACM-Reference-Format}%
\bibliography{bib/references}


\begin{thebibliography}{77}

%%% the Web address.

\ifx \showCODEN    \undefined \def \showCODEN     #1{\unskip}     \fi
\ifx \showDOI      \undefined \def \showDOI       #1{#1}\fi
\ifx \showISBNx    \undefined \def \showISBNx     #1{\unskip}     \fi
\ifx \showISBNxiii \undefined \def \showISBNxiii  #1{\unskip}     \fi
\ifx \showISSN     \undefined \def \showISSN      #1{\unskip}     \fi
\ifx \showLCCN     \undefined \def \showLCCN      #1{\unskip}     \fi
\ifx \shownote     \undefined \def \shownote      #1{#1}          \fi
\ifx \showarticletitle \undefined \def \showarticletitle #1{#1}   \fi
\ifx \showURL      \undefined \def \showURL       {\relax}        \fi
% invisible to TeX
\providecommand\bibfield[2]{#2}
\providecommand\bibinfo[2]{#2}
\providecommand\natexlab[1]{#1}
\providecommand\showeprint[2][]{arXiv:#2}

\bibitem[\protect\citeauthoryear{Allamanis, Barr, Devanbu, and
  Sutton}{Allamanis et~al\mbox{.}}{2018}]%
        {allamanis2018survey}
\bibfield{author}{\bibinfo{person}{Miltiadis Allamanis},
  \bibinfo{person}{Earl~T. Barr}, \bibinfo{person}{Premkumar~T. Devanbu}, {and}
  \bibinfo{person}{Charles~A. Sutton}.} \bibinfo{year}{2018}\natexlab{}.
\newblock \showarticletitle{A Survey of Machine Learning for Big Code and
  Naturalness}.
\newblock \bibinfo{journal}{\emph{Comput. Surveys}} \bibinfo{volume}{51},
  \bibinfo{number}{4} (\bibinfo{year}{2018}), \bibinfo{pages}{81:1--81:37}.
\newblock
\urldef\tempurl%
\url{https://doi.org/10.1145/3212695}
\showDOI{\tempurl}


\bibitem[\protect\citeauthoryear{Alon, Zilberstein, Levy, and Yahav}{Alon
  et~al\mbox{.}}{2019}]%
        {alon2019code2vec}
\bibfield{author}{\bibinfo{person}{Uri Alon}, \bibinfo{person}{Meital
  Zilberstein}, \bibinfo{person}{Omer Levy}, {and} \bibinfo{person}{Eran
  Yahav}.} \bibinfo{year}{2019}\natexlab{}.
\newblock \showarticletitle{code2vec: learning distributed representations of
  code}.
\newblock \bibinfo{journal}{\emph{Proceedings of the ACM on Programming
  Languages}} \bibinfo{volume}{3}, \bibinfo{number}{{POPL}}
  (\bibinfo{year}{2019}), \bibinfo{pages}{40:1--40:29}.
\newblock
\urldef\tempurl%
\url{https://doi.org/10.1145/3290353}
\showDOI{\tempurl}


\bibitem[\protect\citeauthoryear{Bader, Scott, Pradel, and Chandra}{Bader
  et~al\mbox{.}}{2019}]%
        {bader2019getafix}
\bibfield{author}{\bibinfo{person}{Johannes Bader}, \bibinfo{person}{Andrew
  Scott}, \bibinfo{person}{Michael Pradel}, {and} \bibinfo{person}{Satish
  Chandra}.} \bibinfo{year}{2019}\natexlab{}.
\newblock \showarticletitle{Getafix: learning to fix bugs automatically}.
\newblock \bibinfo{journal}{\emph{Proceedings of the ACM on Programming
  Languages}} \bibinfo{volume}{3}, \bibinfo{number}{{OOPSLA}}
  (\bibinfo{year}{2019}), \bibinfo{pages}{159:1--159:27}.
\newblock
\urldef\tempurl%
\url{https://doi.org/10.1145/3360585}
\showDOI{\tempurl}


\bibitem[\protect\citeauthoryear{Barr, Brun, Devanbu, Harman, and Sarro}{Barr
  et~al\mbox{.}}{2014}]%
        {barr2014plastic}
\bibfield{author}{\bibinfo{person}{Earl~T. Barr}, \bibinfo{person}{Yuriy Brun},
  \bibinfo{person}{Premkumar~T. Devanbu}, \bibinfo{person}{Mark Harman}, {and}
  \bibinfo{person}{Federica Sarro}.} \bibinfo{year}{2014}\natexlab{}.
\newblock \showarticletitle{The plastic surgery hypothesis}. In
  \bibinfo{booktitle}{\emph{Proceedings of the 22nd {ACM} {SIGSOFT}
  International Symposium on Foundations of Software Engineering}}. ACM,
  \bibinfo{pages}{306--317}.
\newblock
\urldef\tempurl%
\url{https://doi.org/10.1145/2635868.2635898}
\showDOI{\tempurl}


\bibitem[\protect\citeauthoryear{Chen, Donaldson, Zeller, and Zhang}{Chen
  et~al\mbox{.}}{2017}]%
        {chen2017testing}
\bibfield{author}{\bibinfo{person}{Junjie Chen}, \bibinfo{person}{Alastair~F.
  Donaldson}, \bibinfo{person}{Andreas Zeller}, {and} \bibinfo{person}{Hongyu
  Zhang}.} \bibinfo{year}{2017}\natexlab{}.
\newblock \showarticletitle{Testing and Verification of Compilers (Dagstuhl
  Seminar 17502)}.
\newblock \bibinfo{journal}{\emph{Dagstuhl Reports}} \bibinfo{volume}{7},
  \bibinfo{number}{12} (\bibinfo{year}{2017}), \bibinfo{pages}{50--65}.
\newblock
\urldef\tempurl%
\url{https://doi.org/10.4230/DagRep.7.12.50}
\showDOI{\tempurl}


\bibitem[\protect\citeauthoryear{Cheng, Koc, Harmsen, Shaked, Chandra, Aradhye,
  Anderson, Corrado, Chai, Ispir, et~al\mbox{.}}{Cheng et~al\mbox{.}}{2016}]%
        {cheng2016wide}
\bibfield{author}{\bibinfo{person}{Heng-Tze Cheng}, \bibinfo{person}{Levent
  Koc}, \bibinfo{person}{Jeremiah Harmsen}, \bibinfo{person}{Tal Shaked},
  \bibinfo{person}{Tushar Chandra}, \bibinfo{person}{Hrishi Aradhye},
  \bibinfo{person}{Glen Anderson}, \bibinfo{person}{Greg Corrado},
  \bibinfo{person}{Wei Chai}, \bibinfo{person}{Mustafa Ispir}, {et~al\mbox{.}}}
  \bibinfo{year}{2016}\natexlab{}.
\newblock \showarticletitle{Wide \& deep learning for recommender systems}. In
  \bibinfo{booktitle}{\emph{Proceedings of the 1st workshop on deep learning
  for recommender systems}}. \bibinfo{pages}{7--10}.
\newblock


\bibitem[\protect\citeauthoryear{Compton, Frank, Patros, and Koay}{Compton
  et~al\mbox{.}}{2020}]%
        {compton2020embedding}
\bibfield{author}{\bibinfo{person}{Rhys Compton}, \bibinfo{person}{Eibe Frank},
  \bibinfo{person}{Panos Patros}, {and} \bibinfo{person}{Abigail Koay}.}
  \bibinfo{year}{2020}\natexlab{}.
\newblock \showarticletitle{Embedding Java Classes with code2vec: Improvements
  from Variable Obfuscation}. In \bibinfo{booktitle}{\emph{Proceedings of the
  17th Mining Software Repositories}}. ACM.
\newblock


\bibitem[\protect\citeauthoryear{Csuvik, Horv{\'a}th, Horv{\'a}th, and
  Vid{\'a}cs}{Csuvik et~al\mbox{.}}{2020}]%
        {csuvik2020utilizing}
\bibfield{author}{\bibinfo{person}{Viktor Csuvik}, \bibinfo{person}{D{\'a}niel
  Horv{\'a}th}, \bibinfo{person}{Ferenc Horv{\'a}th}, {and}
  \bibinfo{person}{L{\'a}szl{\'o} Vid{\'a}cs}.}
  \bibinfo{year}{2020}\natexlab{}.
\newblock \showarticletitle{Utilizing Source Code Embeddings to Identify
  Correct Patches}. In \bibinfo{booktitle}{\emph{Proceedings of the 2nd
  International Workshop on Intelligent Bug Fixing}}. IEEE,
  \bibinfo{pages}{18--25}.
\newblock
\urldef\tempurl%
\url{https://doi.org/10.1109/IBF50092.2020.9034714}
\showDOI{\tempurl}


\bibitem[\protect\citeauthoryear{Devlin, Chang, Lee, and Toutanova}{Devlin
  et~al\mbox{.}}{2019}]%
        {devlin2019bert}
\bibfield{author}{\bibinfo{person}{Jacob Devlin}, \bibinfo{person}{Ming{-}Wei
  Chang}, \bibinfo{person}{Kenton Lee}, {and} \bibinfo{person}{Kristina
  Toutanova}.} \bibinfo{year}{2019}\natexlab{}.
\newblock \showarticletitle{{BERT:} Pre-training of Deep Bidirectional
  Transformers for Language Understanding}. In
  \bibinfo{booktitle}{\emph{Proceedings of the 2019 Conference of the North
  American Chapter of the Association for Computational Linguistics: Human
  Language Technologies}}. \bibinfo{pages}{4171--4186}.
\newblock
\urldef\tempurl%
\url{https://doi.org/10.18653/v1/n19-1423}
\showDOI{\tempurl}


\bibitem[\protect\citeauthoryear{Dietterich}{Dietterich}{1998}]%
        {dietterich1998approximate}
\bibfield{author}{\bibinfo{person}{Thomas~G Dietterich}.}
  \bibinfo{year}{1998}\natexlab{}.
\newblock \showarticletitle{Approximate statistical tests for comparing
  supervised classification learning algorithms}.
\newblock \bibinfo{journal}{\emph{Neural computation}} \bibinfo{volume}{10},
  \bibinfo{number}{7} (\bibinfo{year}{1998}), \bibinfo{pages}{1895--1923}.
\newblock


\bibitem[\protect\citeauthoryear{Durieux, Madeiral, Martinez, and
  Abreu}{Durieux et~al\mbox{.}}{2019}]%
        {durieux2019empirical}
\bibfield{author}{\bibinfo{person}{Thomas Durieux}, \bibinfo{person}{Fernanda
  Madeiral}, \bibinfo{person}{Matias Martinez}, {and} \bibinfo{person}{Rui
  Abreu}.} \bibinfo{year}{2019}\natexlab{}.
\newblock \showarticletitle{Empirical Review of Java Program Repair Tools: A
  Large-Scale Experiment on 2,141 Bugs and 23,551 Repair Attempts}. In
  \bibinfo{booktitle}{\emph{Proceedings of the 27th ACM Joint Meeting on
  European Software Engineering Conference and Symposium on the Foundations of
  Software Engineering}}. \bibinfo{publisher}{ACM}, \bibinfo{pages}{302--313}.
\newblock
\urldef\tempurl%
\url{https://doi.org/10.1145/3338906.3338911}
\showDOI{\tempurl}


\bibitem[\protect\citeauthoryear{Fang, Liu, Shi, Huang, and Shi}{Fang
  et~al\mbox{.}}{2020}]%
        {fang2020functional}
\bibfield{author}{\bibinfo{person}{Chunrong Fang}, \bibinfo{person}{Zixi Liu},
  \bibinfo{person}{Yangyang Shi}, \bibinfo{person}{Jeff Huang}, {and}
  \bibinfo{person}{Qingkai Shi}.} \bibinfo{year}{2020}\natexlab{}.
\newblock \showarticletitle{Functional Code Clone Detection with Syntax and
  Semantics Fusion Learning}. In \bibinfo{booktitle}{\emph{Proceedings of the
  29th ACM SIGSOFT International Symposium on Software Testing and Analysis}}
  (Virtual Event, USA). \bibinfo{publisher}{ACM}, \bibinfo{pages}{516–527}.
\newblock
\showISBNx{9781450380089}
\urldef\tempurl%
\url{https://doi.org/10.1145/3395363.3397362}
\showDOI{\tempurl}


\bibitem[\protect\citeauthoryear{Feng, Guo, Tang, Duan, Feng, Gong, Shou, Qin,
  Liu, Jiang, et~al\mbox{.}}{Feng et~al\mbox{.}}{2020}]%
        {feng2020codebert}
\bibfield{author}{\bibinfo{person}{Zhangyin Feng}, \bibinfo{person}{Daya Guo},
  \bibinfo{person}{Duyu Tang}, \bibinfo{person}{Nan Duan},
  \bibinfo{person}{Xiaocheng Feng}, \bibinfo{person}{Ming Gong},
  \bibinfo{person}{Linjun Shou}, \bibinfo{person}{Bing Qin},
  \bibinfo{person}{Ting Liu}, \bibinfo{person}{Daxin Jiang}, {et~al\mbox{.}}}
  \bibinfo{year}{2020}\natexlab{}.
\newblock \showarticletitle{{CodeBERT:} A Pre-Trained Model for Programming and
  Natural Languages}.
\newblock \bibinfo{journal}{\emph{arXiv preprint arXiv:2002.08155}}
  (\bibinfo{year}{2020}).
\newblock
\urldef\tempurl%
\url{https://arxiv.org/abs/2002.08155}
\showURL{%
\tempurl}


\bibitem[\protect\citeauthoryear{Hindle, Barr, Su, Gabel, and Devanbu}{Hindle
  et~al\mbox{.}}{2012}]%
        {hindle2012naturalness}
\bibfield{author}{\bibinfo{person}{Abram Hindle}, \bibinfo{person}{Earl~T.
  Barr}, \bibinfo{person}{Zhendong Su}, \bibinfo{person}{Mark Gabel}, {and}
  \bibinfo{person}{Premkumar~T. Devanbu}.} \bibinfo{year}{2012}\natexlab{}.
\newblock \showarticletitle{On the naturalness of software}. In
  \bibinfo{booktitle}{\emph{Proceedings of the 34th International Conference on
  Software Engineering}}. IEEE, \bibinfo{pages}{837--847}.
\newblock
\urldef\tempurl%
\url{https://doi.org/10.1109/ICSE.2012.6227135}
\showDOI{\tempurl}


\bibitem[\protect\citeauthoryear{Hoang, Kang, Lawall, and Lo}{Hoang
  et~al\mbox{.}}{2020}]%
        {hoang2020cc2vec}
\bibfield{author}{\bibinfo{person}{Thong Hoang}, \bibinfo{person}{Hong~Jin
  Kang}, \bibinfo{person}{Julia Lawall}, {and} \bibinfo{person}{David Lo}.}
  \bibinfo{year}{2020}\natexlab{}.
\newblock \showarticletitle{{CC2Vec:} Distributed Representations of Code
  Changes}. In \bibinfo{booktitle}{\emph{Proceedings of the 42nd International
  Conference on Software Engineering}}. ACM, \bibinfo{pages}{518--529}.
\newblock
\urldef\tempurl%
\url{https://doi.org/10.1145/3377811.3380361}
\showDOI{\tempurl}


\bibitem[\protect\citeauthoryear{Hoang, Lawall, Tian, Oentaryo, and Lo}{Hoang
  et~al\mbox{.}}{2019}]%
        {hoang2019patchnet}
\bibfield{author}{\bibinfo{person}{Thong Hoang}, \bibinfo{person}{Julia
  Lawall}, \bibinfo{person}{Yuan Tian}, \bibinfo{person}{Richard~Jayadi
  Oentaryo}, {and} \bibinfo{person}{David Lo}.}
  \bibinfo{year}{2019}\natexlab{}.
\newblock \showarticletitle{{PatchNet:} Hierarchical Deep Learning-Based Stable
  Patch Identification for the Linux Kernel}.
\newblock \bibinfo{journal}{\emph{CoRR}}  \bibinfo{volume}{abs/1911.03576}
  (\bibinfo{year}{2019}).
\newblock
\urldef\tempurl%
\url{http://arxiv.org/abs/1911.03576}
\showURL{%
\tempurl}


\bibitem[\protect\citeauthoryear{Jiang, Ren, Xiong, and Zhang}{Jiang
  et~al\mbox{.}}{2019}]%
        {jiang2019inferring}
\bibfield{author}{\bibinfo{person}{Jiajun Jiang}, \bibinfo{person}{Luyao Ren},
  \bibinfo{person}{Yingfei Xiong}, {and} \bibinfo{person}{Lingming Zhang}.}
  \bibinfo{year}{2019}\natexlab{}.
\newblock \showarticletitle{Inferring Program Transformations From Singular
  Examples via Big Code}. In \bibinfo{booktitle}{\emph{Proceedings of the 34th
  {IEEE/ACM} International Conference on Automated Software Engineering}}.
  \bibinfo{publisher}{{IEEE}}, \bibinfo{pages}{255--266}.
\newblock
\urldef\tempurl%
\url{https://doi.org/10.1109/ASE.2019.00033}
\showDOI{\tempurl}


\bibitem[\protect\citeauthoryear{Jiang, Xiong, Zhang, Gao, and Chen}{Jiang
  et~al\mbox{.}}{2018}]%
        {jiang2018shaping}
\bibfield{author}{\bibinfo{person}{Jiajun Jiang}, \bibinfo{person}{Yingfei
  Xiong}, \bibinfo{person}{Hongyu Zhang}, \bibinfo{person}{Qing Gao}, {and}
  \bibinfo{person}{Xiangqun Chen}.} \bibinfo{year}{2018}\natexlab{}.
\newblock \showarticletitle{Shaping program repair space with existing patches
  and similar code}. In \bibinfo{booktitle}{\emph{Proceedings of the 27th ACM
  SIGSOFT International Symposium on Software Testing and Analysis}}. ACM,
  \bibinfo{pages}{298--309}.
\newblock
\urldef\tempurl%
\url{https://doi.org/10.1145/3213846.3213871}
\showDOI{\tempurl}


\bibitem[\protect\citeauthoryear{Just, Jalali, and Ernst}{Just
  et~al\mbox{.}}{2014}]%
        {just2014defects4j}
\bibfield{author}{\bibinfo{person}{Ren{\'e} Just}, \bibinfo{person}{Darioush
  Jalali}, {and} \bibinfo{person}{Michael~D Ernst}.}
  \bibinfo{year}{2014}\natexlab{}.
\newblock \showarticletitle{{Defects4J}: A database of existing faults to
  enable controlled testing studies for Java programs}. In
  \bibinfo{booktitle}{\emph{Proceedings of the 23rd International Symposium on
  Software Testing and Analysis}}. ACM, \bibinfo{pages}{437--440}.
\newblock
\urldef\tempurl%
\url{https://doi.org/10.1145/2610384.2628055}
\showDOI{\tempurl}


\bibitem[\protect\citeauthoryear{Karampatsis and Sutton}{Karampatsis and
  Sutton}{2020}]%
        {karampatsis2020how}
\bibfield{author}{\bibinfo{person}{Rafael{-}Michael Karampatsis} {and}
  \bibinfo{person}{Charles~A. Sutton}.} \bibinfo{year}{2020}\natexlab{}.
\newblock \showarticletitle{How Often Do Single-Statement Bugs Occur? The
  ManySStuBs4J Dataset}. In \bibinfo{booktitle}{\emph{Proceedings of the 17th
  Mining Software Repositories}}. IEEE.
\newblock
\urldef\tempurl%
\url{http://arxiv.org/abs/1905.13334}
\showURL{%
\tempurl}


\bibitem[\protect\citeauthoryear{Koyuncu, Liu, Bissyand{\'e}, Kim, Klein,
  Monperrus, and Traon}{Koyuncu et~al\mbox{.}}{2020}]%
        {koyuncu2020fixminer}
\bibfield{author}{\bibinfo{person}{Anil Koyuncu}, \bibinfo{person}{Kui Liu},
  \bibinfo{person}{Tegawend{\'e}~F. Bissyand{\'e}}, \bibinfo{person}{Dongsun
  Kim}, \bibinfo{person}{Jacques Klein}, \bibinfo{person}{Martin Monperrus},
  {and} \bibinfo{person}{Yves~Le Traon}.} \bibinfo{year}{2020}\natexlab{}.
\newblock \showarticletitle{{FixMiner:} Mining relevant fix patterns for
  automated program repair}.
\newblock \bibinfo{journal}{\emph{Empirical Software Engineering}}
  \bibinfo{volume}{25}, \bibinfo{number}{3} (\bibinfo{year}{2020}),
  \bibinfo{pages}{1980--2024}.
\newblock
\urldef\tempurl%
\url{https://doi.org/10.1007/s10664-019-09780-z}
\showDOI{\tempurl}


\bibitem[\protect\citeauthoryear{Koyuncu, Liu, Bissyand{\'e}, Kim, Monperrus,
  Klein, and Le~Traon}{Koyuncu et~al\mbox{.}}{2019}]%
        {koyuncu2019ifixr}
\bibfield{author}{\bibinfo{person}{Anil Koyuncu}, \bibinfo{person}{Kui Liu},
  \bibinfo{person}{Tegawend{\'e}~F. Bissyand{\'e}}, \bibinfo{person}{Dongsun
  Kim}, \bibinfo{person}{Martin Monperrus}, \bibinfo{person}{Jacques Klein},
  {and} \bibinfo{person}{Yves Le~Traon}.} \bibinfo{year}{2019}\natexlab{}.
\newblock \showarticletitle{{iFixR}: Bug Report driven Program Repair}. In
  \bibinfo{booktitle}{\emph{Proceedings of the 27the ACM Joint European
  Software Engineering Conference and Symposium on the Foundations of Software
  Engineering}}. ACM, \bibinfo{pages}{314--325}.
\newblock
\urldef\tempurl%
\url{https://doi.org/10.1145/3338906.3338935}
\showDOI{\tempurl}


\bibitem[\protect\citeauthoryear{Le and Mikolov}{Le and Mikolov}{2014}]%
        {le2014distributed}
\bibfield{author}{\bibinfo{person}{Quoc~V. Le} {and} \bibinfo{person}{Tomas
  Mikolov}.} \bibinfo{year}{2014}\natexlab{}.
\newblock \showarticletitle{Distributed Representations of Sentences and
  Documents}. In \bibinfo{booktitle}{\emph{Proceedings of the 31st
  International Conference on Machine Learning}}.
  \bibinfo{publisher}{JMLR.org}, \bibinfo{pages}{1188--1196}.
\newblock
\urldef\tempurl%
\url{http://proceedings.mlr.press/v32/le14.html}
\showURL{%
\tempurl}


\bibitem[\protect\citeauthoryear{Le, Bao, Lo, Xia, Li, and Pasareanu}{Le
  et~al\mbox{.}}{2019}]%
        {le2019reliability}
\bibfield{author}{\bibinfo{person}{Xuan-Bach~D Le}, \bibinfo{person}{Lingfeng
  Bao}, \bibinfo{person}{David Lo}, \bibinfo{person}{Xin Xia},
  \bibinfo{person}{Shanping Li}, {and} \bibinfo{person}{Corina Pasareanu}.}
  \bibinfo{year}{2019}\natexlab{}.
\newblock \showarticletitle{On reliability of patch correctness assessment}. In
  \bibinfo{booktitle}{\emph{Proceedings of the 41st International Conference on
  Software Engineering}}. IEEE, \bibinfo{pages}{524--535}.
\newblock
\urldef\tempurl%
\url{https://doi.org/10.1109/ICSE.2019.00064}
\showDOI{\tempurl}


\bibitem[\protect\citeauthoryear{Le, Thung, Lo, and Le~Goues}{Le
  et~al\mbox{.}}{2018}]%
        {le2018overfitting}
\bibfield{author}{\bibinfo{person}{Xuan Bach~D Le}, \bibinfo{person}{Ferdian
  Thung}, \bibinfo{person}{David Lo}, {and} \bibinfo{person}{Claire Le~Goues}.}
  \bibinfo{year}{2018}\natexlab{}.
\newblock \showarticletitle{Overfitting in semantics-based automated program
  repair}.
\newblock \bibinfo{journal}{\emph{Empirical Software Engineering}}
  \bibinfo{volume}{23}, \bibinfo{number}{5} (\bibinfo{year}{2018}),
  \bibinfo{pages}{3007--3033}.
\newblock
\urldef\tempurl%
\url{https://doi.org/10.1007/s10664-017-9577-2}
\showDOI{\tempurl}


\bibitem[\protect\citeauthoryear{Le~Goues, Holtschulte, Smith, Brun, Devanbu,
  Forrest, and Weimer}{Le~Goues et~al\mbox{.}}{2015}]%
        {le2015manybugs}
\bibfield{author}{\bibinfo{person}{Claire Le~Goues}, \bibinfo{person}{Neal
  Holtschulte}, \bibinfo{person}{Edward~K Smith}, \bibinfo{person}{Yuriy Brun},
  \bibinfo{person}{Premkumar Devanbu}, \bibinfo{person}{Stephanie Forrest},
  {and} \bibinfo{person}{Westley Weimer}.} \bibinfo{year}{2015}\natexlab{}.
\newblock \showarticletitle{The ManyBugs and IntroClass benchmarks for
  automated repair of {C} programs}.
\newblock \bibinfo{journal}{\emph{IEEE Transactions on Software Engineering}}
  \bibinfo{volume}{41}, \bibinfo{number}{12} (\bibinfo{year}{2015}),
  \bibinfo{pages}{1236--1256}.
\newblock
\urldef\tempurl%
\url{https://doi.org/10.1109/TSE.2015.2454513}
\showDOI{\tempurl}


\bibitem[\protect\citeauthoryear{Le~Goues, Nguyen, Forrest, and
  Weimer}{Le~Goues et~al\mbox{.}}{2012}]%
        {le2012genprog}
\bibfield{author}{\bibinfo{person}{Claire Le~Goues}, \bibinfo{person}{ThanhVu
  Nguyen}, \bibinfo{person}{Stephanie Forrest}, {and} \bibinfo{person}{Westley
  Weimer}.} \bibinfo{year}{2012}\natexlab{}.
\newblock \showarticletitle{{GenProg}: A generic method for automatic software
  repair}.
\newblock \bibinfo{journal}{\emph{IEEE Transactions on Software Engineering}}
  \bibinfo{volume}{38}, \bibinfo{number}{1} (\bibinfo{year}{2012}),
  \bibinfo{pages}{54--72}.
\newblock
\urldef\tempurl%
\url{https://doi.org/10.1109/TSE.2011.104}
\showDOI{\tempurl}


\bibitem[\protect\citeauthoryear{Le~Goues, Pradel, and Roychoudhury}{Le~Goues
  et~al\mbox{.}}{2019}]%
        {le2019automated}
\bibfield{author}{\bibinfo{person}{Claire Le~Goues}, \bibinfo{person}{Michael
  Pradel}, {and} \bibinfo{person}{Abhik Roychoudhury}.}
  \bibinfo{year}{2019}\natexlab{}.
\newblock \showarticletitle{Automated Program Repair}.
\newblock \bibinfo{journal}{\emph{Commun. ACM}} \bibinfo{volume}{62},
  \bibinfo{number}{12} (\bibinfo{year}{2019}), \bibinfo{pages}{56--65}.
\newblock
\urldef\tempurl%
\url{https://doi.org/10.1145/3318162}
\showDOI{\tempurl}


\bibitem[\protect\citeauthoryear{Lin, Koppel, Chen, and Solar-Lezama}{Lin
  et~al\mbox{.}}{2017}]%
        {lin2017quixbugs}
\bibfield{author}{\bibinfo{person}{Derrick Lin}, \bibinfo{person}{James
  Koppel}, \bibinfo{person}{Angela Chen}, {and} \bibinfo{person}{Armando
  Solar-Lezama}.} \bibinfo{year}{2017}\natexlab{}.
\newblock \showarticletitle{{QuixBugs:} A multi-lingual program repair
  benchmark set based on the Quixey Challenge}. In
  \bibinfo{booktitle}{\emph{Proceedings Companion of the 2017 ACM SIGPLAN
  International Conference on Systems, Programming, Languages, and
  Applications: Software for Humanity}}. ACM, \bibinfo{pages}{55--56}.
\newblock
\urldef\tempurl%
\url{https://doi.org/10.1145/3135932.3135941}
\showDOI{\tempurl}


\bibitem[\protect\citeauthoryear{Liu, Kim, Bissyand{\'{e}}, Kim, Kim, Koyuncu,
  Kim, and Traon}{Liu et~al\mbox{.}}{2019a}]%
        {liu2019learning}
\bibfield{author}{\bibinfo{person}{Kui Liu}, \bibinfo{person}{Dongsun Kim},
  \bibinfo{person}{Tegawend{\'{e}}~F. Bissyand{\'{e}}},
  \bibinfo{person}{Tae{-}young Kim}, \bibinfo{person}{Kisub Kim},
  \bibinfo{person}{Anil Koyuncu}, \bibinfo{person}{Suntae Kim}, {and}
  \bibinfo{person}{Yves~Le Traon}.} \bibinfo{year}{2019}\natexlab{a}.
\newblock \showarticletitle{Learning to spot and refactor inconsistent method
  names}. In \bibinfo{booktitle}{\emph{Proceedings of the 41st International
  Conference on Software Engineering}}. IEEE, \bibinfo{pages}{1--12}.
\newblock
\urldef\tempurl%
\url{https://doi.org/10.1109/ICSE.2019.00019}
\showDOI{\tempurl}


\bibitem[\protect\citeauthoryear{Liu, Kim, Bissyand{\'e}, Yoo, and
  Le~Traon}{Liu et~al\mbox{.}}{2018a}]%
        {liu2018mining2}
\bibfield{author}{\bibinfo{person}{Kui Liu}, \bibinfo{person}{Dongsun Kim},
  \bibinfo{person}{Tegawend{\'e}~F Bissyand{\'e}}, \bibinfo{person}{Shin Yoo},
  {and} \bibinfo{person}{Yves Le~Traon}.} \bibinfo{year}{2018}\natexlab{a}.
\newblock \showarticletitle{Mining fix patterns for findbugs violations}.
\newblock \bibinfo{journal}{\emph{IEEE Transactions on Software Engineering}}
  (\bibinfo{year}{2018}).
\newblock
\urldef\tempurl%
\url{https://doi.org/10.1109/TSE.2018.2884955}
\showDOI{\tempurl}


\bibitem[\protect\citeauthoryear{Liu, Kim, Koyuncu, Li, Bissyand{\'e}, and
  Le~Traon}{Liu et~al\mbox{.}}{2018b}]%
        {liu2018closer}
\bibfield{author}{\bibinfo{person}{Kui Liu}, \bibinfo{person}{Dongsun Kim},
  \bibinfo{person}{Anil Koyuncu}, \bibinfo{person}{Li Li},
  \bibinfo{person}{Tegawend{\'e}~F Bissyand{\'e}}, {and} \bibinfo{person}{Yves
  Le~Traon}.} \bibinfo{year}{2018}\natexlab{b}.
\newblock \showarticletitle{A closer look at real-world patches}. In
  \bibinfo{booktitle}{\emph{Proceedings of the 34th International Conference on
  Software Maintenance and Evolution}}. IEEE, \bibinfo{pages}{275--286}.
\newblock
\urldef\tempurl%
\url{https://doi.org/10.1109/ICSME.2018.00037}
\showDOI{\tempurl}


\bibitem[\protect\citeauthoryear{Liu, Koyuncu, Bissyand{\'e}, Kim, Klein, and
  Traon}{Liu et~al\mbox{.}}{2019b}]%
        {liu2019you}
\bibfield{author}{\bibinfo{person}{Kui Liu}, \bibinfo{person}{Anil Koyuncu},
  \bibinfo{person}{Tegawend{\'e}~F Bissyand{\'e}}, \bibinfo{person}{Dongsun
  Kim}, \bibinfo{person}{Jacques Klein}, {and} \bibinfo{person}{Yves~Le
  Traon}.} \bibinfo{year}{2019}\natexlab{b}.
\newblock \showarticletitle{You cannot fix what you cannot find! an
  investigation of fault localization bias in benchmarking automated program
  repair systems}. In \bibinfo{booktitle}{\emph{Proceedings of the 12th IEEE
  International Conference on Software Testing, Verification and Validation}}.
  IEEE, \bibinfo{pages}{102--113}.
\newblock
\urldef\tempurl%
\url{https://doi.org/10.1109/ICST.2019.00020}
\showDOI{\tempurl}


\bibitem[\protect\citeauthoryear{Liu, Koyuncu, Kim, and Bissyand{\'e}}{Liu
  et~al\mbox{.}}{2019c}]%
        {liu2019avatar}
\bibfield{author}{\bibinfo{person}{Kui Liu}, \bibinfo{person}{Anil Koyuncu},
  \bibinfo{person}{Dongsun Kim}, {and} \bibinfo{person}{Tegawend{\'e}~F
  Bissyand{\'e}}.} \bibinfo{year}{2019}\natexlab{c}.
\newblock \showarticletitle{{AVATAR:} Fixing semantic bugs with fix patterns of
  static analysis violations}. In \bibinfo{booktitle}{\emph{Proceedings of the
  26th IEEE International Conference on Software Analysis, Evolution and
  Reengineering}}. IEEE, \bibinfo{pages}{456--467}.
\newblock
\urldef\tempurl%
\url{https://doi.org/10.1109/SANER.2019.8667970}
\showDOI{\tempurl}


\bibitem[\protect\citeauthoryear{Liu, Koyuncu, Kim, and Bissyand{\'e}}{Liu
  et~al\mbox{.}}{2019d}]%
        {liu2019tbar}
\bibfield{author}{\bibinfo{person}{Kui Liu}, \bibinfo{person}{Anil Koyuncu},
  \bibinfo{person}{Dongsun Kim}, {and} \bibinfo{person}{Tegawend{\'e}~F.
  Bissyand{\'e}}.} \bibinfo{year}{2019}\natexlab{d}.
\newblock \showarticletitle{{TBar}: Revisiting Template-based Automated Program
  Repair}. In \bibinfo{booktitle}{\emph{Proceedings of the 28th ACM SIGSOFT
  International Symposium on Software Testing and Analysis}}. ACM,
  \bibinfo{pages}{31--42}.
\newblock
\urldef\tempurl%
\url{https://doi.org/10.1145/3293882.3330577}
\showDOI{\tempurl}


\bibitem[\protect\citeauthoryear{Liu, Koyuncu, Kim, Kim, and Bissyand{\'e}}{Liu
  et~al\mbox{.}}{2018c}]%
        {liu2018lsrepair}
\bibfield{author}{\bibinfo{person}{Kui Liu}, \bibinfo{person}{Anil Koyuncu},
  \bibinfo{person}{Kisub Kim}, \bibinfo{person}{Dongsun Kim}, {and}
  \bibinfo{person}{Tegawend{\'e}~F. Bissyand{\'e}}.}
  \bibinfo{year}{2018}\natexlab{c}.
\newblock \showarticletitle{{LSRepair}: Live search of fix ingredients for
  automated program repair}. In \bibinfo{booktitle}{\emph{Proceedings of the
  25th Asia-Pacific Software Engineering Conference ERA Track}}. {IEEE},
  \bibinfo{pages}{658--662}.
\newblock
\urldef\tempurl%
\url{https://doi.org/10.1109/APSEC.2018.00085}
\showDOI{\tempurl}


\bibitem[\protect\citeauthoryear{Liu, Li, Koyuncu, Kim, Liu, Klein, and
  Bissyand{\'{e}}}{Liu et~al\mbox{.}}{2021}]%
        {liu2021critical}
\bibfield{author}{\bibinfo{person}{Kui Liu}, \bibinfo{person}{Li Li},
  \bibinfo{person}{Anil Koyuncu}, \bibinfo{person}{Dongsun Kim},
  \bibinfo{person}{Zhe Liu}, \bibinfo{person}{Jacques Klein}, {and}
  \bibinfo{person}{Tegawend{\'{e}}~F. Bissyand{\'{e}}}.}
  \bibinfo{year}{2021}\natexlab{}.
\newblock \showarticletitle{A critical review on the evaluation of automated
  program repair systems}.
\newblock \bibinfo{journal}{\emph{Journal of Systems and Software}}
  \bibinfo{volume}{171} (\bibinfo{year}{2021}), \bibinfo{pages}{110817}.
\newblock
\urldef\tempurl%
\url{https://doi.org/10.1016/j.jss.2020.110817}
\showDOI{\tempurl}


\bibitem[\protect\citeauthoryear{Liu, Wang, Koyuncu, Kim, Bissyandé, Kim, Wu,
  Klein, Mao, and Traon}{Liu et~al\mbox{.}}{2020}]%
        {liu2020efficiency}
\bibfield{author}{\bibinfo{person}{Kui Liu}, \bibinfo{person}{Shangwen Wang},
  \bibinfo{person}{Anil Koyuncu}, \bibinfo{person}{Kisub Kim},
  \bibinfo{person}{Tegawendé~F. Bissyandé}, \bibinfo{person}{Dongsun Kim},
  \bibinfo{person}{Peng Wu}, \bibinfo{person}{Jacques Klein},
  \bibinfo{person}{Xiaoguang Mao}, {and} \bibinfo{person}{Yves~Le Traon}.}
  \bibinfo{year}{2020}\natexlab{}.
\newblock \showarticletitle{On the Efficiency of Test Suite based Program
  Repair: A Systematic Assessment of 16 Automated Repair Systems for Java
  Programs}. In \bibinfo{booktitle}{\emph{Proceedings of the 42nd International
  Conference on Software Engineering}}. ACM, \bibinfo{pages}{625--627}.
\newblock
\urldef\tempurl%
\url{https://doi.org/10.1145/3377811.3380338}
\showDOI{\tempurl}


\bibitem[\protect\citeauthoryear{Long and Rinard}{Long and Rinard}{2016}]%
        {long2016automatic}
\bibfield{author}{\bibinfo{person}{Fan Long} {and} \bibinfo{person}{Martin
  Rinard}.} \bibinfo{year}{2016}\natexlab{}.
\newblock \showarticletitle{Automatic patch generation by learning correct
  code}. In \bibinfo{booktitle}{\emph{Proceedings of the 43rd Annual {ACM}
  {SIGPLAN-SIGACT} Symposium on Principles of Programming Languages}},
  Vol.~\bibinfo{volume}{51}. ACM, \bibinfo{pages}{298--312}.
\newblock
\urldef\tempurl%
\url{https://doi.org/10.1145/2837614.2837617}
\showDOI{\tempurl}


\bibitem[\protect\citeauthoryear{Lundberg and Lee}{Lundberg and Lee}{2017}]%
        {NIPS2017_7062}
\bibfield{author}{\bibinfo{person}{Scott~M Lundberg} {and}
  \bibinfo{person}{Su-In Lee}.} \bibinfo{year}{2017}\natexlab{}.
\newblock \showarticletitle{A Unified Approach to Interpreting Model
  Predictions}.
\newblock In \bibinfo{booktitle}{\emph{Advances in Neural Information
  Processing Systems 30}}, \bibfield{editor}{\bibinfo{person}{I.~Guyon},
  \bibinfo{person}{U.~V. Luxburg}, \bibinfo{person}{S.~Bengio},
  \bibinfo{person}{H.~Wallach}, \bibinfo{person}{R.~Fergus},
  \bibinfo{person}{S.~Vishwanathan}, {and} \bibinfo{person}{R.~Garnett}}
  (Eds.). \bibinfo{publisher}{Curran Associates, Inc.},
  \bibinfo{pages}{4765--4774}.
\newblock
\urldef\tempurl%
\url{http://papers.nips.cc/paper/7062-a-unified-approach-to-interpreting-model-predictions.pdf}
\showURL{%
\tempurl}


\bibitem[\protect\citeauthoryear{Madeiral, Durieux, Sobreira, and
  Maia}{Madeiral et~al\mbox{.}}{2018}]%
        {Madeiral2018}
\bibfield{author}{\bibinfo{person}{Fernanda Madeiral}, \bibinfo{person}{Thomas
  Durieux}, \bibinfo{person}{Victor Sobreira}, {and} \bibinfo{person}{Marcelo
  Maia}.} \bibinfo{year}{2018}\natexlab{}.
\newblock \showarticletitle{Towards an automated approach for bug fix pattern
  detection}.
\newblock


\bibitem[\protect\citeauthoryear{Madeiral, Urli, Maia, and Monperrus}{Madeiral
  et~al\mbox{.}}{2019}]%
        {madeiral2019bears}
\bibfield{author}{\bibinfo{person}{Fernanda Madeiral}, \bibinfo{person}{Simon
  Urli}, \bibinfo{person}{Marcelo Maia}, {and} \bibinfo{person}{Martin
  Monperrus}.} \bibinfo{year}{2019}\natexlab{}.
\newblock \showarticletitle{{BEARS:} An Extensible Java Bug Benchmark for
  Automatic Program Repair Studies}. In \bibinfo{booktitle}{\emph{Proceedings
  of the 26th International Conference on Software Analysis, Evolution and
  Reengineering}}. IEEE, \bibinfo{pages}{468--478}.
\newblock
\urldef\tempurl%
\url{https://doi.org/10.1109/SANER.2019.8667991}
\showDOI{\tempurl}


\bibitem[\protect\citeauthoryear{Mann and Whitney}{Mann and Whitney}{1947}]%
        {mann1947test}
\bibfield{author}{\bibinfo{person}{Henry~B Mann} {and}
  \bibinfo{person}{Donald~R. Whitney}.} \bibinfo{year}{1947}\natexlab{}.
\newblock \showarticletitle{On a {{Test}} of {{Whether}} One of {{Two Random
  Variables}} Is {{Stochastically Larger}} than the {{Other}}}.
\newblock \bibinfo{journal}{\emph{The Annals of Mathematical Statistics}}
  \bibinfo{volume}{18}, \bibinfo{number}{1} (\bibinfo{year}{1947}),
  \bibinfo{pages}{50--60}.
\newblock
\urldef\tempurl%
\url{https://doi.org/10.1214/aoms/1177730491}
\showDOI{\tempurl}


\bibitem[\protect\citeauthoryear{Martinez and Monperrus}{Martinez and
  Monperrus}{2015}]%
        {martinez2015mining}
\bibfield{author}{\bibinfo{person}{Matias Martinez} {and}
  \bibinfo{person}{Martin Monperrus}.} \bibinfo{year}{2015}\natexlab{}.
\newblock \showarticletitle{Mining software repair models for reasoning on the
  search space of automated program fixing}.
\newblock \bibinfo{journal}{\emph{Empirical Software Engineering}}
  \bibinfo{volume}{20}, \bibinfo{number}{1} (\bibinfo{year}{2015}),
  \bibinfo{pages}{176--205}.
\newblock
\urldef\tempurl%
\url{https://doi.org/10.1007/s10664-013-9282-8}
\showDOI{\tempurl}


\bibitem[\protect\citeauthoryear{Mikolov, Chen, Corrado, and Dean}{Mikolov
  et~al\mbox{.}}{2013}]%
        {mikolov2013efficient}
\bibfield{author}{\bibinfo{person}{Tomas Mikolov}, \bibinfo{person}{Kai Chen},
  \bibinfo{person}{Greg Corrado}, {and} \bibinfo{person}{Jeffrey Dean}.}
  \bibinfo{year}{2013}\natexlab{}.
\newblock \showarticletitle{Efficient estimation of word representations in
  vector space}.
\newblock \bibinfo{journal}{\emph{arXiv preprint arXiv:1301.3781}}
  (\bibinfo{year}{2013}).
\newblock


\bibitem[\protect\citeauthoryear{Monperrus}{Monperrus}{2018a}]%
        {monperrus2018automatic}
\bibfield{author}{\bibinfo{person}{Martin Monperrus}.}
  \bibinfo{year}{2018}\natexlab{a}.
\newblock \showarticletitle{Automatic software repair: {A} bibliography}.
\newblock \bibinfo{journal}{\emph{Comput. Surveys}} \bibinfo{volume}{51},
  \bibinfo{number}{1} (\bibinfo{year}{2018}), \bibinfo{pages}{17:1--17:24}.
\newblock
\urldef\tempurl%
\url{https://doi.org/10.1145/3105906}
\showDOI{\tempurl}


\bibitem[\protect\citeauthoryear{Monperrus}{Monperrus}{2018b}]%
        {monperrus2018living}
\bibfield{author}{\bibinfo{person}{Martin Monperrus}.}
  \bibinfo{year}{2018}\natexlab{b}.
\newblock \showarticletitle{The living review on automated program repair}. In
  \bibinfo{booktitle}{\emph{HAL/archives-ouvertes. fr, Technical Report}}.
\newblock


\bibitem[\protect\citeauthoryear{Ndichu, Kim, Ozawa, Misu, and
  Makishima}{Ndichu et~al\mbox{.}}{2019}]%
        {ndichu2019machine}
\bibfield{author}{\bibinfo{person}{Samuel Ndichu}, \bibinfo{person}{Sangwook
  Kim}, \bibinfo{person}{Seiichi Ozawa}, \bibinfo{person}{Takeshi Misu}, {and}
  \bibinfo{person}{Kazuo Makishima}.} \bibinfo{year}{2019}\natexlab{}.
\newblock \showarticletitle{A machine learning approach to detection of
  JavaScript-based attacks using {AST} features and paragraph vectors}.
\newblock \bibinfo{journal}{\emph{Applied Soft Computing}}
  \bibinfo{volume}{84} (\bibinfo{year}{2019}).
\newblock
\urldef\tempurl%
\url{https://doi.org/10.1016/j.asoc.2019.105721}
\showDOI{\tempurl}


\bibitem[\protect\citeauthoryear{Pian, Peng, Tang, Sun, Tian, Habib, Klein, and
  Bissyand{\'e}}{Pian et~al\mbox{.}}{2022}]%
        {pian2022metatptrans}
\bibfield{author}{\bibinfo{person}{Weiguo Pian}, \bibinfo{person}{Hanyu Peng},
  \bibinfo{person}{Xunzhu Tang}, \bibinfo{person}{Tiezhu Sun},
  \bibinfo{person}{Haoye Tian}, \bibinfo{person}{Andrew Habib},
  \bibinfo{person}{Jacques Klein}, {and} \bibinfo{person}{Tegawend{\'e}~F
  Bissyand{\'e}}.} \bibinfo{year}{2022}\natexlab{}.
\newblock \showarticletitle{MetaTPTrans: A Meta Learning Approach for
  Multilingual Code Representation Learning}.
\newblock \bibinfo{journal}{\emph{arXiv preprint arXiv:2206.06460}}
  (\bibinfo{year}{2022}).
\newblock


\bibitem[\protect\citeauthoryear{Qi, Mao, Lei, Dai, and Wang}{Qi
  et~al\mbox{.}}{2014}]%
        {qi2014strength}
\bibfield{author}{\bibinfo{person}{Yuhua Qi}, \bibinfo{person}{Xiaoguang Mao},
  \bibinfo{person}{Yan Lei}, \bibinfo{person}{Ziying Dai}, {and}
  \bibinfo{person}{Chengsong Wang}.} \bibinfo{year}{2014}\natexlab{}.
\newblock \showarticletitle{The strength of random search on automated program
  repair}. In \bibinfo{booktitle}{\emph{Proceedings of the 36th International
  Conference on Software Engineering}}. ACM, \bibinfo{pages}{254--265}.
\newblock
\urldef\tempurl%
\url{https://doi.org/10.1145/2568225.2568254}
\showDOI{\tempurl}


\bibitem[\protect\citeauthoryear{Qi, Long, Achour, and Rinard}{Qi
  et~al\mbox{.}}{2015}]%
        {qi2015analysis}
\bibfield{author}{\bibinfo{person}{Zichao Qi}, \bibinfo{person}{Fan Long},
  \bibinfo{person}{Sara Achour}, {and} \bibinfo{person}{Martin Rinard}.}
  \bibinfo{year}{2015}\natexlab{}.
\newblock \showarticletitle{An analysis of patch plausibility and correctness
  for generate-and-validate patch generation systems}. In
  \bibinfo{booktitle}{\emph{Proceedings of the 24th International Symposium on
  Software Testing and Analysis}}. ACM, \bibinfo{pages}{24--36}.
\newblock
\urldef\tempurl%
\url{https://doi.org/10.1145/2771783.2771791}
\showDOI{\tempurl}


\bibitem[\protect\citeauthoryear{Saha, Lyu, Lam, Yoshida, and Prasad}{Saha
  et~al\mbox{.}}{2018}]%
        {saha2018bugs}
\bibfield{author}{\bibinfo{person}{Ripon Saha}, \bibinfo{person}{Yingjun Lyu},
  \bibinfo{person}{Wing Lam}, \bibinfo{person}{Hiroaki Yoshida}, {and}
  \bibinfo{person}{Mukul Prasad}.} \bibinfo{year}{2018}\natexlab{}.
\newblock \showarticletitle{Bugs.jar: A large-scale, diverse dataset of
  real-world java bugs}. In \bibinfo{booktitle}{\emph{Proceedings of the 15th
  IEEE/ACM International Conference on Mining Software Repositories}}. ACM,
  \bibinfo{pages}{10--13}.
\newblock
\urldef\tempurl%
\url{https://doi.org/10.1145/3196398.3196473}
\showDOI{\tempurl}


\bibitem[\protect\citeauthoryear{Saha, Saha, and Prasad}{Saha
  et~al\mbox{.}}{2019}]%
        {saha2019harnessing}
\bibfield{author}{\bibinfo{person}{Seemanta Saha}, \bibinfo{person}{Ripon~K
  Saha}, {and} \bibinfo{person}{Mukul~R Prasad}.}
  \bibinfo{year}{2019}\natexlab{}.
\newblock \showarticletitle{Harnessing evolution for multi-hunk program
  repair}. In \bibinfo{booktitle}{\emph{Proceedings of the 41st International
  Conference on Software Engineering}}. IEEE, \bibinfo{pages}{13--24}.
\newblock
\urldef\tempurl%
\url{https://doi.org/10.1109/ICSE.2019.00020}
\showDOI{\tempurl}


\bibitem[\protect\citeauthoryear{Smith, Barr, Le~Goues, and Brun}{Smith
  et~al\mbox{.}}{2015}]%
        {smith2015cure}
\bibfield{author}{\bibinfo{person}{Edward~K Smith}, \bibinfo{person}{Earl~T
  Barr}, \bibinfo{person}{Claire Le~Goues}, {and} \bibinfo{person}{Yuriy
  Brun}.} \bibinfo{year}{2015}\natexlab{}.
\newblock \showarticletitle{Is the cure worse than the disease? overfitting in
  automated program repair}. In \bibinfo{booktitle}{\emph{Proceedings of the
  10th Joint Meeting on Foundations of Software Engineering}}. ACM,
  \bibinfo{pages}{532--543}.
\newblock
\urldef\tempurl%
\url{https://doi.org/10.1145/2786805.2786825}
\showDOI{\tempurl}


\bibitem[\protect\citeauthoryear{Soto and Le~Goues}{Soto and Le~Goues}{2018}]%
        {soto2018using}
\bibfield{author}{\bibinfo{person}{Mauricio Soto} {and} \bibinfo{person}{Claire
  Le~Goues}.} \bibinfo{year}{2018}\natexlab{}.
\newblock \showarticletitle{Using a probabilistic model to predict bug fixes}.
  In \bibinfo{booktitle}{\emph{Proceedings of the 25th International Conference
  on Software Analysis, Evolution and Reengineering}}. IEEE,
  \bibinfo{pages}{221--231}.
\newblock
\urldef\tempurl%
\url{https://doi.org/10.1109/SANER.2018.8330211}
\showDOI{\tempurl}


\bibitem[\protect\citeauthoryear{Tian, Li, Pian, Kabore, Liu, Habib, Klein, and
  Bissyand{\'e}}{Tian et~al\mbox{.}}{2022a}]%
        {tian2022predicting}
\bibfield{author}{\bibinfo{person}{Haoye Tian}, \bibinfo{person}{Yinghua Li},
  \bibinfo{person}{Weiguo Pian}, \bibinfo{person}{Abdoul~Kader Kabore},
  \bibinfo{person}{Kui Liu}, \bibinfo{person}{Andrew Habib},
  \bibinfo{person}{Jacques Klein}, {and} \bibinfo{person}{Tegawend{\'e}~F
  Bissyand{\'e}}.} \bibinfo{year}{2022}\natexlab{a}.
\newblock \showarticletitle{Predicting Patch Correctness Based on the
  Similarity of Failing Test Cases}.
\newblock \bibinfo{journal}{\emph{ACM Transactions on Software Engineering and
  Methodology}} (\bibinfo{year}{2022}).
\newblock


\bibitem[\protect\citeauthoryear{Tian, Liu, Kabor{\'e}, Koyuncu, Li, Klein, and
  Bissyand{\'e}}{Tian et~al\mbox{.}}{2020}]%
        {tian2020evaluating}
\bibfield{author}{\bibinfo{person}{Haoye Tian}, \bibinfo{person}{Kui Liu},
  \bibinfo{person}{Abdoul~Kader Kabor{\'e}}, \bibinfo{person}{Anil Koyuncu},
  \bibinfo{person}{Li Li}, \bibinfo{person}{Jacques Klein}, {and}
  \bibinfo{person}{Tegawend{\'e}~F Bissyand{\'e}}.}
  \bibinfo{year}{2020}\natexlab{}.
\newblock \showarticletitle{Evaluating representation learning of code changes
  for predicting patch correctness in program repair}. In
  \bibinfo{booktitle}{\emph{Proceedings of the 35th IEEE/ACM International
  Conference on Automated Software Engineering}}. IEEE,
  \bibinfo{pages}{981--992}.
\newblock


\bibitem[\protect\citeauthoryear{Tian, Tang, Habib, Wang, Liu, Xia, Klein, and
  Bissyand{\'e}}{Tian et~al\mbox{.}}{2022b}]%
        {tian2022change}
\bibfield{author}{\bibinfo{person}{Haoye Tian}, \bibinfo{person}{Xunzhu Tang},
  \bibinfo{person}{Andrew Habib}, \bibinfo{person}{Shangwen Wang},
  \bibinfo{person}{Kui Liu}, \bibinfo{person}{Xin Xia},
  \bibinfo{person}{Jacques Klein}, {and} \bibinfo{person}{Tegawend{\'e}~F
  Bissyand{\'e}}.} \bibinfo{year}{2022}\natexlab{b}.
\newblock \showarticletitle{Is this Change the Answer to that Problem?
  Correlating Descriptions of Bug and Code Changes for Evaluating Patch
  Correctness}. In \bibinfo{booktitle}{\emph{Proceedings of the 37th IEEE/ACM
  International Conference on Automated Software Engineering}}. IEEE.
\newblock


\bibitem[\protect\citeauthoryear{Wang, Wen, Lin, Wu, Qin, Zou, Mao, and
  Jin}{Wang et~al\mbox{.}}{2020}]%
        {wang2020automated}
\bibfield{author}{\bibinfo{person}{Shangwen Wang}, \bibinfo{person}{Ming Wen},
  \bibinfo{person}{Bo Lin}, \bibinfo{person}{Hongjun Wu},
  \bibinfo{person}{Yihao Qin}, \bibinfo{person}{Deqing Zou},
  \bibinfo{person}{Xiaoguang Mao}, {and} \bibinfo{person}{Hai Jin}.}
  \bibinfo{year}{2020}\natexlab{}.
\newblock \showarticletitle{Automated Patch Correctness Assessment: How Far are
  We?}. In \bibinfo{booktitle}{\emph{Proceedings of the 35th IEEE/ACM
  International Conference on Automated Software Engineering}}.
  \bibinfo{publisher}{ACM}.
\newblock


\bibitem[\protect\citeauthoryear{Wei and Li}{Wei and Li}{2017}]%
        {wei2017supervised}
\bibfield{author}{\bibinfo{person}{Huihui Wei} {and} \bibinfo{person}{Ming
  Li}.} \bibinfo{year}{2017}\natexlab{}.
\newblock \showarticletitle{Supervised Deep Features for Software Functional
  Clone Detection by Exploiting Lexical and Syntactical Information in Source
  Code}. In \bibinfo{booktitle}{\emph{Proceedings of the 26th International
  Joint Conference on Artificial Intelligence}}. \bibinfo{publisher}{Morgan
  Kaufmann}, \bibinfo{pages}{3034--3040}.
\newblock
\urldef\tempurl%
\url{https://doi.org/10.24963/ijcai.2017/423}
\showDOI{\tempurl}


\bibitem[\protect\citeauthoryear{Weimer, Fry, and Forrest}{Weimer
  et~al\mbox{.}}{2013}]%
        {weimer2013leveraging}
\bibfield{author}{\bibinfo{person}{Westley Weimer}, \bibinfo{person}{Zachary~P
  Fry}, {and} \bibinfo{person}{Stephanie Forrest}.}
  \bibinfo{year}{2013}\natexlab{}.
\newblock \showarticletitle{Leveraging program equivalence for adaptive program
  repair: Models and first results}. In \bibinfo{booktitle}{\emph{Proceedings
  of the 28th IEEE/ACM International Conference on Automated Software
  Engineering}}. IEEE, \bibinfo{pages}{356--366}.
\newblock
\urldef\tempurl%
\url{https://doi.org/10.1109/ASE.2013.6693094}
\showDOI{\tempurl}


\bibitem[\protect\citeauthoryear{Weimer, Nguyen, Le~Goues, and Forrest}{Weimer
  et~al\mbox{.}}{2009}]%
        {weimer2009automatically}
\bibfield{author}{\bibinfo{person}{Westley Weimer}, \bibinfo{person}{ThanhVu
  Nguyen}, \bibinfo{person}{Claire Le~Goues}, {and} \bibinfo{person}{Stephanie
  Forrest}.} \bibinfo{year}{2009}\natexlab{}.
\newblock \showarticletitle{Automatically finding patches using genetic
  programming}. In \bibinfo{booktitle}{\emph{Proceedings of the 31st
  International Conference on Software Engineering}}. IEEE,
  \bibinfo{pages}{364--374}.
\newblock
\urldef\tempurl%
\url{https://doi.org/10.1109/ICSE.2009.5070536}
\showDOI{\tempurl}


\bibitem[\protect\citeauthoryear{Wen, Chen, Wu, Hao, and Cheung}{Wen
  et~al\mbox{.}}{2018}]%
        {wen2018context}
\bibfield{author}{\bibinfo{person}{Ming Wen}, \bibinfo{person}{Junjie Chen},
  \bibinfo{person}{Rongxin Wu}, \bibinfo{person}{Dan Hao}, {and}
  \bibinfo{person}{Shing-Chi Cheung}.} \bibinfo{year}{2018}\natexlab{}.
\newblock \showarticletitle{Context-aware patch generation for better automated
  program repair}. In \bibinfo{booktitle}{\emph{Proceedings of the 40th
  International Conference on Software Engineering}}. ACM,
  \bibinfo{pages}{1--11}.
\newblock
\urldef\tempurl%
\url{https://doi.org/10.1145/3180155.3180233}
\showDOI{\tempurl}


\bibitem[\protect\citeauthoryear{Wilcoxon}{Wilcoxon}{1945}]%
        {wilcoxon1945individual}
\bibfield{author}{\bibinfo{person}{F. Wilcoxon}.}
  \bibinfo{year}{1945}\natexlab{}.
\newblock \showarticletitle{Individual Comparisons by Ranking Methods}.
\newblock \bibinfo{journal}{\emph{Biometrics Bulletin}} \bibinfo{volume}{1},
  \bibinfo{number}{6} (\bibinfo{year}{1945}), \bibinfo{pages}{80--83}.
\newblock


\bibitem[\protect\citeauthoryear{Xin and Reiss}{Xin and Reiss}{2017}]%
        {xin2017identifying}
\bibfield{author}{\bibinfo{person}{Qi Xin} {and} \bibinfo{person}{Steven~P
  Reiss}.} \bibinfo{year}{2017}\natexlab{}.
\newblock \showarticletitle{Identifying test-suite-overfitted patches through
  test case generation}. In \bibinfo{booktitle}{\emph{Proceedings of the 26th
  ACM SIGSOFT International Symposium on Software Testing and Analysis}}. ACM,
  \bibinfo{pages}{226--236}.
\newblock
\urldef\tempurl%
\url{https://doi.org/10.1145/3092703.3092718}
\showDOI{\tempurl}


\bibitem[\protect\citeauthoryear{Xiong, Liu, Zeng, Zhang, and Huang}{Xiong
  et~al\mbox{.}}{2018}]%
        {xiong2018identifying}
\bibfield{author}{\bibinfo{person}{Yingfei Xiong}, \bibinfo{person}{Xinyuan
  Liu}, \bibinfo{person}{Muhan Zeng}, \bibinfo{person}{Lu Zhang}, {and}
  \bibinfo{person}{Gang Huang}.} \bibinfo{year}{2018}\natexlab{}.
\newblock \showarticletitle{Identifying patch correctness in test-based program
  repair}. In \bibinfo{booktitle}{\emph{Proceedings of the 40th International
  Conference on Software Engineering}}. ACM, \bibinfo{pages}{789--799}.
\newblock
\urldef\tempurl%
\url{https://doi.org/10.1145/3183519.3183540}
\showDOI{\tempurl}


\bibitem[\protect\citeauthoryear{Xiong, Wang, Yan, Zhang, Han, Huang, and
  Zhang}{Xiong et~al\mbox{.}}{2017}]%
        {xiong2017precise}
\bibfield{author}{\bibinfo{person}{Yingfei Xiong}, \bibinfo{person}{Jie Wang},
  \bibinfo{person}{Runfa Yan}, \bibinfo{person}{Jiachen Zhang},
  \bibinfo{person}{Shi Han}, \bibinfo{person}{Gang Huang}, {and}
  \bibinfo{person}{Lu Zhang}.} \bibinfo{year}{2017}\natexlab{}.
\newblock \showarticletitle{Precise condition synthesis for program repair}. In
  \bibinfo{booktitle}{\emph{Proceedings of the 39th IEEE/ACM International
  Conference on Software Engineering}}. IEEE, \bibinfo{pages}{416--426}.
\newblock
\urldef\tempurl%
\url{https://doi.org/10.1109/ICSE.2017.45}
\showDOI{\tempurl}


\bibitem[\protect\citeauthoryear{Yan, Liu, Niu, Li, Liu, Liu, Klein, and
  Bissyand{\'{e}}}{Yan et~al\mbox{.}}{2022}]%
        {yan2022crex}
\bibfield{author}{\bibinfo{person}{Dapeng Yan}, \bibinfo{person}{Kui Liu},
  \bibinfo{person}{Yuqing Niu}, \bibinfo{person}{Li Li}, \bibinfo{person}{Zhe
  Liu}, \bibinfo{person}{Zhiming Liu}, \bibinfo{person}{Jacques Klein}, {and}
  \bibinfo{person}{Tegawend{\'{e}}~F. Bissyand{\'{e}}}.}
  \bibinfo{year}{2022}\natexlab{}.
\newblock \showarticletitle{Crex: Predicting patch correctness in automated
  repair of {C} programs through transfer learning of execution semantics}.
\newblock \bibinfo{journal}{\emph{Information and Software Technology}}
  \bibinfo{volume}{152} (\bibinfo{year}{2022}), \bibinfo{pages}{107043}.
\newblock
\urldef\tempurl%
\url{https://doi.org/10.1016/j.infsof.2022.107043}
\showDOI{\tempurl}


\bibitem[\protect\citeauthoryear{Yang and Yang}{Yang and Yang}{2020}]%
        {yang2020exploring}
\bibfield{author}{\bibinfo{person}{Bo Yang} {and} \bibinfo{person}{Jinqiu
  Yang}.} \bibinfo{year}{2020}\natexlab{}.
\newblock \showarticletitle{Exploring the Differences between Plausible and
  Correct Patches at Fine-Grained Level}. In
  \bibinfo{booktitle}{\emph{Proceedings of the 2nd International Workshop on
  Intelligent Bug Fixing}}. IEEE, \bibinfo{pages}{1--8}.
\newblock
\urldef\tempurl%
\url{https://doi.org/10.1109/IBF50092.2020.9034821}
\showDOI{\tempurl}


\bibitem[\protect\citeauthoryear{Yang, Zhikhartsev, Liu, and Tan}{Yang
  et~al\mbox{.}}{2017}]%
        {yang2017better}
\bibfield{author}{\bibinfo{person}{Jinqiu Yang}, \bibinfo{person}{Alexey
  Zhikhartsev}, \bibinfo{person}{Yuefei Liu}, {and} \bibinfo{person}{Lin Tan}.}
  \bibinfo{year}{2017}\natexlab{}.
\newblock \showarticletitle{Better test cases for better automated program
  repair}. In \bibinfo{booktitle}{\emph{Proceedings of the 11th Joint Meeting
  on Foundations of Software Engineering}}. ACM, \bibinfo{pages}{831--841}.
\newblock
\urldef\tempurl%
\url{https://doi.org/10.1145/3106237.3106274}
\showDOI{\tempurl}


\bibitem[\protect\citeauthoryear{Ye, Gu, Martinez, Durieux, and Monperrus}{Ye
  et~al\mbox{.}}{2021}]%
        {ye2019automated}
\bibfield{author}{\bibinfo{person}{He Ye}, \bibinfo{person}{Jian Gu},
  \bibinfo{person}{Matias Martinez}, \bibinfo{person}{Thomas Durieux}, {and}
  \bibinfo{person}{Martin Monperrus}.} \bibinfo{year}{2021}\natexlab{}.
\newblock \showarticletitle{Automated Classification of Overfitting Patches
  with Statically Extracted Code Features}.
\newblock \bibinfo{journal}{\emph{IEEE Transactions on Software Engineering}}
  (\bibinfo{year}{2021}).
\newblock


\bibitem[\protect\citeauthoryear{Ye, Martinez, Durieux, and Monperrus}{Ye
  et~al\mbox{.}}{2019b}]%
        {ye2019comprehensive}
\bibfield{author}{\bibinfo{person}{He Ye}, \bibinfo{person}{Matias Martinez},
  \bibinfo{person}{Thomas Durieux}, {and} \bibinfo{person}{Martin Monperrus}.}
  \bibinfo{year}{2019}\natexlab{b}.
\newblock \showarticletitle{A Comprehensive Study of Automatic Program Repair
  on the QuixBugs Benchmark}. In \bibinfo{booktitle}{\emph{Proceedings of the
  1st International Workshop on Intelligent Bug Fixing}}. IEEE,
  \bibinfo{pages}{1--10}.
\newblock
\urldef\tempurl%
\url{https://doi.org/10.1109/IBF.2019.8665475}
\showDOI{\tempurl}


\bibitem[\protect\citeauthoryear{Ye, Martinez, and Monperrus}{Ye
  et~al\mbox{.}}{2019a}]%
        {ye2019automated-dataset}
\bibfield{author}{\bibinfo{person}{He Ye}, \bibinfo{person}{Matias Martinez},
  {and} \bibinfo{person}{Martin Monperrus}.} \bibinfo{year}{2019}\natexlab{a}.
\newblock \showarticletitle{Automated patch assessment for program repair at
  scale}.
\newblock \bibinfo{journal}{\emph{arXiv preprint arXiv:1909.13694}}
  (\bibinfo{year}{2019}).
\newblock


\bibitem[\protect\citeauthoryear{Yu, Cao, Tang, Nie, Huang, and Wu}{Yu
  et~al\mbox{.}}{2020}]%
        {yu2020order}
\bibfield{author}{\bibinfo{person}{Zeping Yu}, \bibinfo{person}{Rui Cao},
  \bibinfo{person}{Qiyi Tang}, \bibinfo{person}{Sen Nie},
  \bibinfo{person}{Junzhou Huang}, {and} \bibinfo{person}{Shi Wu}.}
  \bibinfo{year}{2020}\natexlab{}.
\newblock \showarticletitle{Order Matters: Semantic-Aware Neural Networks for
  Binary Code Similarity Detection}. In \bibinfo{booktitle}{\emph{Proceedings
  of the AAAI Conference on Artificial Intelligence}}.
  \bibinfo{publisher}{{AAAI}}, \bibinfo{pages}{1145--1152}.
\newblock
\urldef\tempurl%
\url{https://doi.org/10.1609/aaai.v34i01.5466}
\showDOI{\tempurl}


\bibitem[\protect\citeauthoryear{Yu, Martinez, Danglot, Durieux, and
  Monperrus}{Yu et~al\mbox{.}}{2019}]%
        {yu2019alleviating}
\bibfield{author}{\bibinfo{person}{Zhongxing Yu}, \bibinfo{person}{Matias
  Martinez}, \bibinfo{person}{Benjamin Danglot}, \bibinfo{person}{Thomas
  Durieux}, {and} \bibinfo{person}{Martin Monperrus}.}
  \bibinfo{year}{2019}\natexlab{}.
\newblock \showarticletitle{Alleviating patch overfitting with automatic test
  generation: a study of feasibility and effectiveness for the Nopol repair
  system}.
\newblock \bibinfo{journal}{\emph{Empirical Software Engineering}}
  \bibinfo{volume}{24}, \bibinfo{number}{1} (\bibinfo{year}{2019}),
  \bibinfo{pages}{33--67}.
\newblock
\urldef\tempurl%
\url{https://doi.org/10.1007/s10664-018-9619-4}
\showDOI{\tempurl}


\bibitem[\protect\citeauthoryear{Zhao and Huang}{Zhao and Huang}{2018}]%
        {zhao2018deepsim}
\bibfield{author}{\bibinfo{person}{Gang Zhao} {and} \bibinfo{person}{Jeff
  Huang}.} \bibinfo{year}{2018}\natexlab{}.
\newblock \showarticletitle{DeepSim: deep learning code functional similarity}.
  In \bibinfo{booktitle}{\emph{Proceedings of the 2018 {ACM} Joint Meeting on
  European Software Engineering Conference and Symposium on the Foundations of
  Software Engineering}}. \bibinfo{pages}{141--151}.
\newblock


\bibitem[\protect\citeauthoryear{Zhou, Shen, and Zhong}{Zhou
  et~al\mbox{.}}{2019}]%
        {zhou2019lancer}
\bibfield{author}{\bibinfo{person}{Shufan Zhou}, \bibinfo{person}{Beijun Shen},
  {and} \bibinfo{person}{Hao Zhong}.} \bibinfo{year}{2019}\natexlab{}.
\newblock \showarticletitle{Lancer: Your Code Tell Me What You Need}. In
  \bibinfo{booktitle}{\emph{Proceedings of the 34th {IEEE/ACM} International
  Conference on Automated Software Engineering}}. \bibinfo{publisher}{{IEEE}},
  \bibinfo{pages}{1202--1205}.
\newblock
\urldef\tempurl%
\url{https://doi.org/10.1109/ASE.2019.00137}
\showDOI{\tempurl}


\end{thebibliography}

\end{document}